\documentclass[aps,prd,twocolumn,preprintnumbers,nofootinbib,superscriptaddress,amsmath]{revtex4-2}

\usepackage{graphicx}
\usepackage{amsmath}
\usepackage{amssymb}
\usepackage{amsfonts}
\usepackage[colorlinks]{hyperref}
\usepackage{url}
\usepackage[dvipsnames]{xcolor}
\usepackage{bm}
\usepackage{array}
\usepackage{placeins}
\usepackage{bbold}
\usepackage{algorithm}
\usepackage{algpseudocode}
\usepackage[normalem]{ulem}
\usepackage{xspace}
\usepackage{orcidlink}
\usepackage{subcaption}
\usepackage[font=small]{caption}
\usepackage{ragged2e}

\definecolor{dodgerblue}{HTML}{1E90FF}
\definecolor{crimson}{HTML}{DC143C}
\definecolor{auwien}{HTML}{8B64DD}

\hypersetup{
    colorlinks=true,
    linkcolor=auwien,
    filecolor=auwien,
    citecolor = auwien,      
    urlcolor=auwien,
}

\newcommand{\ssout}[1]{}

\def\d{\mathrm{d}}
\newcommand{\rmi}{\mathrm{i}}
\newcommand{\rme}{\operatorname{e}}
\allowdisplaybreaks[4]
\newcommand{\sub}[1]{_{\text{#1}}}

\newcommand{\uvec}[1]{\bm{\hat{#1}}}

\newcommand{\ord}[1]{\mathcal{O} \left( #1 \right)}
\newcommand{\D}{\mathcal{D}}
\newcommand{\av}[1]{\left\langle #1 \right\rangle}

\newcommand{\pyEFPE}{\texttt{pyEFPE}\xspace}
\newcommand{\pyEFPEHM}{\texttt{pyEFPEHM}\xspace}
\newcommand{\STfour}{\texttt{SpinTaylorT4}\xspace}

\newcommand{\vfive}{\texttt{SEOBNRv5}\xspace}
\newcommand{\vfiveEHM}{\texttt{SEOBNRv5EHM}\xspace}
\newcommand{\vfivePHM}{\texttt{SEOBNRv5PHM}\xspace}
\newcommand{\TEOBDali}{\texttt{TEOBResumS-Dali}\xspace}

\newcommand{\AEI}{\affiliation{Max Planck Institute for Gravitational Physics (Albert Einstein Institute), D-14476 Potsdam, Germany}}

\newcommand{\IFT}{\affiliation{Instituto de F\'isica Te\'orica UAM/CSIC, Universidad Aut\'onoma de Madrid, Cantoblanco 28049 Madrid, Spain}}

\newcommand{\Bham}{\affiliation{School of Physics and Astronomy and Institute for Gravitational Wave Astronomy, University of Birmingham, Edgbaston, Birmingham, B15 2TT, United Kingdom}}

\newcommand{\Maryland}{\affiliation{Department of Physics, University of Maryland, College Park, MD 20742, USA}
}

\begin{document}
\makeatother

\title{Post-Newtonian inspiral waveform model for eccentric precessing binaries with higher-order modes and matter effects}

\author{Gonzalo Morras \orcidlink{0000-0002-9977-8546}}
\email{gonzalo.morras@aei.mpg.de}
\AEI
\IFT

\author{Geraint Pratten \orcidlink{0000-0003-4984-0775}}
\Bham

\author{Patricia Schmidt \orcidlink{0000-0003-1542-1791}}
\Bham

\author{Alessandra Buonanno \orcidlink{0000-0002-5433-1409}}
\AEI
\Maryland


\date{\today}

\begin{abstract}

We introduce \pyEFPEHM, a post-Newtonian (PN) inspiral waveform model for eccentric and spin-precessing compact binaries that includes higher-order modes and matter effects. Accurate and efficient waveform models capturing these effects are essential for probing compact-binary formation channels and exploiting current and future gravitational-wave (GW) observations.
\pyEFPEHM extends \pyEFPE, significantly improving its physical content and accuracy. In particular, we show that above 2.5PN order the quasi-circular contributions to the orbital phasing dominate at each PN order, and we incorporate all available higher-order quasi-circular PN corrections to the phasing, including adiabatic tidal effects. We generalize the multiple-scale analysis solution of the spin-precession equations, extending it to higher PN orders and including all available quasi-circular corrections. Finally, we add eccentric corrections up to 1PN order in the waveform amplitudes, including the GW multipoles $(l,|m|) = (2,2),(2,1),(2,0),(3,3),(3,2),(3,1),(3,0),(4,4),(4,2),(4,0)$.
We validate \pyEFPEHM against existing analytical waveform models and numerical relativity simulations, showing that it provides a robust and computationally efficient description of the inspiral, with good agreement across a broad region of parameter space and up to close to merger. The accuracy degrades in the late inspiral for systems with very unequal masses ($m_2/m_1 \lesssim 0.1$), significant spins aligned with the orbital angular momentum ($|\chi_\mathrm{eff}| \gtrsim 0.5$), and high eccentricities ($e \gtrsim 0.6$), where the PN expansion is expected to break down.
\pyEFPEHM represents a significant step toward physically complete and efficient waveform modeling of eccentric and precessing binaries, providing a foundation for future extensions including higher-order corrections, calibration to numerical relativity, and merger–ringdown modeling.

\end{abstract}
\maketitle

\section{Introduction}
\label{sec:intro}

Over the coming years, the central goals of GW astronomy include probing strong-field gravity, matter at extreme densities, and the astrophysical formation channels of black holes and neutron stars~\cite{Gair:2012nm,Evans:2021gyd,LISA:2022yao,Harry:2022zey,LISAConsortiumWaveformWorkingGroup:2023arg,ET:2025xjr}. Achieving these objectives in a robust manner requires accurate waveform models that capture the rich physical effects present during the coalescence of compact binaries, such as orbital eccentricity, spin precession, higher-order modes, and matter effects, among others. At the same time, waveform models must be sufficiently computationally efficient to be employed in large-scale data-analysis applications.

As the LIGO-Virgo-KAGRA (LVK) Collaboration~\cite{LIGOScientific:2014pky,VIRGO:2014yos,KAGRA:2018plz} continues to observe with improving sensitivity~\cite{KAGRA:2013rdx} and to expand the catalog of detected compact-binary coalescences into the several hundreds~\cite{LIGOScientific:2018mvr,LIGOScientific:2020ibl,KAGRA:2021vkt,LIGOScientific:2025slb}, the importance of modeling these physical effects is becoming increasingly evident. Evidence for dynamical formation channels has emerged from the observation of very massive black holes with large spins~\cite{LIGOScientific:2020iuh,LIGOScientific:2025rsn} and from lower-mass asymmetric systems with large primary black-hole spins~\cite{LIGOScientific:2025brd}, both suggestive of hierarchical mergers~\cite{Gerosa:2021mno,Alvarez:2024dpd,Li:2025iux}, as well as from evidence of orbital eccentricity in some high-mass binary black hole mergers~\cite{Gayathri:2020coq,Gamba:2021gap,Romero-Shaw:2022xko,Gupte:2024jfe,Planas:2025jny,Romero-Shaw:2025vbc}, based on analyses using aligned-spin eccentric models. Similarly, the identification of an eccentric neutron star--black hole merger~\cite{Morras:2025xfu,Planas:2025plq,Jan:2025fps,Kacanja:2025kpr,Tiwari:2025fua,Phukon:2025cky}, together with population studies showing evidence for spin orientations misaligned with the orbital plane~\cite{LIGOScientific:2025pvj}, suggests that hierarchical triples~\cite{Stegmann:2025clo,Romero-Shaw:2025otx,Stegmann:2025zkb} undergoing Lidov--Kozai oscillations~\citep{Zeipel:1910,Lidov:1962,Kozai:1962} may contribute significantly to the population of compact-binary mergers observed by the LVK.

Looking ahead, next-generation ground-based detectors such as the Einstein Telescope~\cite{ET:2019dnz,Branchesi:2023mws,ET:2025xjr} and Cosmic Explorer~\cite{Reitze:2019iox,Evans:2021gyd}, as well as space-based observatories such as LISA~\cite{LISA:2017pwj,LISA:2024hlh}, will extend the accessible frequency band toward much lower frequencies. This will allow compact binaries to be observed much earlier in their inspiral, where orbital eccentricity is expected to be more prominent, as the binary will have had less time to circularize through GW emission. At the same time, the much higher signal-to-noise ratios anticipated for events observed by these future detectors will make physical effects that are subdominant in current LVK observations, such as spin precession and higher-order modes, increasingly important.

Despite substantial progress in waveform modeling for quasi-circular precessing~\cite{Pompili:2023tna,Ramos-Buades:2023ehm,Estelles:2021gvs,Estelles:2025zah,Hamilton:2025xru,Yoo:2023spi} and eccentric spin-aligned binaries~\cite{Gamboa:2024hli,Paul:2024ujx,Planas:2025feq,Nee:2025nmh,Ramos-Buades:2026kbq}, only a few models currently describe fully general eccentric and precessing systems~\cite{Liu:2023ldr,Morras:2025nlp,Albanesi:2025txj}, and these remain limited in accuracy, physical completeness, and computational efficiency. 

In this context, we introduce \pyEFPEHM, a PN inspiral waveform model for eccentric and precessing binaries with higher-order modes and adiabatic tidal effects. This model extends \pyEFPE~\cite{Morras:2025nlp}, which, despite its more limited physical completeness and accuracy, has already enabled a wide range of applications, including finding the first evidence for orbital eccentricity in a neutron star–black hole merger~\cite{Morras:2025xfu}, simulations of stellar-origin black hole binaries for LISA data challenges~\cite{MojitoLight_description}, and studies of the astrophysical implications of eccentricity in neutron star-black hole binaries~\cite{Romero-Shaw:2025otx}. In addition, \pyEFPE has been adapted by independent groups to explore extensions of general relativity~\cite{Roy:2025xih} and to probe intrinsic neutron star ellipticity in eccentric binaries~\cite{Miao:2025tqu}.

Both \pyEFPE and \pyEFPEHM are based on the Efficient Fully Precessing Eccentric (EFPE) framework developed in Refs.~\cite{Klein:2018ybm,Klein:2021jtd,Arredondo:2024nsl}, which we review in Sec.~\ref{sec:EFPE_primer}. Compared to its predecessor, \pyEFPEHM incorporates several key improvements:

\begin{itemize}
    \item In Sec.~\ref{sec:HighPNQC}, we incorporate high-order quasi-circular PN corrections to the orbital phasing motivated by the observation that, above 2.5PN order, quasi-circular contributions dominate at each PN order. Following this strategy, we include all currently available quasi-circular PN corrections, namely up to 4.5PN order in the non-spinning sector~\cite{Blanchet:2023bwj}, up to 4PN order in the fully spinning spin-orbit and spin-spin sectors~\cite{Cho:2022syn,Khalil:2023kep} (with partial results at 3.5PN order in the fully spinning case), 3.5PN aligned cubic-in-spin effects~\cite{Marsat:2014xea}, 7.5PN adiabatic tidal effects~\cite{Dones:2024odv}, and 6.5PN adiabatic spin-tidal effects~\cite{Abdelsalhin:2018reg}. These additions significantly improve the phasing in the late inspiral and enable applications to binaries containing neutron stars.
    \item In Sec.~\ref{sec:HighPN_MSA} we extend the multiple-scale analysis (MSA) solution of the spin-precession equations, previously available only up to 2PN order, to higher PN orders. Following the same strategy as for the orbital phasing, we further supplement the eccentric 2PN equations used in \pyEFPE with all known quasi-circular corrections, which are currently available up to 4PN order.
    \item In Sec.~\ref{sec:HMs} we include eccentric waveform amplitude corrections up to 1PN order following Ref.~\cite{Morras:2025nbp}. In particular, \pyEFPEHM incorporates the GW multipoles $(l,|m|) = (2,2)$, $(2,1)$, $(2,0)$, $(3,3)$, $(3,2)$, $(3,1)$, $(3,0)$, $(4,4)$, $(4,2)$, and $(4,0)$.
\end{itemize}

In Sec.~\ref{sec:validate} we thoroughly test and validate \pyEFPEHM. We assess its computational performance and compare it to other analytical waveform models, including \pyEFPE, \STfour~\cite{lalsuite_code,Buonanno:2009zt,Sturani:2015STA,Isoyama:2020lls}, \vfiveEHM~\cite{Gamboa:2024hli,Gamboa:2024imd}, \vfivePHM~\cite{Pompili:2023tna,Khalil:2023kep,Ramos-Buades:2023ehm,Estelles:2025zah}, and \TEOBDali~\cite{Nagar:2024oyk,Nagar:2024dzj,Albanesi:2025txj}, through timing and mismatch studies (Sec.~\ref{sec:validate:wf}). We further benchmark the model against numerical relativity simulations in Sec.~\ref{sec:validate:NR}, and perform full Bayesian parameter estimation on simulated data from current ground-based detector networks using both \pyEFPEHM and \vfivePHM injections in Sec.~\ref{sec:validate:PE}. These studies show that \pyEFPEHM provides a robust and computationally efficient description of the inspiral of eccentric and precessing compact binaries, with good agreement with existing analytical waveform models and numerical relativity simulations across a broad region of parameter space and up to close to merger. We find that the accuracy in the late inspiral degrades for systems with very unequal masses ($m_2/m_1 \lesssim 0.1$), significant aligned spins ($|\chi_\mathrm{eff}| \gtrsim 0.5$), and high eccentricities ($e \gtrsim 0.6$), where the PN expansion is expected to break down.

Finally, in Sec.~\ref{sec:conclusion} we conclude, summarizing our main results and discussing possible directions for future work.

Unless otherwise specified, we use geometric units ($G=c=1$) throughout. Vectors are denoted in boldface, with hats indicating unit vectors. Angular momenta are expressed in dimensionless form by rescaling with the total mass squared, $M^2$, so that, for example, $\bm{L} = \bm{L}^\mathrm{physical}/M^2$ and $\bm{S}_i = \bm{S}_i^\mathrm{physical}/M^2$.

\section{Summary of EFPE Waveform Modeling}
\label{sec:EFPE_primer}

The \pyEFPEHM model presented in this work is an extension of the \pyEFPE~\cite{Morras:2025nlp} model, itself based on the Efficient Fully Precessing Eccentric (EFPE) waveform introduced in Ref.~\cite{Klein:2021jtd}. Here, we extend \pyEFPE by incorporating higher-order PN quasi-circular corrections to the phasing and precession equations, as well as higher-order multipoles in the GW emission. These additions significantly improve the accuracy of the model while utilizing the same theoretical framework, described in detail in Ref.~\cite{Morras:2025nlp} and briefly summarized below.

EFPE models are PN in nature. The evolution of the orbital positions and spins of the binary components is obtained by solving the PN equations of motion~\cite{Blanchet:2013haa}, while the GW signal is computed using the multipolar post-Minkowskian formalism matched to the PN source~\cite{Thorne:1980ru}. The computational efficiency of EFPE models relies on the separation of time-scales present during the inspiral of compact binaries. In particular,

\begin{equation}
    T_\mathrm{orb} \! \sim \! \mathcal{O}\left(\!\frac{M}{v^3}\!\right) \! \ll \! T_\mathrm{PA} \! \sim \! T_\mathrm{SP} \! \sim \! \mathcal{O}\left(\!\frac{M}{v^5}\!\right) \! \ll \! T_\mathrm{RR} \! \sim \! \mathcal{O}\left(\!\frac{M}{v^8}\!\right) \! ,
    \label{eq:TimeScales}
\end{equation}

\noindent where $T_\mathrm{orb}$, $T_\mathrm{PA}$, $T_\mathrm{SP}$, and $T_\mathrm{RR}$ denote the orbital, periastron-advance, spin-precession, and radiation-reaction time-scales, respectively. Here $v$ is the characteristic orbital velocity, which satisfies $v \ll 1$ during the inspiral.

Within this framework, the orbital dynamics are approximated using the quasi-Keplerian parametrization. 
A multiple-scale analysis (MSA)~\cite{Bender:1999,Gerosa:2015tea,Chatziioannou:2017tdw,Morras:2025nlp} is then employed to describe the precession of these quasi-Keplerian orbits induced by spin components misaligned with the orbital angular momentum. Finally, radiation reaction is incorporated by computing the slow evolution of the parameters entering the quasi-Keplerian and MSA solutions, such as the eccentricity and the orbital and precession frequencies. While this separation of time-scales is an approximation, couplings between the different scales can be consistently included as perturbative corrections within the EFPE framework.

Building on this conceptual overview, we now turn to the explicit construction of the GW signal. In the EFPE framework, the GW polarizations $h_{+,\times}$ are decomposed in terms of spin-weighted spherical harmonics as~\cite{Thorne:1980ru,Kidder:2007}

\begin{equation}
  h_+ - i h_\times = \sum_{l=2}^\infty \sum_{m=-l}^l h^{lm} ~{}_{-2} Y^{lm}(\Theta, \Phi) \, ,
  \label{eq:hlm_def}
\end{equation}

\noindent where $(\Theta, \Phi)$ are the spherical angles of the wave propagation vector measured in the inertial binary source frame, and ${}_{-2} Y^{lm}$ are the spin-weighted spherical harmonics of spin weight $-2$. For precessing binaries, the inertial-frame modes $h^{lm}$ have a complicated time-dependent structure, and we therefore write them in terms of the simpler co-precessing modes $H^{l m}$~\cite{Schmidt:2012rh}, as 

\begin{equation}
  h^{l m'} \equiv \sum_{m=-l}^l D^l_{m' m}(\phi_z,\theta_L,\zeta) H^{l m} \, ,
  \label{eq:twist-up}
\end{equation}

\noindent where $D^l_{m'm}(\phi_z,\theta_L,\zeta)$ are the Wigner $D$-matrices and $\phi_z$, $\theta_L$ and $\zeta$ are the three Euler angles describing the rotation from the co-precessing frame, that is instantaneously aligned with the direction of the Newtonian orbital angular momentum, to the inertial frame~\cite{Schmidt:2010it}. 
These Euler angles are obtained from the MSA approximation (see Sec.~\ref{sec:HighPN_MSA} for more details). Note that, while the co-precessing-frame modes $H^{lm}$ closely resemble the GW modes of an aligned-spin binary, precession effects are also present in these modes~\cite{Schmidt:2012rh,Boyle:2014ioa,Ramos-Buades:2020noq}. At the 1PN order considered in \pyEFPEHM, the residual precession-induced corrections to the co-precessing-frame modes are negligible. We therefore approximate the modes $H^{lm}$ by those of an aligned-spin binary, which satisfy the mode symmetry

\begin{align}
    H^{l -m} = (-1)^l (H^{l m})^{*}  \, .
    \label{eq:Hl-m_from_Hlm}
\end{align}

Putting together Eqs.~\eqref{eq:hlm_def} and~\eqref{eq:twist-up}, the waveform polarizations in the inertial frame can be written as

\begin{equation}
    h_+ - \rmi h_\times = \sum_{l=2}^\infty \sum_{m=-l}^l A_{l,m} \, \hat{H}^{lm} \, ,
    \label{eq:hpmihc}
\end{equation}

\noindent where we have defined

\begin{subequations}
\begin{align}
    A_{l,m} \equiv & h_0 \sum_{m'=-l}^l {}_{-2}Y^{l m'}(\Theta, \Phi)  D^l_{m'm}(\phi_z,\theta_L,\zeta) \, , \\
    H^{l m} \equiv & h_0 \Hat{H}^{l m} \, , \label{eq:HofHhat} \\
    h_0 \equiv & 4\sqrt{\frac{\pi}{5}} \frac{M\nu}{d_L} (M \omega)^{2/3} \, ,
\end{align}
\end{subequations}

\noindent with $d_L$ the luminosity distance to the binary, $M = m_1 + m_2$ the total mass, $\omega$ the mean orbital angular velocity, and $\nu = m_1 m_2/M^2$ the symmetric mass ratio.

Since the GW polarizations $h_+$ and $h_\times$ are real valued, combining Eq.~\eqref{eq:hpmihc} with its complex conjugate and using the mode symmetry in Eq.~\eqref{eq:Hl-m_from_Hlm}, the two polarizations can be written as

\begin{equation}
  h_{+,\times} = \sum_{l=2}^\infty \sum_{m=-l}^l \mathsf{A}^{+,\times}_{l,m} \hat{H}^{lm} \, ,
  \label{eq:hpc_AH}
\end{equation}

\noindent where

\begin{subequations}
\begin{align}
  \mathsf{A}^+_{l,m} =& \frac{1}{2} \left[ A_{l,m} + (-1)^l (A_{l,-m})^* \right] \, , \\
  \mathsf{A}^\times_{l,m} =& \frac{\rmi}{2} \left[ A_{l,m} - (-1)^l (A_{l,-m})^* \right] \, ,
\end{align}
\label{eq:Apc_lm_def}
\end{subequations}

\noindent which have the property

\begin{equation}
    \mathsf{A}^{+,\times}_{l,-m} = (-1)^l (\mathsf{A}^{+,\times}_{l,m})^{*}
    \label{eq:Apc_lm_mode_symtry}
\end{equation}

To compute the GW polarizations as a function of time, we need expressions for the co-precessing-frame modes $\hat{H}^{lm}$ as functions of time. For a system following quasi-Keplerian orbits, and after factoring out the secular effect of periastron advance, these modes are periodic with the orbital period and can therefore be expressed as a Fourier series,

\begin{equation}
    \hat{H}^{l m} = \rme^{-\rmi m \delta\lambda} \sum_{p=-\infty}^\infty N_p^{l m} \rme^{-\rmi p \ell} \, ,
    \label{eq:Hlm_Nlmp}
\end{equation}

\noindent where $N^{lm}_p$ are the amplitudes of the eccentric harmonics~\cite{Arredondo:2024nsl,Morras:2025nbp}, $\delta\lambda$ is the argument of periastron and $\ell$ the mean anomaly, defined as 

\begin{subequations}
\label{eq:qKforHlm}
\begin{align}
    \delta\lambda \equiv & \lambda - \ell = k \ell \, , \label{eq:qKforHlm:dl} \\
    \ell \equiv & n (t - t_0) \, , \label{eq:qKforHlm:ell}
\end{align}
\end{subequations}

\noindent with $k$ the periastron advance, $\lambda = (1 + k) \ell$ the mean orbital phase, $n = 2 \pi/P$ the mean motion with $P$ the orbital period, and $t_0$ a constant of integration. From Eqs.~\eqref{eq:Hl-m_from_Hlm} and~\eqref{eq:Hlm_Nlmp} we can deduce that 

\begin{align}
    N^{l -m}_p = (-1)^l (N^{l m}_{-p})^{*} \, .  \label{eq:Nl-m_from_Nlm}
\end{align}

Substituting Eq.~\eqref{eq:Hlm_Nlmp} into Eq.~\eqref{eq:hpc_AH}, the GW polarizations can finally be written as
\begin{equation}
  h_{+,\times} = \sum_{l=2}^\infty \sum_{m=-l}^l \sum_{n=-\infty}^\infty \mathcal{A}^{+,\times}_{l,m,n} \rme^{-\rmi(n\lambda + (m-n)\delta\lambda)} \, ,
  \label{eq:hpc_decomp}
\end{equation}

\noindent with 

\begin{equation}
  \mathcal{A}^{+,\times}_{l,m,n} = N^{lm}_{n} \mathsf{A}^{+,\times}_{l,m}  \, .
  \label{eq:Almn}
\end{equation}

From Eqs.~\eqref{eq:Apc_lm_mode_symtry} and~\eqref{eq:Nl-m_from_Nlm}, we can deduce that $\mathcal{A}^{+,\times}_{l,m,n}$ have the property

\begin{equation}
  \mathcal{A}^{+,\times}_{l,-m,n} = (\mathcal{A}^{+,\times}_{l,m,-n})^{*} \, ,
  \label{eq:Almn_mode_symtry}
\end{equation}

\noindent that very simply relates the amplitudes of negative $m$ modes with negative $n$ modes.

Eq.~\eqref{eq:hpc_decomp} provides a convenient representation of the time-domain GW polarizations, in which the waveform is decomposed into modes with slowly varying amplitudes $\mathcal{A}^{+,\times}_{l,m,n}(t)$, driven by spin precession and radiation reaction, and rapidly growing phases $n\lambda + (m-n)\delta\lambda$,that generate oscillations on the orbital time-scale, with frequencies that themselves evolve slowly on the radiation-reaction time-scale. This separation of time-scales allows the frequency-domain waveform to be obtained analytically by approximating the Fourier transform of Eq.~\eqref{eq:hpc_decomp}, given by

\begin{equation}
    \tilde{h}_{+,\times}(f) = \int_{-\infty}^\infty \d t \, h_{+,\times}(t) \rme^{2 \pi \rmi f t} \, .
    \label{eq:FourierTransform_def}
\end{equation}

In our case, the amplitude $\mathcal{A}^{+,\times}_{l,m,n}(t)$ oscillates on the spin-precession time-scale $T_\mathrm{SP} \sim \ord{M/v^5}$, which is slightly shorter than the time-scale of the stationary phase approximation (SPA) of each mode

\begin{equation}
    T^\mathrm{SPA}_{m,n} = \frac{1}{\sqrt{|n\ddot{\lambda} + (m-n)\ddot{\delta\lambda}|}} \, , \label{eq:TmnSPA}
\end{equation}

\noindent which is $\ord{M/v^{5.5}}$. This violates a key assumption of the SPA, namely that the amplitude can be treated as constant over a time interval $T \gtrsim T^\mathrm{SPA}_{m,n}$ around the stationary point, making the SPA potentially inaccurate. To account for the time dependence of the amplitudes, we employ the Shifted Uniform Asymptotics (SUA) method introduced in Ref.~\cite{Klein:2014bua}. Within the SUA framework, the frequency-domain GW polarizations are approximated as

\begin{subequations}
\label{eq:h_SUA}
\begin{align}
    & \tilde{h}_{+,\times} = \sqrt{2\pi}\sum_{l=2}^\infty \sum_{m=-l}^l \sum_{n=-\infty}^\infty \Big\{ T_{m,n} \mathcal{A}^{\mathrm{corr} \, +,\times}_{l,m,n}(t_{m,n})\nonumber\\
    & \qquad \qquad\qquad \times \rme^{\rmi (2 \pi f t_{m,n} - \phi_{m,n}(t_{m,n}) - \pi/4)} \Big\} \, \label{eq:h_SUA:h} \\
    & \phi_{m, n}(t) = n \lambda(t) + (m - n) \delta\lambda(t) \, , \label{eq:h_SUA:phi} \\
    & \dot{\phi}_{m, n}(t_{m,n}) = 2 \pi f \, , \label{eq:h_SUA:t} \\
    & T_{m,n} = \frac{1}{\sqrt{\left|\ddot{\phi}_{m, n}(t_{m,n})\right|}} \, , \label{eq:h_SUA:T} \\
    & \mathcal{A}^{\mathrm{corr} \, +,\times}_{l,m,n}(t_{m,n}) = \sum_{k=-k_\mathrm{max}}^{k_\mathrm{max}} a_{k,k_\mathrm{max}} \mathcal{A}^{+,\times}_{l,m,n}(t_{m,n} + k T_{m,n}) \, ,\label{eq:h_SUA:Acorr}
\end{align}
\end{subequations}

The SUA expression in Eq.~\eqref{eq:h_SUA:h} closely resembles the SPA, with $t_{m,n}$ the usual stationary time obtained by solving Eq.~\eqref{eq:h_SUA:t}, and $T_{m,n}$ the associated SPA time-scale. The key difference is that, in the SUA, the amplitude $\mathcal{A}^{+,\times}_{l,m,n}(t_{m,n})$ is replaced by a corrected amplitude $\mathcal{A}^{\mathrm{corr}\,+,\times}_{l,m,n}(t_{m,n})$, defined in Eq.~\eqref{eq:h_SUA:Acorr}, which accounts for the variation of $\mathcal{A}^{+,\times}_{l,m,n}(t)$ over the SPA time-scale region by sampling it at shifted times $t_{m,n} + k T_{m,n}$. The coefficients $a_{k,k_\mathrm{max}}$ are determined by solving the following linear system of equations~\cite{Klein:2014bua,Morras:2025nlp},

\begin{subequations}
\begin{align}
    \frac{1}{2} a_{0,k_\mathrm{max}} + \sum_{k=1}^{k_\mathrm{max}} a_{k, k_\mathrm{max}} & = \frac{1}{2} & \, ,  & \; p = 0,  \\
    \sum_{k=1}^{k_\mathrm{max}} \frac{(\rmi \, k^2)^p}{(2 p-1)!!} a_{k, k_\mathrm{max}} & = \frac{1}{2} & \, , & \; 1 \leq p \leq k_\mathrm{max} \, , \label{eq:akkmax_eqs_Klein:linearsystem_k>=1}\\
    a_{-k,k_\mathrm{max}}  = a_{k, k_\mathrm{max}} \, . & &  &
\end{align}
\label{eq:akkmax_eqs_Klein}
\end{subequations}

In summary, EFPE models exploit the separation of time-scales present during the inspiral of compact binaries to decompose the GW signal into modes with slowly varying amplitudes and frequencies. This structure enables the efficient construction of waveforms in both the time and frequency domains. Both domains are implemented in \pyEFPEHM.

\section{Adding high-order post-Newtonian quasi-circular corrections to the phasing}
\label{sec:HighPNQC}

PN results for the phasing of eccentric binaries are currently available only up to 2.5PN order for fully spinning binaries and 3PN order for systems with spins aligned with the orbital angular momentum~\cite{Arun:2009mc,Henry:2023tka}. In contrast, for quasi-circular (QC) binaries, the problem simplifies considerably, allowing higher-order PN corrections to be derived. In particular, the phasing of QC binaries is known up to 4.5PN in the non-spinning sector~\cite{Blanchet:2023bwj}, 4PN in the fully spinning spin-orbit and spin-spin sectors~\cite{Cho:2022syn,Khalil:2023kep} (with partial results at 3.5PN in the fully spinning case), 3.5PN for aligned cubic-in-spin effects~\cite{Marsat:2014xea}, 7.5PN in the adiabatic tidal sector~\cite{Dones:2024odv} and 6.5PN in the adiabatic spin-tidal sector~\cite{Abdelsalhin:2018reg}. These QC corrections, which have been included in \pyEFPEHM, are detailed in Appendix~\ref{sec:appendix:PN_QC}. Note that, during the preparation of this work, eccentric tidal corrections up to 7.5PN order were derived in Refs.~\cite{Henry:2025uta,Henry:2026bqh}, however, they are not included in \pyEFPEHM.

In this section, we show that incorporating the quasi-circular limit of the higher-order PN corrections listed above captures the dominant contribution at the corresponding PN order in the phasing of an eccentric binary. Neglecting $\log y$ terms, which vary slowly over the inspiral, the evolution of the mean orbital phase $\lambda$ with respect to the PN parameter $y$ is given by

\begin{equation}
    \frac{\d \lambda}{\d y} = \sum_{n=0}^\infty \sum_{m=0}^\infty \underbrace{c_{n, m} y^{n - 6} e^{2 m}(y)}_{\d\lambda_{n,m}/\d y},
    \label{eq:dlamda_dy_ecc_expanded}
\end{equation}

\noindent where $c_{n, m}$ are coefficients that encode the PN dynamics and $e(y)$ is the eccentricity as a function of the PN parameter

\begin{equation}
    y = \frac{(M \omega)^{1/3}}{\sqrt{1 - e^2}} \, , \label{eq:y_def}
\end{equation}

\noindent which is related to the norm of the Newtonian angular momentum ($L_N = \nu/y$). We describe the evolution of the binary in terms of $y$, since it simplifies the PN evolution equations, removing coordinate singularities when $e \to 1$.

Due to radiation reaction, eccentricity decreases rapidly as the PN parameter increases during the inspiral~\cite{Peters:1963ux}. 
Therefore, eccentric effects in Eq.~\eqref{eq:dlamda_dy_ecc_expanded} are most relevant during the early inspiral, where the PN parameter $y$ is smaller and high-order PN terms are suppressed. As a result, the contribution from eccentric corrections at higher PN orders is suppressed throughout the entire inspiral and cannot accumulate to yield an $\ord{1}$ correction to the phase. To make this argument more precise, we examine how $e$ evolves with $y$.

Although GW emission tends to circularize the orbit, a small residual eccentricity can be induced if the binary components have spins perpendicular to the orbital angular momentum, $\bm{s}_{i \perp}$. This residual eccentricity is approximately given by~\cite{Klein:2010ti,Klein:2018ybm}

\begin{equation}
    e_\mathrm{min}^2 = \frac{5}{304} \left\Vert \bm{s}_{1\perp} - \bm{s}_{2\perp}  \right\Vert^2 y^4  \, .
    \label{eq:emin_spin}
\end{equation}

\noindent For aligned-spin systems (i.e., $\bm{s}_{i \perp} = 0$), eccentricity is expected to decay to zero. At leading (0PN) order, $e$ and $y$ are related by~\cite{Morras:2025nlp}

\begin{equation}
    e^2 = e_0^2 \left(\frac{1 + \frac{121}{304} e^2}{1 + \frac{121}{304} e_0^2} \right)^{-145/121} \left(\frac{y}{y_0}\right)^{-19/3}   \, ,
    \label{eq:e2_0PN_decay}
\end{equation}

\noindent where $e_0$ and $y_0$ are the eccentricity and PN parameter at some initial reference time. 

\begin{figure}[t!]
\centering  
\includegraphics[width=0.5\textwidth]{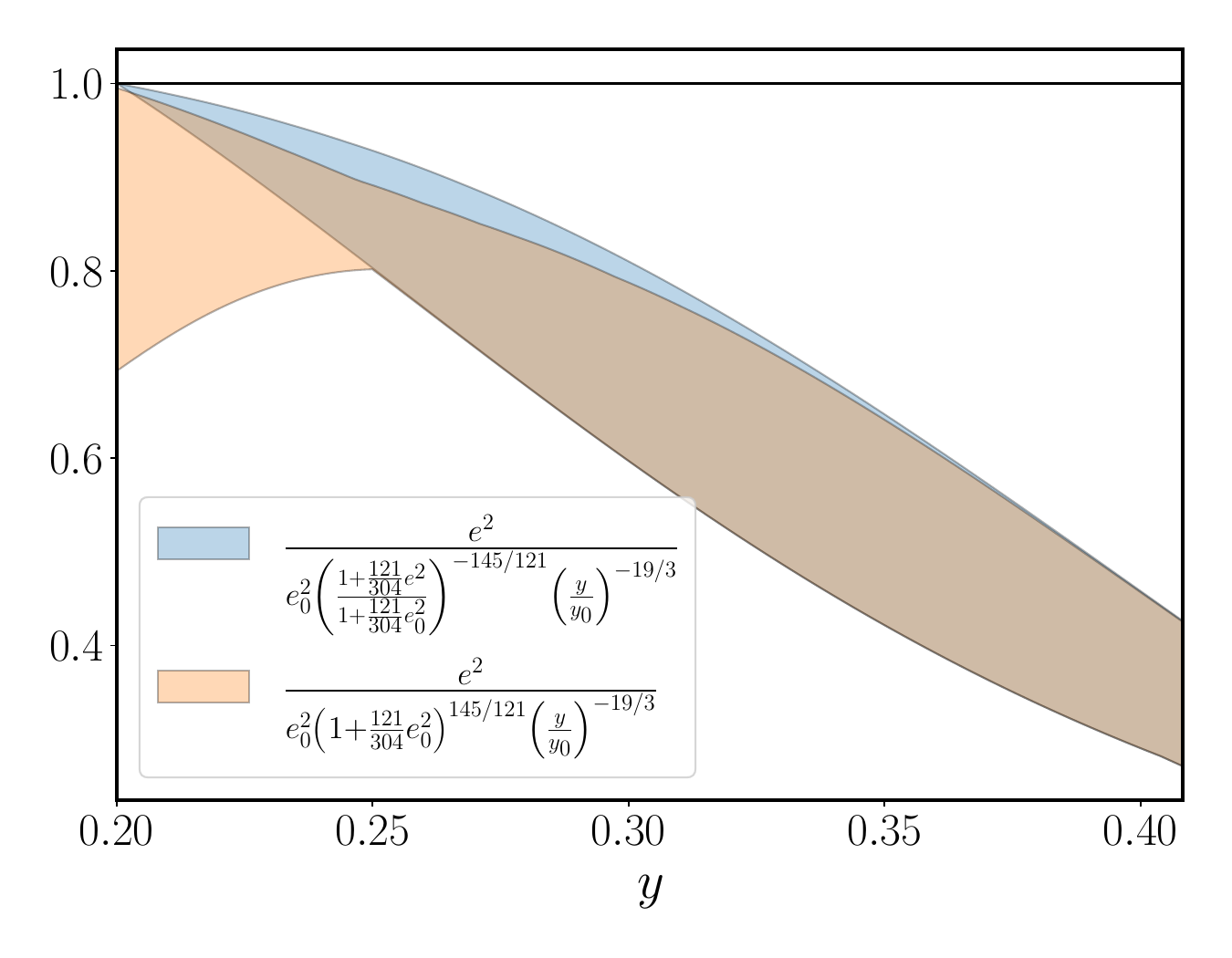}
\caption{\justifying Ratio of $e^2$ obtained from numerical integration of the 3PN evolution equations to (i) the 0PN analytical prediction of Eq.~\eqref{eq:e2_0PN_decay} (blue) and (ii) the upper bound of Eq.~\eqref{eq:e2_decay_UpperBound} (orange). Shaded regions represent the spread over $10^4$ simulations with randomly sampled parameters $e_0^2 \in [0.01, 0.9]$, $q = m_2/m_1 \in [0, 1]$, and $s_{i,z} \in [-1, 1]$.}
\label{fig:e2_decay_3PN_vs_0PN}
\end{figure}

In Fig.~\ref{fig:e2_decay_3PN_vs_0PN} , we compare the 3PN evolution of eccentricity for aligned-spin binaries~\cite{Henry:2023tka} with the 0PN prediction from Eq.~\eqref{eq:e2_0PN_decay}. We find that higher-order PN corrections accelerate the decay of eccentricity relative to the 0PN estimate. Therefore, Eq.~\eqref{eq:e2_0PN_decay} can be used to construct an upper bound on $e^2$ during the inspiral,

\begin{align}
    e^2 \lesssim e_0^2 \left(1 + \frac{121}{304} e_0^2 \right)^{145/121}  \left(\frac{y}{y_0}\right)^{-19/3}  \, ,
    \label{eq:e2_decay_UpperBound}    
\end{align}

\noindent which we also compare against the exact eccentricity evolution in Fig.~\ref{fig:e2_decay_3PN_vs_0PN}. Substituting this upper bound in Eq.~\eqref{eq:dlamda_dy_ecc_expanded}, we obtain an upper bound on the evolution of each term contributing to the total orbital phase

\begin{equation}
    \left| \frac{\d \lambda_{n, m}}{\d y} \right| \lesssim |c'_{n,m}| e^{2 m}_0 y_0^{\frac{19}{3} m} y^{ n - \frac{19}{3} m - 6}  ,
    \label{eq:dlamdanm_dy_ecc_expanded}
\end{equation}

\noindent where $c'_{n,m}$ absorbs $\ord{1}$ numerical constants from the eccentricity bound. This can be directly integrated to give

\begin{equation}
    \left| \Delta \lambda_{n,m} \right| \lesssim \frac{|c'_{n,m}| e^{2 m}_0 y_0^{n - 5}}{5 + \frac{19}{3} m - n}  \left[1 - \left( \frac{y_0}{y_f} \right)^{5 + \frac{19}{3} m  - n} \right] \, ,
    \label{eq:lambda_nm}
\end{equation}

\noindent where $y_f$ is the final PN parameter, which in typical observations satisfies $y_f \gg y_0$. Under this assumption, the contribution to the phase from each term $\Delta \lambda_{n,m}$ behaves as

\begin{align}
    \left| \Delta \lambda_{n,m} \right| \lesssim 
    \begin{cases}
        \frac{|c'_{n,m}|}{5 + \frac{19}{3} m - n} e^{2 m}_0 y_0^{n - 5} & n < 5 + \frac{19}{3} m  \\[0.5em]
        |c'_{n,m}| e^{2 m}_0 y_0^{\frac{19}{3} m} \log{\frac{y_f}{y_0}} & n = 5 + \frac{19}{3} m \\[0.5em]
        \frac{|c'_{n,m}|}{n - 5 - \frac{19}{3} m} e^{2 m}_0 y_0^{\frac{19}{3} m} y_f^{n - 5 - \frac{19}{3} m} & n > 5 + \frac{19}{3} m  
    \end{cases}  \, .
    \label{eq:lambda_nm_y0_ll_yf}
\end{align}

Therefore, for quasi-circular corrections ($m = 0$), terms below 2.5PN order ($n < 5$) scale as $y_0^{n - 5}$. Since $y_0 \ll 1$, these contributions to the phase can be large. Meanwhile, QC terms above 2.5PN order ($n > 5$) scale as $y_f^{n - 5}$ and thus can become significant near merger, where $y_f \sim 0.4$~\cite{Cabero:2016ayq}.

On the other hand, for eccentric corrections ($m \geq 1$), terms up to 5.5PN ($n < 11$) scale as $y_0^{n - 5}$. As before, those below 2.5PN order can have large contributions, while those at 3PN and higher are now heavily suppressed due to the smallness of $y_0$. For terms above 5.5PN, which are required to accurately model adiabatic tidal deformability effects~\cite{Flanagan:2007ix,Dones:2024odv}, their contribution to the phase scales as $y_0^{n - 5}$ if $n < 5 + \frac{19}{3} m$, and as $y_0^{\frac{19}{3} m} y_f^{n - 5 - \frac{19}{3} m}$ if $n > 5 + \frac{19}{3} m$. In both cases, the phase contribution is strongly suppressed by a large power of $y_0$, scaling as $y_0^{\min\left(n - 5, \frac{19}{3} m\right)}$.

\section{Extending the precession description to higher post-Newtonian orders}
\label{sec:HighPN_MSA}

\subsection{High post-Newtonian order precession equations}
\label{sec:HighPN_MSA:Prec_Eqs}

To describe the spin-induced precession of a compact binary, we track the evolution of the component spins, $\bm{s}_1$ and $\bm{s}_2$, and of the direction of the Newtonian orbital angular momentum, $\uvec{l}_N$~\cite{Schmidt:2010it,OShaughnessy:2011pmr,Boyle:2011gg,Schmidt:2012rh}. 
Neglecting tidal torquing effects~\cite{Poisson:1994yf,Tagoshi:1997jy,Alvi:2001mx,Chatziioannou:2012gq,Saketh:2022xjb}, the norms of the component spins, $s_1$ and $s_2$, are conserved. 
Since $\uvec{l}_N$ is a unit vector, its norm is also conserved by definition. 
It then follows that their evolution equations can be written as~\cite{Barker:1975ae,Racine:2008qv,Klein:2018ybm,Klein:2021jtd,Colin:2026iry}

\begin{subequations}
\label{eq:raw_prec_eqs}
\begin{align}
\D \uvec{l}_N &= y^6 \bm{\Omega}_{l_N} \times \uvec{l}_N \, , \label{eq:raw_prec_eqs:lN} \\
\D \bm{s}_1 &= \mu_2 y^5 \bm{\Omega}_1 \times \bm{s}_1 \, ,  \label{eq:raw_prec_eqs:s1} \\
\D \bm{s}_2 &= \mu_1 y^5 \bm{\Omega}_2  \times \bm{s}_2 \, ,  \label{eq:raw_prec_eqs:s2}
\end{align}
\end{subequations}

\noindent where~\cite{Klein:2021jtd}

\begin{subequations}
\label{eq:raw_prec_eqs_defs}
\begin{align}
\D &= \frac{M}{\left(1-e^2 \right)^{3/2}} \frac{\d}{\d t} \, , \\
\mu_i &= \frac{m_i}{M} \, , \\
\bm{s}_i &= \frac{\bm{S}_i}{\mu_i} \, .
\end{align}
\end{subequations}

In the eccentric case, the precession frequencies in Eq.~\eqref{eq:raw_prec_eqs} are currently known only up to next-to-leading order (NLO)~\cite{Racine:2008qv}, which we denote as 2PN in order to match the PN counting used for the orbital phase evolution.

For quasi-circular binaries, however, the precession frequencies are known up to 4PN order~\cite{Bohe:2012mr,Sturani:2015STA,Akcay:2020qrj,Khalil:2023kep}. 
Using the same arguments as in Sec.~\ref{sec:HighPNQC}, eccentric corrections are relatively suppressed at high PN orders, and the quasi-circular contributions provide the dominant terms. 
Nevertheless, the 2.5PN eccentric corrections accumulate over long inspirals, as shown in Sec.~\ref{sec:HighPNQC}, and the quasi-circular approximation therefore becomes less accurate. 
For this reason, deriving the 2.5PN and 3PN eccentric corrections to the precession equations is important. 

To ensure that the discussion in this section remains applicable when higher-order PN corrections become available, or when calibration parameters are introduced, we keep the PN coefficients general and provide the explicit expressions in Appendix~\ref{sec:appendix:PN_prec_expressions}. 
In particular, the explicit form of the precession frequencies in Eq.~\eqref{eq:raw_prec_eqs} is given in Eq.~\eqref{eq:prec_freqs}.

At 2.5PN order and beyond, one finds in the PN literature~\cite{Akcay:2020qrj,Khalil:2023kep} that

\begin{equation}
    \uvec{l}_N \cdot (\D \uvec{l}_N)^\mathrm{PN} \neq 0 \, ,
    \label{eq:Dlnraw}
\end{equation}

\noindent implying an apparent time variation of the norm of $\uvec{l}_N$, in contradiction with its definition as a unit vector. 
Following Ref.~\cite{Akcay:2020qrj}, we resolve this inconsistency by removing the component of $(\D \uvec{l}_N)^\mathrm{PN}$ parallel to $\uvec{l}_N$, i.e.

\begin{equation}
    \D \uvec{l}_N = (\D \uvec{l}_N)^\mathrm{PN} - \uvec{l}_N \cdot (\D \uvec{l}_N)^\mathrm{PN} \, .
    \label{eq:Dln_of_Dlnraw}
\end{equation}

The Multiple Scale Analysis (MSA) solution to the precession equations employed in \pyEFPE~\cite{Morras:2025nlp,Klein:2021jtd} strongly relies on the structure of the 2PN spin-precession equations. 
Here, we generalize this framework to higher PN orders.

We assume that one can construct a combination of $\bm{s}_1$, $\bm{s}_2$, and $\uvec{l}_N$ that is approximately conserved and has the form

\begin{equation}
    \bm{J} = a_J \uvec{l}_N  + b_J (\bm{s}_1 + \bm{s}_2) + c_J (\bm{s}_1 - \bm{s}_2) \, .
    \label{eq:J_def}
\end{equation}

Different choices of this vector are possible, and determining the optimal one is left for future work. 
In \pyEFPEHM, we choose $\bm{J}$ to be the PN total angular momentum~\cite{Khalil:2023kep}, and the coefficients $a_J$, $b_J$, and $c_J$ are given in Eq.~\eqref{eq:J_vec_coefs} of Appendix~\ref{sec:appendix:PN_prec_expressions}.

As in \pyEFPE, it is convenient to describe the system using scalar quantities rather than vectorial degrees of freedom. In particular, we use the effective inspiral spin parameter~\cite{Damour:2001tu,Racine:2008qv,Ajith:2009bn}

\begin{equation}
    \chi_\mathrm{eff} = \uvec{l}_N \cdot(\bm{s}_1 + \bm{s}_2) \, ,
    \label{eq:chi_eff_def}
\end{equation}

\noindent the reduced aligned-spin difference

\begin{align}
\delta\chi &= \uvec{l}_N \cdot \left( \bm{s}_1 - \bm{s}_2 \right) \, ,
\label{eq:dchi_def}
\end{align}

\noindent and the azimuthal angle $\phi_z$ of $\uvec{l}_N$ in a frame aligned with $\bm{J}$, that can be computed as

\begin{equation}
   \D \phi_z = \frac{(\D \uvec{l}_N ) \cdot ( \uvec{\jmath} \times \uvec{l}_N )}{1 - (\uvec{\jmath} \cdot \uvec{l}_N)^2} \, ,  \label{eq:Dphiz_raw}
\end{equation}

\noindent where $\uvec{\jmath} = \bm{J}/J$ is the unit vector along $\bm{J}$, which is approximately conserved. Given $\bm{J}$, $\chi_\mathrm{eff}$, $\delta\chi$, and $\phi_z$, the six degrees of freedom of $\bm{s}_1$, $\bm{s}_2$, and $\uvec{l}_N$ can be reconstructed, since their norms are conserved. The evolution equations for $\chi_\mathrm{eff}$ and $\delta\chi$ follow from differentiating Eqs.~(\ref{eq:chi_eff_def},\ref{eq:dchi_def}) and substituting Eq.~\eqref{eq:raw_prec_eqs}, yielding

\begin{subequations}
\label{eq:eff_spins_prec_eqs}
\begin{align}
\D \chi_\mathrm{eff} & = 3 y^8 A_\chi \; \uvec{l}_N \cdot (\bm{s}_1 \times \bm{s}_2) + \nu y^{11} B_\chi \, , \label{eq:eff_spins_prec_eqs:chi} \\
\D \delta\chi & = 3 y^6 A_{\delta \chi} \; \uvec{l}_N \cdot (\bm{s}_1 \times \bm{s}_2) + \nu y^{11} B_{\delta \chi} \, , \label{eq:eff_spins_prec_eqs:dchi} 
\end{align}
\end{subequations}

\noindent where $A_\chi$, $B_\chi$, $A_{\delta\chi}$, and $B_{\delta\chi}$ are $\ord{y^0}$ constants. Their explicit PN expressions are given in Eq.~\ref{eq:eff_spins_prec_eqs_coefs} of Appendix~\ref{sec:appendix:PN_prec_expressions}.

\subsection{Approximate solutions for $\chi_\mathrm{eff}$ and $\delta\chi$}
\label{sec:HigherPN_MSA:chi_eff_dchi}

Beyond 2PN order, $\chi_\mathrm{eff}$ is no longer conserved. Nevertheless, from Eq.~\eqref{eq:eff_spins_prec_eqs:dchi} one finds that, at 3.5PN order,

\begin{align}
    \frac{\d \chi_\mathrm{eff}}{\d \delta \chi} =& \frac{\D \chi_\mathrm{eff}}{\D \delta \chi} = y^2 \frac{A_\chi}{A_{\delta\chi}} + \ord{y^5} \, .
    \label{eq:Dchi_Ddchi}
\end{align}

This equation can be integrated analytically if $y$, $A_\chi$, and $A_{\delta \chi}$ are treated as constants. This constitutes an approximation, since these quantities evolve due to radiation reaction and because $A_\chi$ and $A_{\delta \chi}$ depend on the component spins, which themselves evolve through spin precession. 
Unless otherwise stated, we neglect these effects in this section and assume that all quantities are constant except for the explicitly evolving $\chi_\mathrm{eff}$ and $\delta\chi$. 
Spin-dependent terms appearing inside these ``constants'' are replaced by their precession-averaged values, described in Sec.~\ref{sec:HighPN_MSA:RR:SpinAvg}. 
We find that this approximation has a small impact on the final solution and allows us to extend the results of Refs.~\cite{Morras:2025nlp,Klein:2021jtd} to higher PN orders. 
Under these assumptions, Eq.~\eqref{eq:Dchi_Ddchi} integrates to

\begin{subequations}
\label{eq:chi_eff_of_dchi}
\begin{align}
\chi_\mathrm{eff} & =\chi_{\mathrm{eff},0} + k_\chi (\delta \chi - \delta \chi_0) \, , \label{eq:chi_eff_of_dchi:chi_eff_of_dchi} \\
k_\chi & = y^2 \frac{A_\chi}{A_{\delta\chi}} \, .\label{eq:chi_eff_of_dchi:kchi} 
\end{align}
\end{subequations}

Thus, once $\delta\chi$ is known, $\chi_\mathrm{eff}$ follows directly. We therefore focus on Eq.~\eqref{eq:eff_spins_prec_eqs:dchi}. 
We again neglect the 4PN contribution proportional to $B_{\delta \chi}$. 
Although the triple product $\uvec{l}_N \cdot (\bm{s}_1 \times \bm{s}_2)$ is in general difficult to compute, its square can be expressed using the identity

\begin{align}
    \big[\bm{v}_1 \cdot (\bm{v}_2 \times \bm{v}_3) \big]^2 = \det \left\{ \begin{pmatrix}
        \bm{v}_1 \cdot \bm{v}_1 & \bm{v}_1 \cdot \bm{v}_2 & \bm{v}_1 \cdot \bm{v}_3 \\
        \bm{v}_2 \cdot \bm{v}_1 & \bm{v}_2 \cdot \bm{v}_2 & \bm{v}_2 \cdot \bm{v}_3 \\
        \bm{v}_3 \cdot \bm{v}_1 & \bm{v}_3 \cdot \bm{v}_2 & \bm{v}_3 \cdot \bm{v}_3
    \end{pmatrix} \right\} \, ,
    \label{eq:triple_prod_identity}
\end{align}

\noindent to write

\begin{align}
    & (\D \delta\chi)^2 = 9 y^{12} A_{\delta \chi}^2 \; \big[\uvec{l}_N \cdot (\bm{s}_1 \times \bm{s}_2) \big]^2 \nonumber \\
    &= 9 y^{12} A_{\delta \chi}^2  \Big[s_1^2 s_2^2 - (\bm{s}_1 \cdot \bm{s}_2)^2 - \frac{s_1^2 + s_2^2 - 2 (\bm{s}_1 \cdot \bm{s}_2)}{4} \chi_\mathrm{eff}^2  \nonumber\\
    &  \quad\quad\quad\quad\quad + \frac{s_1^2 - s_2^2}{2} \chi_\mathrm{eff} \delta \chi - \frac{s_1^2 + s_2^2 + 2 (\bm{s}_1 \cdot \bm{s}_2)}{4} \delta\chi^2 \Big] \, , \label{eq:Dchi2_scalars}
\end{align}

To solve this equation, we express $\bm{s}_1 \cdot \bm{s}_2$ in terms of $\delta\chi$. This relation is obtained by squaring Eq.~\eqref{eq:J_def} and solving for $\bm{s}_1 \cdot \bm{s}_2$. Following Ref.~\cite{Morras:2025nlp}, and in order to avoid large numerical cancellations, we introduce $\Delta_{J^2}$ through

\begin{align}
    J^2 =& (a_J + b_J \chi_{\mathrm{eff},0} + c_J \delta\chi_0)^2 + (b_J + c_J)^2 s_{1\perp,0}^2 \nonumber\\
     &+ (b_J - c_J)^2 s_{2\perp,0}^2 + (b_J^2 - c_J^2) \Delta_{J^2}
    \label{eq:J2_of_DJ2}    
\end{align}

\noindent where

\begin{align}
s_{i\perp,0}^2 & = \left\Vert \bm{s}_{i\perp,0}\right\Vert^2 = \left\Vert \bm{s}_{i,0} - (\uvec{l}_{N,0} \cdot \bm{s}_{i,0})\uvec{l}_{N,0} \right\Vert^2 \, ,
\end{align}

\noindent and the subscript ``0'' denotes initial values. Consequently,

\begin{equation}
    \Delta_{J^2,\mathrm{ini}} = 2 \bm{s}_{1\perp,0} \cdot \bm{s}_{2\perp,0} \, .
    \label{eq:DJ20_def}
\end{equation}

Equating the square of Eq.~\eqref{eq:J_def} to Eq.~\eqref{eq:J2_of_DJ2} and solving for $\bm{s}_1 \cdot \bm{s}_2$, one finds

\begin{align}
    \bm{s}_1 \cdot \bm{s}_2 =& \frac{\chi_{\mathrm{eff},0}^2 - \delta \chi_0^2}{4} + \frac{\Delta_{J^2}}{2} \nonumber\\
    & - \frac{a_J}{b_J^2 - c_J^2} \big[b_J (\chi_\mathrm{eff} - \chi_{\mathrm{eff},0}) + c_J (\delta\chi - \delta\chi_0) \big] \, .
    \label{eq:s1s2_of_DJ2}    
\end{align}

\noindent substituting Eq.~\eqref{eq:chi_eff_of_dchi} and Eq.~\eqref{eq:s1s2_of_DJ2} into Eq.~\eqref{eq:Dchi2_scalars} yields

\begin{align}
    (\D \delta\chi)^2 =& \frac{9}{4} y^{12} A_{\delta \chi}^2 \big[a_{c T} (\delta\chi - \delta\chi_0)^3 - b_{cT} (\delta\chi - \delta\chi_0)^2 + \nonumber\\
    & + c_{c \perp} (\delta\chi - \delta\chi_0) + d_{c \perp} \big] \, ,
    \label{eq:Ddchi2_ddchi}
\end{align}

\noindent where the coefficients of the cubic polynomial are defined as

\begin{subequations}
\label{eq:cubic_coef_defs}
\begin{align}
    a_{cT} =& (1 - k_\chi^2) \kappa_c \label{eq:cubic_coef_defs:acT} \, ,\\
    b_{cT} =& \kappa_c^2 - 2 \kappa_c \delta \Tilde{\chi}_0 + \Tilde{\chi}_{\mathrm{eff},0}^2 + (1 - k_\chi)^2 s_{1 \perp,0}^2  \nonumber\\
    & + (1 + k_\chi)^2 s_{2 \perp,0}^2 + (1 - k_\chi^2) \Delta_{J^2} \, , \label{eq:cubic_coef_defs:bcT} \\
    c_{c\perp} =& 2 \big[(\kappa_c - \delta \Tilde{\chi}_0) \Delta_{J^2} + (1 - k_\chi) (\chi_0 - \delta \chi_0) s_{1 \perp,0}^2 \nonumber\\
    &- (1 + k_\chi) (\chi_0 + \delta\chi_0) s_{2 \perp,0}^2 \big] \, , \label{eq:cubic_coef_defs:ccp} \\
    d_{c\perp} =& 4 s_{1 \perp,0}^2 s_{2 \perp,0}^2 - \Delta_{J^2}^2 \label{eq:cubic_coef_defs:dcp} \, ,
\end{align}
\end{subequations}

\noindent and we have introduced the following variables to simplify the expression

\begin{subequations}
\label{eq:supp_cubic_coef_defs}
\begin{align}
    \kappa_c =& \frac{2 a_J \Tilde{c}_J}{b_J^2 - c_J^2} \label{eq:supp_cubic_coef_defs:kappac} \, , \\
    \Tilde{c}_J =& c_J + k_\chi b_J \, , \\
    \Tilde{b}_J =& b_J + k_\chi c_J \, , \\
    \Tilde{\chi}_{\mathrm{eff},0} =& \chi_{\mathrm{eff},0} - k_\chi \delta\chi_0 \, , \\
    \delta\Tilde{\chi}_0 =& \delta\chi_0 - k_\chi \chi_{\mathrm{eff},0} \, .
\end{align}
\end{subequations}

The right-hand side of Eq.~\eqref{eq:Ddchi2_ddchi} is a cubic polynomial, which can be factorized as

\begin{align}
    \left( \D \delta \chi \right)^2 & = \left( \frac{3}{2} A_{\delta\chi} y^{6} \right)^2 \nonumber\\
& \quad \times ( \delta \chi - \delta\chi_+ ) ( \delta \chi - \delta\chi_- ) (a_{cT} \delta\chi  - \delta \chi_3) \, ,   \label{eq:Dchidsq_poly} \\
    \delta\chi_- &\leq \delta\chi_+ \leq \frac{\delta\chi_3}{a_{cT}} \, , \label{eq:Dchidsq_roots}
\end{align}

Assuming the coefficients to be constant, the solution of this differential equation is~\cite{Klein:2021jtd}

\begin{align}
 \delta \chi &= \delta \chi_- + (\delta \chi_+ - \delta \chi_-) \text{sn}^2 (\psi\sub{p}; m) \, , \label{eq:dchi_sol}
\end{align}

\noindent where 

\begin{subequations}
\label{eq:m_Dpsip_defs}
\begin{align}
 m &= \frac{a_{cT} (\delta \chi_+ - \delta \chi_-)}{\delta \chi_3 - a_{cT} \delta \chi_-} \, , \\
 \D \psi\sub{p} &= \frac{3 A y^6}{4} \sqrt{\delta \chi_3 - a_{cT} \delta \chi_-} \label{eq:Dpsi_p}\, ,
\end{align}    
\end{subequations}

\noindent and $\text{sn}(\psi\sub{p}; m) = \sin(\mathrm{am}(\psi\sub{p}; m))$ is the Jacobi elliptic sine function, with $\mathrm{am}(\psi\sub{p}; m)$ being the Jacobi amplitude. We use the same conventions for the elliptic functions and integrals as in Ref.~\cite{Klein:2021jtd}. In Eq.~\eqref{eq:dchi_sol} we observe that we obtain a solution for $\delta\chi$ that is exactly the same as in \pyEFPE~\cite{Klein:2021jtd,Morras:2025nlp}, with the higher PN corrections modifying the values of the roots of the polynomial $\delta \chi_+$, $\delta \chi_-$ and $\delta \chi_3$ and the prefactor $A_{\delta \chi}$. To find the actual roots of the cubic polynomial we start by computing

\begin{subequations}
\label{eq:depressed_cubic}
\begin{align}
p & =  \frac{b_{cT}^2}{3} - a_{cT} c_{c \perp}  \, , \label{eq:depressed_cubic:p} \\
q &= -\frac{2 b_{cT}^3}{27} + \frac{a_{cT} b_{cT} c_{c \perp}}{3}  + a_{cT}^2 d_{c \perp} \, ,\label{eq:depressed_cubic:q}
\end{align}
\end{subequations}

\noindent and then, the roots of the cubic polynomial can be written as

\begin{subequations}    
\begin{align}
 \delta \chi_3 &= Y_3 - dY + a_{cT}\delta \chi_0 \, , \\
 \delta \chi_\pm &= \frac{Y_\pm - dY}{a_{cT}} + \delta \chi_0 \, ,
\end{align}
\end{subequations}

\noindent with

\begin{subequations}
\label{eq:Ydefs}
\begin{align}
    Y_3 =& 2 \sqrt{\frac{p}{3}} \cos\left[ \frac{\arg(G)}{3} \right] \, , \\
    Y_\pm =& 2 \sqrt{\frac{p}{3}} \cos\left[ \frac{\arg(G) \mp 2 \pi}{3} \right] \, , \\
    G =& -\frac{q}{2} + \rmi \left[ \left( \frac{p}{3} \right)^3 - \left( \frac{q}{2} \right)^2 \right]^{1/2} \nonumber \\
      =& -\frac{q}{2} + \rmi a_{c T} \Bigg[\frac{b_{cT}^3 d_{c\perp}}{27}+\frac{b_{cT}^2 c_{c\perp}^2}{108}-\frac{b_{cT} a_{cT} c_{c\perp} d_{c\perp}}{6} \nonumber\\
      & -\frac{a_{cT}^2 d_{c\perp}^2}{4}-\frac{a_{cT} c_{c\perp}^3}{27}\Bigg]^{1/2} \, , \label{eq:Ydefs:G_simplified} \\
    dY =& \frac{b_{cT}}{3} \, .
\end{align}
\end{subequations}

Note that, similarly as in Ref.~\cite{Morras:2025nlp}, in Eq.~\eqref{eq:Ydefs:G_simplified} we have simplified the square root term to avoid the large numerical cancellations. Using the definitions of Eq.~\eqref{eq:Ydefs}, we can simplify Eqs.~\eqref{eq:m_Dpsip_defs}, to obtain

\begin{subequations}
\label{eq:m_Dpsip_simple}
\begin{align}
 m &= \frac{Y_+ - Y_-}{Y_3 - Y_-}  = \frac{\sin\left[ \frac{\arg\left(G \right)}{3}\right]}{\cos\left[\frac{\arg\left(G \right)}{3} - \frac{\pi}{6}\right]}. \label{eq:ellip_m_of_argG}  \\
 \D \psi\sub{p} &= \frac{3}{4} A_{\delta\chi} y^6 \sqrt{ Y_3 - Y_- },  \label{eq:Dpsip_Y}
\end{align}    
\end{subequations}

As in \pyEFPE, it is convenient to write the solution of $\delta\chi$ of Eq.~\eqref{eq:dchi_sol} as
\begin{equation}
    \delta \chi = \delta\chi\sub{av} - \delta\chi\sub{diff} \left(1 - 2 \text{sn}^2 (\psi\sub{p} , m)\right) \, ,
    \label{eq:dchi_sol_avdiff}
\end{equation}
\noindent where
\begin{subequations}
\label{eq:dchiavdiff}
\begin{align}
\delta\chi\sub{av}   &= \frac{\delta\chi_+ + \delta\chi_-}{2}  = \delta \chi_0 + \frac{Y_+ + Y_- - 2 dY}{2 a_{cT}} \, , \label{eq:dchiavdiff:dchiav} \\
\delta\chi\sub{diff} & = \frac{\delta\chi_+ - \delta\chi_-}{2} =  \frac{Y_+ - Y_-}{2 a_{cT}}\, . \label{eq:dchiavdiff:dchidiff}
\end{align}
\end{subequations}

\subsection{Approximate solutions for the Euler angles}

As in \pyEFPE, we parametrize the rotation between the co-precessing frame aligned with $\uvec{l}_N$ and the approximately inertial frame aligned with $\uvec{\jmath}$ using the Euler angles $\phi_z$, $\theta_L$, and $\zeta$. The latter frame is approximately inertial since $\uvec{\jmath}$ is approximately conserved. The first Euler angle, $\phi_z$, was introduced in Eq.~\eqref{eq:Dphiz_raw}, while the second Euler angle, $\theta_L$, is the angle between $\uvec{l}_N$ and $\uvec{\jmath}$, i.e.

\begin{equation}
    \cos{\theta_L} = \uvec{l}_N \cdot \uvec{\jmath} = \frac{a_J + b_J \chi_\mathrm{eff} + c_J \delta\chi}{J} \, ,
    \label{eq:costhL}
\end{equation}

\noindent and the third Euler angle, $\zeta$, is determined by the minimal rotation condition~\cite{Buonanno:2002fy,Boyle:2011gg}

\begin{equation}
    \D \zeta = - \cos \theta_L \D \phi_z \, .
    \label{eq:Dzeta_minrot}
\end{equation}

Substituting Eq.~\eqref{eq:raw_prec_eqs:lN} into Eq.~\eqref{eq:Dphiz_raw} yields

\begin{equation}
    \D\phi_z = y^6 J \frac{(\bm{\Omega}_{l_N}\cdot \bm{J}) - (\bm{\Omega}_{l_N} \cdot \uvec{l}_N)(\bm{J} \cdot \uvec{l}_N)}{J^2 - (\bm{J} \cdot \uvec{l}_N)^2} \, ,
    \label{eq:Dphiz_OmegalN}
\end{equation}

\noindent which can be evaluated using Eq.~\eqref{eq:J_def} together with

\begin{equation}
    \bm{\Omega}_{l_N} = \omega_s (\bm{s}_1 + \bm{s}_2) + \omega_\delta (\bm{s}_1 - \bm{s}_2) \, ,
    \label{eq:OmegalN_for}
\end{equation}

\noindent where $\omega_s$ and $\omega_\delta$ follow from Eq.~\eqref{eq:prec_freqs:lN} in Appendix~\ref{sec:appendix:PN_prec_expressions}, neglecting the 4PN terms proportional to $(\bm{s}_1 + \bm{s}_2) \times \uvec{l}_N$ and $(\bm{s}_1 - \bm{s}_2) \times \uvec{l}_N$. One then finds

\begin{subequations}
\label{eq:Dphiz}
\begin{align}
\D \phi_z =& y^6 \left\{ A_{\D\phi,J} J +  A_{\D\phi,\Pi} \left[ \frac{N_-}{D_-} + \frac{N_+}{D_+} \right] \right\} \, , \\
A_{\D\phi,J} = & \frac{\omega_\delta + k_\chi \omega_s}{\tilde{c}_J} \\
A_{\D\phi,\Pi} =& \frac{1}{2} \frac{c_J \omega_s - b_J \omega_\delta}{(b_J^2 - c_J^2) \tilde{c}_J} \\
N_\pm =& (J \pm J_{\parallel,0}) \bigg\{\tilde{b}_J (J \pm J_{\parallel,0}) \nonumber\\
& \mp \frac{1}{2} (1 + k_\chi) (b_J + c_J)^2 (\chi_{\mathrm{eff},0} + \delta\chi_0) \nonumber\\
& \mp \frac{1}{2} (1 - k_\chi) (b_J - c_J)^2 (\chi_{\mathrm{eff},0} - \delta\chi_0) \bigg\} \nonumber\\
& - \tilde{c}_J \big[ (b_J + c_J)^2 s_{1\perp,0}^2 - (b_J - c_J)^2 s_{2\perp,0}^2\big]\, , \\
D_\pm =& 2 J (1 \pm \cos{\theta_L}) = 2[J \pm (a_J + b_J \chi_\mathrm{eff} + c_J \delta\chi)]  \nonumber \\
=& B_\pm - C_\pm \left[1 - 2 \text{sn}^2 (\psi\sub{p}, m) \right] \, , \label{eq:Dphiz:Dpm} \\
B_\pm =& J \pm J_{\parallel,0} \pm \tilde{c}_J (\delta\chi_\mathrm{av} - \delta \chi_0) \, , \\
C_\pm =& \pm \tilde{c}_J \delta\chi\sub{diff} \, , \\
J_{\parallel,0} =& a_J + b_J \chi_{\mathrm{eff},0} + c_J \delta\chi_0 \, ,
\end{align}
\end{subequations}

As in Refs.~\cite{Klein:2021jtd,Morras:2025nlp}, this equation can be integrated analytically in the absence of radiation reaction and decomposed into a secular and a periodic contribution,

\begin{subequations}
\label{eq:phiz_sol}
\begin{align}
\phi_z &= \phi_{z,0} + \delta\phi_z, \\
\D \phi_{z,0} &= \av{\D \phi_z} \nonumber\\
&= y^6 \left\{ A_{\D\phi,J} J  + A_{\D\phi,\Pi} \frac{P_- + P_+}{K(m)}\right\} \label{eq:phiz_sol:zeta0} \\
\delta \phi_z &= \int_{t_0}^{t_0 +\delta t}  \D \phi_z - \av{\D \phi_z}  \, \d t \nonumber\\
&= \frac{4}{3 \sqrt{Y_3 - Y_-}} \frac{A_{\D\phi,\Pi}}{A_{\delta\chi}} \big\{\delta P_-(\hat{\psi}\sub{p}) + \delta P_+(\hat{\psi}\sub{p})\big\} , \label{eq:phiz_sol:dphiz}
\end{align}
\end{subequations}

\noindent where

\begin{subequations}
\label{eq:P_pm_def}
\begin{align}
P_\pm & = \frac{N_\pm}{B_\pm - C_\pm}  \Pi\left( \frac{-2 C_\pm}{B_\pm - C_\pm}, m \right) \, ,  \\
\delta P_\pm(\hat{\psi}\sub{p}) & = \frac{N_\pm}{B_\pm - C_\pm} \Bigg\{ \Pi \left[ \frac{- 2 C_+}{B_+ - C_+} ; \text{am}(\hat{\psi}\sub{p}; m) ; m \right] \nonumber\\ 
& \qquad \qquad - \frac{\Pi\left( \frac{- 2 C_+}{B_+ - C_+}, m \right)}{K(m)} \hat{\psi}\sub{p} \Bigg\} \, ,
\end{align}
\end{subequations}

\noindent with $\Pi(n, m)$ and $\Pi(n; \phi; m)$ the complete and incomplete elliptic integral of the third kind, $t_0$ the start of a precession cycle, $Y_3$ and $Y_-$ the quantities defined in Eq.~\eqref{eq:Ydefs}, and $\delta t \in (0, T\sub{p}]$ and $\hat{\psi}\sub{p} \in (-K(m), K(m)]$ the time and phase within the cycle.

An analogous procedure yields the solution for the third Euler angle $\zeta$. Substituting Eqs.~\eqref{eq:Dphiz} and \eqref{eq:costhL} into Eq.~\eqref{eq:Dzeta_minrot} and integrating, one finds

\begin{subequations}
\label{eq:zeta_sol}
\begin{align}
\zeta = & \zeta_0 + \delta\zeta, \\
\D \zeta_0 = &  - y^6\Bigg\{ A_{\D\phi,J} \big[ a_J + b_J \tilde{\chi}_{\mathrm{eff},0} + \tilde{c}_J \av{\delta\chi} \big] \nonumber \\
& + A_{\D\phi,\Pi} \left[ 4 a_J \tilde{b}_J + 2 (b_J^2 - c_J^2) \tilde{\chi}_{\mathrm{eff},0} + \frac{P_- - P_+}{K(m)} \right] \Bigg\}, \label{eq:zeta_sol:zeta0}  \\
\delta\zeta = &  \frac{2}{3} \frac{b_J^2 - c_J^2}{(1 - k_\chi^2) a_J} \sqrt{Y_3 - Y_-} \frac{A_{\D\phi,J}}{A_{\delta\chi}} \delta E(\hat{\psi}\sub{p}) \nonumber \\
& - \frac{4}{3 \sqrt{Y_3 - Y_-}} \frac{A_{\D\phi,\Pi}}{A_{\delta\chi}} \big\{\delta P_-(\hat{\psi}\sub{p}) - \delta P_+(\hat{\psi}\sub{p})\big\} \, . \label{eq:zeta_sol:dzeta}
\end{align}
\end{subequations}

\noindent where

\begin{subequations}
\begin{align}
    \av{\delta\chi} &= \delta\chi\sub{av} - \frac{2\delta\chi\sub{diff}}{m} \left[ \frac{E(m)}{K(m)} - 1 + \frac{m}{2} \right] \, , \label{eq:dchi_prec_avg} \\
    \delta E(\hat{\psi}\sub{p}) &= E \left[ \text{am} (\hat{\psi}\sub{p}; m) ; m \right] - \frac{E(m)}{K(m)} \hat{\psi}\sub{p} \, ,
\end{align}    
\end{subequations}

\noindent with $E(m)$ and $E(\phi; m)$ the complete and incomplete elliptic integrals of the second kind.

\subsection{Addition of radiation reaction effects}
\label{sec:HighPN_MSA:RR}

In the previous description of the spin-precession dynamics, we have neglected the effect of radiation reaction. This is justified because the radiation-reaction timescale is much longer than the spin-precession timescale, so its impact within a single precession cycle is small. Nonetheless, radiation reaction accumulates over many cycles, primarily through the secular evolution of the orbital parameters, which in turn modifies the spin-precession dynamics. In this section, we incorporate radiation-reaction effects by closely following Refs.~\cite{Klein:2021jtd,Morras:2025nlp}, since our PN-extended solution has a structure very similar to the 2PN solution presented there.

When radiation reaction is included, the PN parameter $y$ increases as the binary becomes more compact, while the eccentricity $e$ decreases as the orbit circularizes, approaching the residual eccentricity induced by spin effects of Eq.~\eqref{eq:emin_spin}.

The PN evolution equations for the PN parameter $y$, the squared eccentricity $e^2$, the mean orbital phase $\lambda$, and the argument of periastron $\delta\lambda$ can be written as

\begin{subequations}
\label{eq:RR_eqs_no_SP}
\begin{align}
 \D y &= \nu y^9 \sum_{n \geq 0} a_n \left(y, e^2, \uvec{l}_N, \bm{s}_1, \bm{s}_2 \right) y^n \, , \\
 \D e^2 &= - \nu y^8 \sum_{n \geq 0} b_n \left(y, e^2, \uvec{l}_N, \bm{s}_1, \bm{s}_2 \right) y^n \, , \\
 \D \lambda &= y^3 \, , \\
 \D \delta\lambda &= \frac{k y^3}{1 + k}, \quad k = y^2 \sum_{n \geq 0} k_n\left(y, e^2, \uvec{l}_N, \bm{s}_1, \bm{s}_2 \right) y^n  \label{eq:RR_eqs_no_SP:dl}\, ,
\end{align}
\end{subequations}

\noindent where $k$ is the periastron advance. The coefficients $a_n$, $b_n$, and $k_n$ are given up to 3PN order in Ref.~\cite{Morras:2025nlp}, while Appendix~\ref{sec:appendix:PN_QC} provides higher-PN-order quasi-circular expressions for $a_n$, as discussed in Sec.~\ref{sec:HighPNQC}.

In practice, we integrate the radiation-reaction equations numerically using Runge--Kutta methods~\cite{Press:2007nr,Scipy:2020}. Since the PN coefficients depend on the angular momenta $\uvec{l}_N$, $\bm{s}_1$, and $\bm{s}_2$, which vary on the spin-precession timescale, a direct integration of Eq.~\eqref{eq:RR_eqs_no_SP} would require resolving this short timescale, leading to prohibitively small time steps. To avoid this issue, we replace the quantities in Eq.~\eqref{eq:RR_eqs_no_SP} that vary on the spin-precession timescale by their averages over one precession cycle. This removes the explicit dependence of the radiation-reaction equations on the fast precessional dynamics and allows for an efficient numerical integration on the radiation-reaction timescale.

\subsubsection{Spin couplings averaging}
\label{sec:HighPN_MSA:RR:SpinAvg}

Inspecting the dependence of Eq.~\eqref{eq:RR_eqs_no_SP} on the angular momenta, we find that the spin-orbit terms depend on $\chi_\mathrm{eff}$ and $\delta\chi$. $\av{\delta\chi}$ was already computed in Eq.~\eqref{eq:dchi_prec_avg} and from Eq.~\eqref{eq:chi_eff_of_dchi}, then 

\begin{equation}
    \av{\chi_\mathrm{eff}}  =\chi_{\mathrm{eff},0} + k_\chi (\av{\delta \chi} - \delta \chi_0) \, .
    \label{eq:chi_eff_avg}
\end{equation}

As discussed in Sec.~\ref{sec:HigherPN_MSA:chi_eff_dchi}, in the MSA we assumed some parameters, such as $k_\chi$, to be constant by replacing the angular momenta entering them with their precession-averaged values. Consequently, in Eqs.~\eqref{eq:chi_eff_avg} and \eqref{eq:dchi_prec_avg}, both sides depend on the precession-averaged quantities to be determined. We resolve this implicit dependence using a fixed-point iteration scheme. We initialize the MSA solution of Sec.~\ref{sec:HigherPN_MSA:chi_eff_dchi} by setting each precession-averaged quantity equal to its initial value (e.g., $\av{\delta\chi}=\delta\chi_0$ and $\av{\chi_\mathrm{eff}}=\chi_{\mathrm{eff},0}$). This provides a first estimate for the precession-averaged quantities, which is then reinserted into the equations and iterated. Since the differences between the precession-averaged and initial values are small, and these averaged quantities enter only at high PN order (3PN and above), the iteration converges rapidly.

As in Ref.~\cite{Morras:2025nlp}, for the spin-spin couplings in Eq.~\eqref{eq:RR_eqs_no_SP} we must compute the precession averages of

\begin{subequations}
\begin{align}
    \sigma_i^{(1)} &= \bm{s}_i^2, \\
    \sigma_i^{(2)} &= \left(\uvec{l}_N \cdot \bm{s}_i\right)^2, \\
    \sigma_i^{(3)} &= \left|\uvec{l}_N \times \bm{s}_i\right|^2 \cos 2\psi_i,
\end{align}
\end{subequations}

\noindent where $i \in \{0,1,2\}$, $\bm{s}_0=\bm{s}_1+\bm{s}_2$, and $\psi_i$ denotes the angle between the periastron line and the projection of $\bm{s}_i$ onto the orbital plane. Since tidal torques are neglected, $\sigma^{(1)}_{1,2}=s_{1,2}^2$ are constant. Moreover, as shown in Ref.~\cite{Klein:2021jtd}, the angles $\psi_i$ evolve on the periastron-advance timescale ($\D\psi_i \sim y^5$). As a result, $\sigma_i^{(3)} \propto \cos 2\psi_i$ oscillates rapidly compared to the radiation-reaction timescale ($T_\mathrm{RR} \sim y^{-8}$) and averages to zero,
\begin{equation}
    \av{\sigma_i^{(3)}} = 0 \, \quad \text{for $i=\{0,1,2\}$}\, .
\end{equation}

For $\sigma_0^{(1)}$ we write
\begin{align}
    & \av{\sigma_0^{(1)}} = \av{\left( \bm{s}_1 + \bm{s}_2 \right)^2}  = s_1^2 + s_2^2 + 2 \av{\bm{s}_1 \cdot \bm{s}_2 } \nonumber\\ 
    & = \chi_{\mathrm{eff},0}^2 + s_{1\perp,0}^2 + s_{2\perp,0}^2 + \Delta_{J^2}^2 - \frac{a_J \tilde{c}_J}{b_J^2 - c_J^2} \left(\av{\delta\chi} - \delta\chi_0 \right) \, ,
\end{align}

\noindent where we have substituted Eq.~\eqref{eq:s1s2_of_DJ2} for $\bm{s}_1 \cdot \bm{s}_2$ and then taken the precession average. A similar procedure applies to the squares of the spin components aligned with the orbital angular momentum,

\begin{subequations}
\begin{align}
    \av{\sigma_{0}^{(2)}} &=  \av{\chi_\mathrm{eff}^2} \\
    \av{\sigma_{1,2}^{(2)} } &= \av{\left(\bm{\hat{L}} \cdot \bm{s}_{1,2}\right)^2} = \av{\left(\frac{\chi_\mathrm{eff} \pm \delta \chi}{2}\right)^2} \nonumber\\
    & = \frac{\av{\chi_\mathrm{eff}^2} \pm 2 \av{\chi_\mathrm{eff}\delta\chi} + \av{\delta\chi^2}}{4}  \, ,    
\end{align}
\end{subequations}

\noindent where the terms that contain $\chi_\mathrm{eff}$ can be always written in terms of $\delta\chi$ by substituting Eq.~\eqref{eq:chi_eff_of_dchi}, and assuming that $k_\chi$ is constant, we only need to compute $\av{\delta \chi^2}$.
An analogous situation occurs for the aligned cubic-in-spin terms listed in Appendix~\ref{sec:appendix:PN_QC}, which depend on $\av{\chi_\mathrm{eff}^n \delta\chi^{3-n}}$ with $n\in\{0,1,2,3\}$. These can likewise be rewritten in terms of $\av{\delta\chi^n}$ using Eq.~\eqref{eq:chi_eff_of_dchi}. Following Refs.~\cite{Klein:2021jtd,Morras:2025nlp}, the required precession averages are

\begin{subequations}
    \label{eq:prec_avg_defs}
    \begin{align}
        \av{\delta\chi} =& \delta\chi\sub{av} +  F_1(m) \delta\chi\sub{diff}  \label{eq:prec_avg_defs:dchi} \\
        \av{\delta \chi^2} =& \av{\delta\chi}^2 + \left(\frac{1}{2} - F_2 (m) \right) \delta\chi_\mathrm{diff}^2  \, , \label{eq:prec_avg_defs:dchi2} \\
   \av{\delta \chi^3}  =& \av{\delta\chi} \left(3 \av{\delta \chi^2} - 2 \av{\delta\chi}^2\right) - F_3(m) \delta\chi\sub{diff}^3  \, , \label{eq:prec_avg_defs:dchi3} \\
        F_1(m) =& -\frac{2}{m} \left[ \frac{E(m)}{K(m)} - 1 + \frac{m}{2} \right] \, , \label{eq:prec_avg_defs:F1} \\
        F_2(m) =& \frac{4}{m^2} \bigg[ \left(\frac{E(m)}{K(m)}\right)^2 + \frac{2}{3} \frac{E(m)}{K(m)} (m - 2) \nonumber\\
        & + \frac{m^2}{8} - \frac{m}{3} + \frac{1}{3} \bigg] \, , \label{eq:prec_avg_defs:F2} \\
        F_3(m) =& \frac{16}{m^3} \bigg[\left(\frac{E(m)}{K(m)}\right)^3+\left(\frac{E(m)}{K(m)}\right)^2 (m-2) \nonumber\\
        & + \frac{E(m)}{K(m)} \left(\frac{4}{15} m^2 - \frac{19}{15} m + \frac{19}{15}\right) \nonumber\\
        & - \frac{2}{15} m^2 + \frac{2}{5} m - \frac{4}{15}\bigg] \, , \label{eq:prec_avg_defs:F3}
        \end{align}
\end{subequations}

\noindent where in the limit $m \to 0$,

\begin{subequations}
\begin{align}
    F_1(m) =& \frac{m}{8} + \ord{m^2}  \, , \label{eq:F_dchi_prec_acg_m_to_0:1} \\
    F_2(m) =& \frac{m^2}{256} + \ord{m^3} \, , \label{eq:F_dchi_prec_acg_m_to_0:2} \\
    F_3(m) =& \frac{3}{32} m + \ord{m^2} \, . \label{eq:F_dchi_prec_acg_m_to_0:3}
\end{align}
\end{subequations}

\subsubsection{Evolution of the total angular momentum}

Up to this point, we have treated the vector $\bm{J}$, defined in Eq.~\eqref{eq:J_def}, as constant. In general, however, $\bm{J}$ evolves in time, and its derivative can be written as

\begin{align}
    \D \bm{J} =& a_J \D \uvec{l}_N  + b_J \D (\bm{s}_1 + \bm{s}_2) + c_J \D (\bm{s}_1 - \bm{s}_2) \nonumber \\
               & + (\D a_J) \uvec{l}_N  + (\D b_J) (\bm{s}_1 + \bm{s}_2) + (\D c_J) (\bm{s}_1 - \bm{s}_2) \, , \label{eq:DJ_vec}
\end{align}

\noindent where the derivatives of $\uvec{l}_N$, $\bm{s}_1$, and $\bm{s}_2$ are given by Eq.~\eqref{eq:raw_prec_eqs}. This induces a change in the direction of $\bm{J}$, which can be computed as

\begin{align}
    \D \uvec{\jmath} =& \frac{1}{J^3} (\bm{J} \times \D \bm{J}) \times \bm{J} \, , \label{eq:Dj_vec_of_DJ_vec}
\end{align}

In principle, this change in the direction of $\bm{J}$ would require an additional rotation, beyond that relating the co-precessing frame to the $\bm{J}$-aligned frame, in order to reach a truly inertial frame. In this work, however, we neglect this effect, since it is typically very small except for systems undergoing strong nutational resonances or transitional precession~\cite{Zhao:2017tro}, where the spins have large components anti-aligned with the orbital angular momentum and the direction of $\bm{J}$ can vary significantly.

We do, however, account for the evolution of the magnitude $J$. Most of this evolution is captured by writing $J$ as in Eq.~\eqref{eq:J2_of_DJ2}, since in the absence of spin precession that equation describes the evolution of $J$ with $\Delta_{J^2}$ constant. When the binary precesses, however, $\Delta_{J^2}$ varies in time according to

\begin{equation}
    \D J^2 = 2 \bm{J} \cdot \D \bm{J} \, .
    \label{eq:DJ2_of_DJ_vec}
\end{equation}

Substituting Eq.~\eqref{eq:J2_of_DJ2} for $J^2$ on the left-hand side and Eqs.~\eqref{eq:J_def} and \eqref{eq:DJ_vec} for $\bm{J}$ and $\D\bm{J}$ on the right-hand side, we can solve for $\D \Delta_{J^2}$. The result depends on the specific choice of $a_J$, $b_J$, and $c_J$, as well as on the precession equations of Eq.~\eqref{eq:raw_prec_eqs}, and can be written as

\begin{align}
    \D \Delta_{J^2} = - 2 \tilde{c}_J  \frac{\delta\chi - \delta\chi_0}{y^2} \kappa_{\D \Delta_{J^2}} \D y + y \delta_{\D \Delta_{J^2}} \D\delta\chi \, ,
    \label{eq:DDJ2_SP}
\end{align}

\noindent where the term proportional to $\kappa_{\D \Delta_{J^2}}$ arises from radiation-reaction effects, while the term proportional to $\delta_{\D \Delta_{J^2}}$ originates from the fact that, since Eq.~\eqref{eq:Dln_of_Dlnraw} removes the components of $\D \uvec{l}_N$ parallel to $\uvec{l}_N$, one has $\D \bm{J} \neq 0$ even in the absence of radiation reaction. The coefficients $\kappa_{\D \Delta_{J^2}}$ and $\delta_{\D \Delta_{J^2}}$ for \pyEFPEHM are given in Eq.~\eqref{eq:coefs_J2_PN} of Appendix~\ref{sec:appendix:PN_prec_expressions}.

Eq.~\eqref{eq:DDJ2_SP} depends explicitly on the spin-precession timescale, making a direct numerical integration computationally expensive. As for the Euler angles $\phi_z$ and $\zeta$ in Eqs.~\eqref{eq:phiz_sol} and \eqref{eq:zeta_sol}, we therefore decompose $\D \Delta_{J^2}$ into secular and periodic parts,

\begin{subequations}
\label{eq:DDJ2_prec_avg}
\begin{align}
\Delta_{J^2} = & \Delta_{J^2,0} + \delta \Delta_{J^2} \, , \label{eq:DDJ2_prec_avg:separation}\\
\D \Delta_{J^2,0} = & - 2 \tilde{c}_J  \frac{\av{\delta\chi} - \delta\chi_0}{y^2} \kappa_{\D \Delta_{J^2}} \D y, \,  \label{eq:DDJ2_prec_avg:secular}\\
\delta \Delta_{J^2} = & \frac{4}{3} \frac{b_J^2 - c_J^2}{(1 - k_\chi^2) a_J} \sqrt{Y_3 - Y_-} \frac{\kappa_{\D \Delta_{J^2}}}{A_{\delta\chi}}\frac{\D y}{y^8} \delta E(\hat{\psi}\sub{p}) \nonumber \\
& - 2 y \delta_{\D \Delta_{J^2}} \delta\chi_\mathrm{diff} \left[ 1 - \text{sn}^2(\psi\sub{p}; m) \right] , \label{eq:DDJ2_prec_avg:periodic}
\end{align}    
\end{subequations}

\noindent where, as in the rest of the MSA solution, we have treated $\kappa_{\D \Delta_{J^2}}$, $\delta_{\D \Delta_{J^2}}$, and $\D y$ as constants by replacing the angular momenta entering them with their precession-averaged values. Since $\delta \Delta_{J^2}$ is small and rapidly oscillatory on the spin-precession timescale, we neglect its contribution to the binary evolution through the radiation-reaction time-scale. Nevertheless, we retain $\delta \Delta_{J^2}$ when setting the initial condition for $\Delta_{J^2,0}$, so that Eq.~\eqref{eq:DDJ2_prec_avg:separation} agrees with Eq.~\eqref{eq:DJ20_def} at the initial time.

\subsubsection{Evolution of the elliptic parameter}

Due to radiation reaction, the elliptic parameter $m$ evolves slowly in time, so that the spin-precession period $2K(m)$ is no longer constant. Consequently, a fixed increment of the phase $\psi_p$ no longer corresponds to a fixed number of precession cycles. Following Refs.~\cite{Klein:2021jtd,Morras:2025nlp}, we account for this effect by defining a new phase $\bar{\psi}_p$ that is proportional to the accumulated number of precession cycles,
\begin{equation}
    \D\bar{\psi}_p = \frac{\pi}{2 K(m)} \D \psi\sub{p} \, .
    \label{eq:barpsip_def}
\end{equation}
From this, the correct phase of the Jacobi elliptic functions is recovered as

\begin{equation}
    \hat{\psi}\sub{p}(t) = \frac{2 K[m(t)]}{\pi} \hat{\bar{\psi}}_p(t) \, ,
    \label{eq:hatpsip_hatbarpsip}
\end{equation}

\noindent where $\hat{\bar{\psi}}_p$ satisfies
\begin{subequations}
\label{eq:hatbarpsip_def}
\begin{align}
&\bar{\psi}_p - \hat{\bar{\psi}}_p = n \pi \text{, for some }  n \in \mathbb{Z} \, , \\
&-\pi < \hat{\bar{\psi}}_p \leq \pi \, ,
\end{align}
\end{subequations}

\noindent and tracks the phase within the precession cycle also when including radiation reaction.

\section{Adding Gravitational Wave Higher Order Modes}
\label{sec:HMs}

In \pyEFPEHM, we include GW amplitude corrections up to 1PN order. This means we incorporate the  $(l,|m|) = (2,2)$, $(2,1)$, $(2,0)$, $(3,3)$, $(3,2)$, $(3,1)$, $(3,0)$, $(4,4)$, $(4,2)$, and $(4,0)$ multipoles. For these modes, we use the analytic expressions for the 1PN eccentric harmonic amplitudes $N^{lm}_p$ derived in Ref.~\cite{Morras:2025nbp}, which are not expanded in eccentricity and are thus valid at arbitrary eccentricity.

As in \pyEFPE, rather than fixing the number of included modes a priori, which can be computationally inefficient and may lead to the inclusion of negligible contributions in certain regions of parameter space, \pyEFPEHM selects the relevant modes dynamically during the inspiral. The mode content is updated at the time-steps chosen by the adaptive Runge-Kutta scheme used to integrate the equations of motion, ensuring that changes in the waveform multipolar structure track the evolving orbital dynamics. At each update, the model determines the minimal set of harmonics, $\bm{p}_{lm}^\mathrm{sel}$, required to represent the waveform to a user-specified accuracy $\epsilon_N$, such that

\begin{subequations}
\begin{align}
    & \Delta_h \equiv \frac{\Vert \hat{h} \Vert^2 - 2 \sum_{l} \sum_{m \geq 0} \sum_{p \in \bm{p}_{l m}^\mathrm{sel}} |N^{l m}_p|^2}{\Vert \hat{h} \Vert^2} \leq \epsilon_N \label{eq:strain_error_def}  \, , \\
    & \Vert \hat{h} \Vert^2 = \sum_{l=2}^\infty\left(\Vert \hat{H}^{l 0}\Vert^2 + 2 \sum_{m=1}^l \Vert \hat{H}^{l m}\Vert^2 \right)  \, , \label{eq:normh} \\
    & \Vert \hat{H}^{l m}\Vert^2 = \sum_{p=-\infty}^\infty |N^{l m}_p|^2 \, ,
\end{align}
\end{subequations}

\noindent where $\Vert \hat{H}^{l m}\Vert^2$ can be computed in closed form and is given in Ref.~\cite{Morras:2025nbp} for the 1PN modes. The factor of 2 in Eq.~\eqref{eq:strain_error_def} comes from using the mode symmetry of Eq.~\eqref{eq:Nl-m_from_Nlm}.

In practice, to minimize the number of included harmonics, we order the squared amplitudes $|N^{l m}_p|^2$ from largest to smallest and take as many terms as necessary for the strain error $\Delta_h$ to fall below the specified tolerance $\epsilon_N$. We refer the reader to Ref.~\cite{Morras:2025nbp} for an in-depth analysis of the number of modes that need to be included across the parameter space.

\begin{figure}[h!]
\centering  
\includegraphics[width=0.45\textwidth]{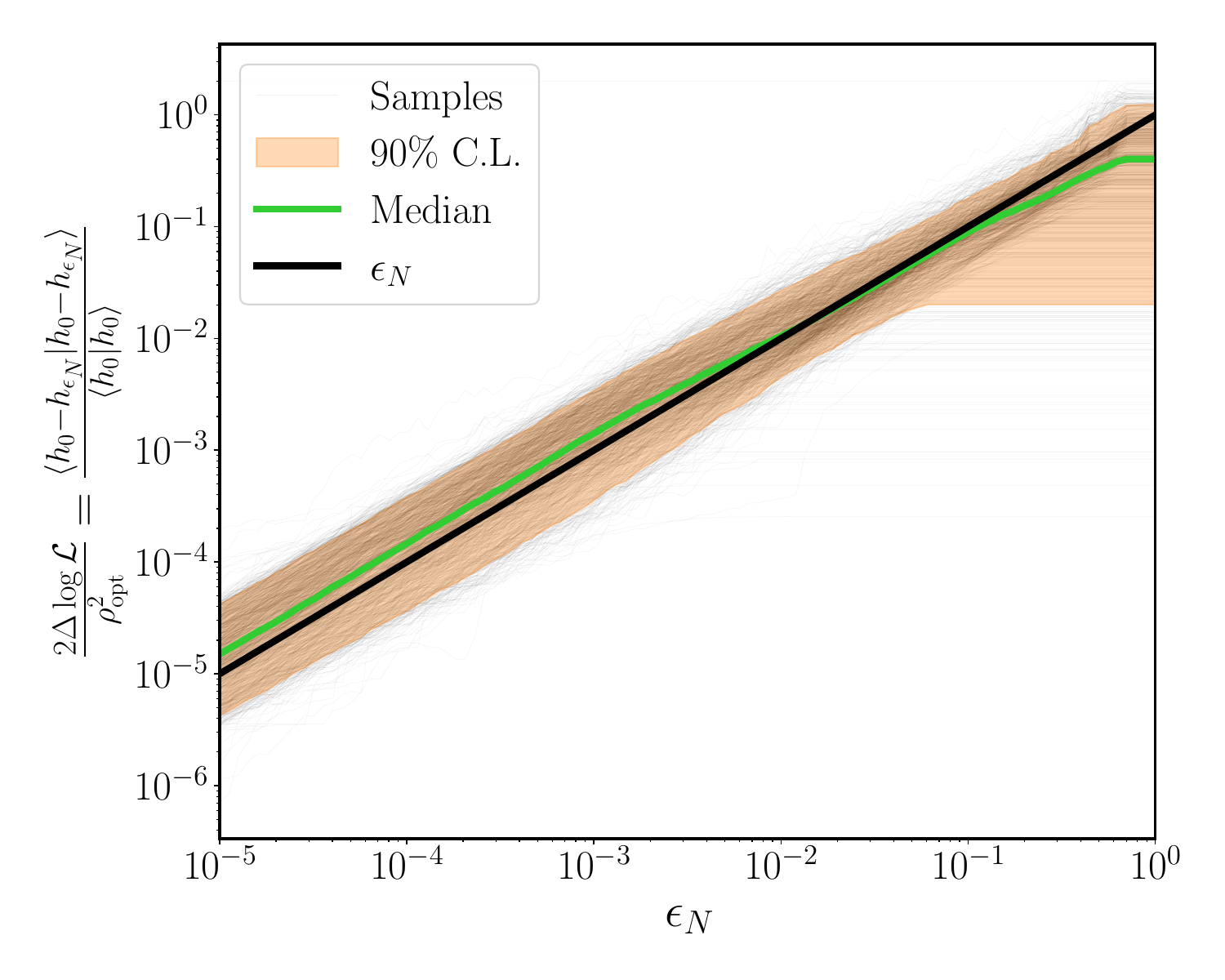}
\caption{\justifying Relative log-likelihood error, $2 \log\mathcal{L}/\rho_\mathrm{opt}^2$, as a function of the amplitude tolerance $\epsilon_N$ for 1000 random \pyEFPEHM waveform realizations. Source parameters are drawn assuming isotropic binary and spin orientations, with uniform distributions for the chirp mass $\mathcal{M}_c \in [0.95, 20]\,M_\odot$, initial eccentricity $e_0 \in [0, 0.7]$, mass ratio $q \in [0.05, 1]$, and dimensionless spin magnitudes $\chi_i \in [0, 0.9]$.
The reference waveform $h_0$ is generated using $\epsilon_N = 10^{-9}$ and is compared to waveforms $h_{\epsilon_N}$ constructed with tolerances $\epsilon_N \in [10^{-5}, 1]$. Inner products are computed using the Advanced LIGO A+ noise power spectral density~\cite{KAGRA:2013rdx,ObservingScenariosPSDs}, over the frequency range $20\,\mathrm{Hz}$--$4096\,\mathrm{Hz}$, where $20\,\mathrm{Hz}$ is the starting frequency of the waveform, and assuming a $256\,\mathrm{s}$ data segment.
In addition to showing the log-likelihood error as a function of $\epsilon_N$ for each sample, we display the median and 90\% confidence interval. The diagonal reference line $\epsilon_N$ is also shown to guide the expected scaling of the log-likelihood error.
}
\label{fig:logLrelerr_atol}
\end{figure}

To quantify the impact of the amplitude tolerance $\epsilon_N$ on actual waveform applications, it is useful to relate the strain error $\Delta_h$ introduced by truncating the mode content to something that is more relevant to data analysis. In Ref.~\cite{Morras:2025nlp}, it was argued that the tolerance $\epsilon_N$ appearing in Eq.~\eqref{eq:strain_error_def} is closely related to the error in the log-likelihood according to

\begin{equation}
    \Delta \log\mathcal{L} \lesssim \rho_\mathrm{opt}^2 \epsilon_N \, , \label{eq:DeltalogL_approx}
\end{equation}

\noindent where we have defined

\begin{subequations}
\begin{align}
    &\Delta \log\mathcal{L} = \log \mathcal{L}(h_0, h_0) - \log \mathcal{L}(h_0, h_{\epsilon_N}) \, , \\
    &\log \mathcal{L}(h_A, h_B) = - \frac{1}{2} \langle h_A - h_B| h_A - h_B \rangle \, , \\
    &\rho_\mathrm{opt}^2 = \langle h_0| h_0 \rangle \, ,
\end{align}
\end{subequations}

\noindent with $h_0$ denoting the waveform obtained by including all modes, and $h_{\epsilon_N}$ the approximation constructed by retaining the minimal set of modes such that $\Delta_h \leq \epsilon_N$. Here $\rho_\mathrm{opt}$ is the optimal signal-to-noise ratio of $h_0$, and $\langle \cdot | \cdot \rangle$ is the noise-weighted inner product, defined as~\cite{Sathyaprakash:1991mt,Finn:1992xs}

\begin{equation}
    \langle a | b \rangle = 4 \mathrm{Re}\left\{ \int_{f_\mathrm{min}}^{f_\mathrm{max}} \frac{\tilde{a}^{*}(f) \tilde{b}(f)}{S_n(f)} \d f  \right\} \, ,
    \label{eq:InnerProdDef}
\end{equation}

\noindent where $S_n(f)$ is the one-sided noise power spectral density (PSD) of the detector. For most data-analysis applications, one requires the log-likelihood error to satisfy $\Delta \log\mathcal{L} \lesssim 1$. If Eq.~\eqref{eq:DeltalogL_approx} holds, this implies
\begin{equation}
    \epsilon_N \lesssim \frac{1}{\rho_\mathrm{opt}^2} \, , \label{eq:sensible_epsilonN}
\end{equation}
\noindent providing a simple estimate for the amplitude tolerance $\epsilon_N$ needed to analyze a signal with optimal signal-to-noise ratio $\rho_\mathrm{opt}$. Since Eq.~\eqref{eq:DeltalogL_approx} is only an approximate relation, we explicitly test its validity for \pyEFPEHM. To this end, in Fig.~\ref{fig:logLrelerr_atol} we show the relative log-likelihood error, $2 \log\mathcal{L}/\rho_\mathrm{opt}^2$, for random waveform realizations and realistic detector noise, as described in the figure caption. Comparing Fig.~\ref{fig:logLrelerr_atol} with the analogous plot presented in Appendix~B of Ref.~\cite{Morras:2025nlp}, we find that the inclusion of higher-order multipoles weakens the correlation between the log-likelihood error and the amplitude tolerance $\epsilon_N$. As a consequence, and given that Ref.~\cite{Morras:2025nbp} showed that the number of modes in the waveform scales only logarithmically with $\epsilon_N$, we adopt a more conservative default value of $\epsilon_N = 10^{-4}$ in \pyEFPEHM (compared to $\epsilon_N = 10^{-3}$ in \pyEFPE). Unless otherwise specified, this choice is assumed throughout the remainder of this work.

\section{Testing and validation}
\label{sec:validate}

In this section we test and validate the \pyEFPEHM waveform, assessing its accuracy, robustness, and computational efficiency for GW data analysis of inspiral signals.

\subsection{Waveform comparisons}
\label{sec:validate:wf}

In this subsection we test the accuracy and computational speed of \pyEFPEHM by comparing it with other available waveform models. The relevant quantity for data analysis is the detector strain, which, in the long-wavelength approximation~\cite{Schutz:1987xok} can be written as

\begin{equation}
    h = F_+(\alpha,\delta,\psi) h_+ + F_\times(\alpha,\delta,\psi) h_\times \, ,
    \label{eq:h_detector}
\end{equation}

\noindent where $F_+$ and $F_\times$ are the detector antenna pattern functions, which depend on the sky location $(\alpha,\delta)$ and polarization angle $\psi$. For simplicity, during waveform comparison, we use that this detector strain can be expressed in terms of an effective polarization angle $\kappa(\alpha, \delta, \psi)$ as~\cite{Ramos-Buades:2023ehm}

\begin{equation}
    h = A(\alpha,\delta,\psi) \big[ \cos{(2\kappa)} h_+ + \sin{(2\kappa)} h_\times \big]\, ,
    \label{eq:h_from_pols}
\end{equation}

To quantify the difference between two strains $h_1$ and $h_2$, we compute their mismatch $\mathcal{MM}$~\cite{Sathyaprakash:1991mt,Finn:1992xs} as
\begin{equation}
    \mathcal{MM}(h_1, h_2) = 1 - \mathcal{M}(h_1, h_2) = 1 - \frac{\langle h_1 | h_2 \rangle}{\sqrt{\langle h_1 | h_1 \rangle \langle h_2 | h_2 \rangle}} \, .
    \label{eq:SimpleMisMatchDef}
\end{equation}

\noindent where $\mathcal{M}(h_1, h_2)$ is the match, and the noise-weighted inner product $\langle \cdot | \cdot \rangle$ is defined in Eq.~\eqref{eq:InnerProdDef}.

Even in the quasi-circular case, comparing different waveform approximants can be nontrivial, as different models may adopt distinct conventions for quantities associated with trivial symmetries of the system, such as the reference orbital phase $\phi_0$, the reference time $t_0$, the polarization angle $\kappa$, or rigid rotations of the in-plane spins by an angle $\phi_S$. Since these parameters carry little direct astrophysical information and their specific definitions can differ across waveform models, it is convenient and standard practice to minimize the mismatch over them when performing waveform comparisons. We therefore define

\begin{equation}
    \overline{\mathcal{MM}}(h_1, h_2) = \min_{\phi_0, t_0, \kappa, \phi_S} \mathcal{MM}(h_1, h_2)  \, .
    \label{eq:MinMisMatchDef_QC}
\end{equation}

Following Ref.~\cite{Harry:2016ijz}, the minimization over the polarization angle $\kappa$ is performed analytically, while the minimization over the reference time $t_0$ is efficiently carried out using the fast Fourier transform (FFT). The minimization over the reference phase $\phi_0$ and the in-plane spin rotation angle $\phi_S$ is performed numerically using a differential evolution algorithm~\cite{Storn:1997uea}.

Since the eccentricity is a gauge-dependent quantity, with different waveform models adopting different definitions and parametrizations of the radial phasing, comparing eccentric waveforms introduces additional ambiguities. In \pyEFPEHM, the orbit is specified by the initial PN time eccentricity $e_0$ and the initial mean anomaly $\ell_0$. When comparing against other eccentric waveform models, we therefore also minimize over these parameters, defining

\begin{equation}
    \overline{\mathcal{MM}}(h_1, h_2) = \min_{\substack{\phi_0, t_0, \kappa, \\ \phi_S, e_0, \ell_0}} \mathcal{MM}(h_1, h_2)  \, ,
    \label{eq:MinMisMatchDef_ecc}
\end{equation}

\noindent where the additional minimization over $e_0$ and $\ell_0$ is also performed numerically using the differential evolution algorithm. Since differences in eccentricity due to distinct definitions are expected to be fractional and because $e=0$ corresponds to the quasi-circular limit in all cases, we restrict the range of $e_0$ in the minimization to $\pm 30\%$ about the reference value unless otherwise stated.

To mimic the treatment of GW signals in data-analysis applications, we approximate the frequency integral in Eq.~\eqref{eq:InnerProdDef} by a sum over discrete Fourier frequencies corresponding to a data segment of duration $T$~\cite{Thrane:2018qnx},

\begin{equation}
    \langle a | b \rangle = 4 \mathrm{Re}\left\{ \sum_{k= \lceil f_\mathrm{min}/\Delta f \rceil}^{\lfloor f_\mathrm{max}/\Delta f \rfloor} \frac{\tilde{a}^{*}(k \Delta f) \tilde{b}(k\Delta f)}{S_n(k \Delta f)} \Delta f  \right\} \, , 
    \label{eq:InnerProdDiscrete}
\end{equation}

\noindent where $\Delta f = 1/T$ is the frequency resolution of the discrete Fourier transform.

\subsubsection{Parameter space explored}
\label{sec:validate:wf:params}

For the mismatch studies presented in this section, we adopt a common set of analysis choices. All waveform comparisons are performed starting at a minimum frequency of $f_\mathrm{min} = 20\,\mathrm{Hz}$, while waveform generation begins at a lower GW frequency $f_0^{\mathrm{GW},22} = 16\,\mathrm{Hz}$. We choose $f_\mathrm{min}=20\,\mathrm{Hz}$ to remain consistent with Ref.~\cite{Morras:2025nlp} and with standard choices in LIGO analyses. The lower starting frequency $f_0^{\mathrm{GW},22}$ provides a buffer for tapering when numerically Fourier transforming time-domain waveforms, thereby reducing spectral artifacts near $f_\mathrm{min}$. Although a lower starting frequency would capture a larger fraction of the higher-order modes, we verified that repeating a subset of comparisons at $f_0^{\mathrm{GW},22} = 10\,\mathrm{Hz}$ yields nearly identical mismatches, at a significantly higher computational cost. We use the same data-segment durations and corresponding chirp mass ranges as in Ref.~\cite{Morras:2025nlp}, listed in Table~\ref{table:ChirpMassRanges}, and assume the projected Advanced LIGO A+ noise power spectral density~\cite{KAGRA:2013rdx,ObservingScenariosPSDs}.

While we always simulate \pyEFPEHM until the PN parameter reaches its value at the innermost stable circular orbit (ISCO) of a test particle in the Schwarzschild metric ($y=y_\mathrm{ISCO} =1/\sqrt{6} = 0.408 \ldots$)~\cite{Bardeen:1972fi}, the maximum waveform generation frequency $f_\mathrm{max}$ is analysis dependent. Sometimes we consider a constant number, or a fraction of either the ISCO frequency $f_\mathrm{ISCO} = c^3/(6 \sqrt{6} \pi G M)$, or the frequency of the minimum energy circular orbit $f_\mathrm{MECO}$~\cite{Cabero:2016ayq} as computed in Ref.~\cite{Pratten:2020fqn}.

\begin{table}[t!]
\centering
\begin{tabular}{c | c  c  }
$T~[\mathrm{s}]$ &  \multicolumn{2}{c}{$\mathcal{M}_c~[M_\odot]$} \\
{}   & Min.   & Max.  \\ 
\hline
\hline
4 & 12 & 20  \\
8 & 8 & 12  \\
16 & 5 & 8  \\
32 & 3.3 & 5  \\
64 & 2.2 & 3.3  \\
128 & 1.4 & 2.2  \\
256 & 0.95 & 1.4  \\
\hline
\end{tabular}
\caption{Segment durations $T$ and corresponding chirp mass ($\mathcal{M}_c$) ranges used in the mismatch studies~\cite{Morras:2025nlp}.}
\label{table:ChirpMassRanges}
\end{table}

For each segment duration $T$, we evaluate waveforms at parameters randomly drawn from uniform distributions in chirp mass, mass ratio $q = m_2/m_1 \in [0.05, 1]$, and component spin magnitudes $\chi_i \in [0, 0.9]$, with isotropic binary orientations and spin directions. For aligned-spin analyses, we draw spins from the same distributions but set the components perpendicular to the orbital angular momentum to zero. In all comparisons, we restrict to black-hole binaries by setting the tidal parameters to zero. When eccentricity is included, we further assume a uniform distribution in the initial eccentricity $e_0 = e_{16\mathrm{Hz}} \in [0, 0.4]$.

These choices are designed to cover the region of parameter space relevant for current LVK observations while maintaining consistency with previous studies.

\subsubsection{Comparing \pyEFPEHM with \pyEFPE}
\label{sec:validate:wf:pyEFPE}

Since \pyEFPEHM introduces new features and code optimizations, we begin by comparing it to its predecessor, \pyEFPE. To ensure that the two waveform models are compared under identical conditions, we restrict the PN expansion to the same orders in both cases, 3PN for the non-spinning phasing, 2PN for the spin-dependent terms, and 0PN for the GW amplitudes. We also fix the SUA parameter $k_\mathrm{max}$ and the amplitude tolerance $\epsilon_N$ to the same values for both models, namely $k_\mathrm{max}=1$ and $\epsilon_N = 10^{-4}$, corresponding to the default \pyEFPEHM settings.

\begin{figure}[t!]
\centering  
\includegraphics[width=0.5\textwidth]{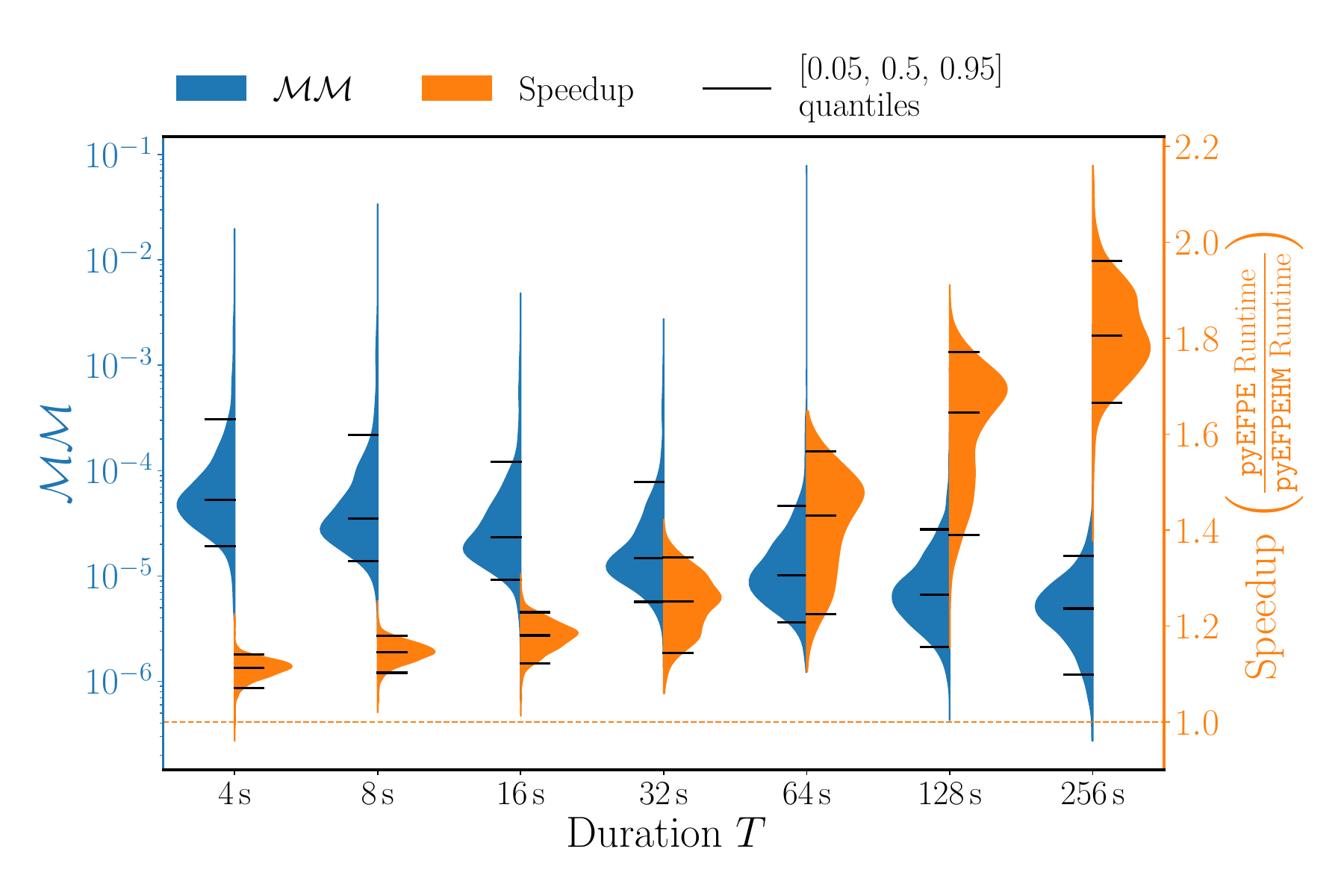}
\caption{\justifying Violin plots showing the distribution of mismatches (left) and speedup (right) when comparing \pyEFPEHM and \pyEFPE as a function of the segment duration $T$ or, equivalently, the chirp mass range $\mathcal{M}_c$ specified in Table~\ref{table:ChirpMassRanges}. Each distribution consists of 2000 random samples drawn from the parameter distributions described in Sec.~\ref{sec:validate:wf:params}, considering eccentricity and precession, and using a maximum frequency of $f_\mathrm{max} = 4096\,\mathrm{Hz}$.
The waveforms are not generated using their default configurations, but instead employ a common set of settings chosen to enforce agreement between the two models. In particular, the non-spinning phasing, spin-dependent effects, and GW amplitudes are computed at 3PN, 2PN, and 0PN order, respectively. We use $k_\mathrm{max}=1$ for the SUA and an amplitude tolerance of $\epsilon_N = 10^{-4}$.
}
\label{fig:equalized_pyEFPEHM_vs_pyEFPE_MM_Speedup_violins}
\end{figure}

In Fig.~\ref{fig:equalized_pyEFPEHM_vs_pyEFPE_MM_Speedup_violins} we show the distributions of mismatch and speedup between \pyEFPEHM and \pyEFPE for 2000 samples at each segment duration (and corresponding chirp mass range) listed in Table~\ref{table:ChirpMassRanges}. Although these comparisons include both eccentricity and precession, we do not minimize the mismatch over intrinsic parameters, since \pyEFPEHM and \pyEFPE use identical conventions.

From Fig.~\ref{fig:equalized_pyEFPEHM_vs_pyEFPE_MM_Speedup_violins}, we observe that the mismatch decreases for increasing signal duration (and decreasing chirp mass), ranging from $\sim 10^{-4}$ down to $\sim 10^{-6}$. The dominant source of this mismatch is that, in \pyEFPEHM, the SUA expansion is applied only to the precessing part of the amplitude, since corrections arising from applying the SUA to the non-precessing part enter at $\mathcal{O}(1.5\mathrm{PN})$, and are therefore neglected, as they lie beyond the 1PN accuracy at which the amplitudes are computed. These higher-PN corrections are more relevant for short signals, where only the late inspiral is observed.

We also observe that the mismatch distributions of Fig.~\ref{fig:equalized_pyEFPEHM_vs_pyEFPE_MM_Speedup_violins} exhibit small tails extending to higher mismatches. The outliers in these tails correspond to systems for which the MSA approximation develops a discontinuity, causing the spline interpolation of the precession amplitudes to behave poorly. As discussed in Ref.~\cite{Morras:2025nlp}, such discontinuities can arise during very abrupt transitional precession~\cite{Zhao:2017tro} or when the total and orbital angular momenta become exactly aligned. In the latter case, the physical (coordinate) discontinuities in $\D\phi_z$ and $\D\zeta$ (see Eq.~\eqref{eq:Dphiz}) become spurious discontinuities in $\delta\phi_z$ and $\delta\zeta$ when doing the MSA (see Eq.~\eqref{eq:phiz_sol:dphiz} and Eq.~\eqref{eq:zeta_sol:dzeta}).

Fig.~\ref{fig:equalized_pyEFPEHM_vs_pyEFPE_MM_Speedup_violins} also shows that the speedup, defined as the ratio of the \pyEFPE runtime to the \pyEFPEHM runtime, increases with signal duration, ranging from $\sim 1.1$ to $\sim 2$. This implies that \pyEFPEHM is between $\sim 10\%$ and $\sim 100\%$ faster than \pyEFPE. The primary reason for this improvement is a set of code optimizations in the computation of the GW polarizations, including improved vectorisation and reduced Python overhead through more efficient array handling, as well as replacing the polynomial interpolation of the eccentric harmonic amplitudes $N^{lm}_p$ with spline interpolation. These optimizations are less impactful for short signals, where the overall computational cost is dominated by the Runge-Kutta integration of the equations of motion, which is implemented in Python and therefore subject to significant interpreter overhead.

\subsubsection{Comparing \pyEFPEHM with \STfour}
\label{sec:validate:wf:SpinTaylorT4}

As noted in Ref.~\cite{Morras:2025nlp}, \texttt{EFPE} models and \texttt{TaylorT4} models~\cite{Buonanno:2009zt} share the same phasing in the quasi-circular, spin-aligned limit. Nonetheless, \pyEFPEHM and \STfour differ in several respects, allowing us to test the accuracy of specific modelling choices adopted in \pyEFPEHM.

Most notably, for binaries with spins misaligned with the orbital angular momentum, \STfour~\cite{lalsuite_code,Buonanno:2009zt,Sturani:2015STA,Isoyama:2020lls} models spin precession by numerically integrating the spin-precession equations, whereas \pyEFPEHM employs the MSA to approximate their solution, as described in Sec.~\ref{sec:HighPN_MSA}. The equations solved by the two models are also not identical, as they adopt different spin supplementary conditions (SSCs). \pyEFPEHM uses the Newton-Wigner~\cite{Khalil:2023kep,Pryce:1948pf,Newton:1949cq} SSC, while \STfour employs the covariant Tulczyjew-Dixon~\cite{Tulczyjew:1959zza,Dixon:1970zza} SSC. This difference leads to small modifications in the spin-precession equations starting at 2.5PN order.

Furthermore, \STfour is a time-domain waveform model and must be numerically Fourier transformed, using a fast Fourier transform, to compute the overlap integral of Eq.~\eqref{eq:InnerProdDiscrete}. In contrast, \pyEFPEHM analytically approximates the frequency-domain waveform using the SUA method. Comparing the two models therefore also provides a test of the accuracy of the SUA approximation.

\begin{figure}[t!]
\centering  
\includegraphics[width=0.5\textwidth]{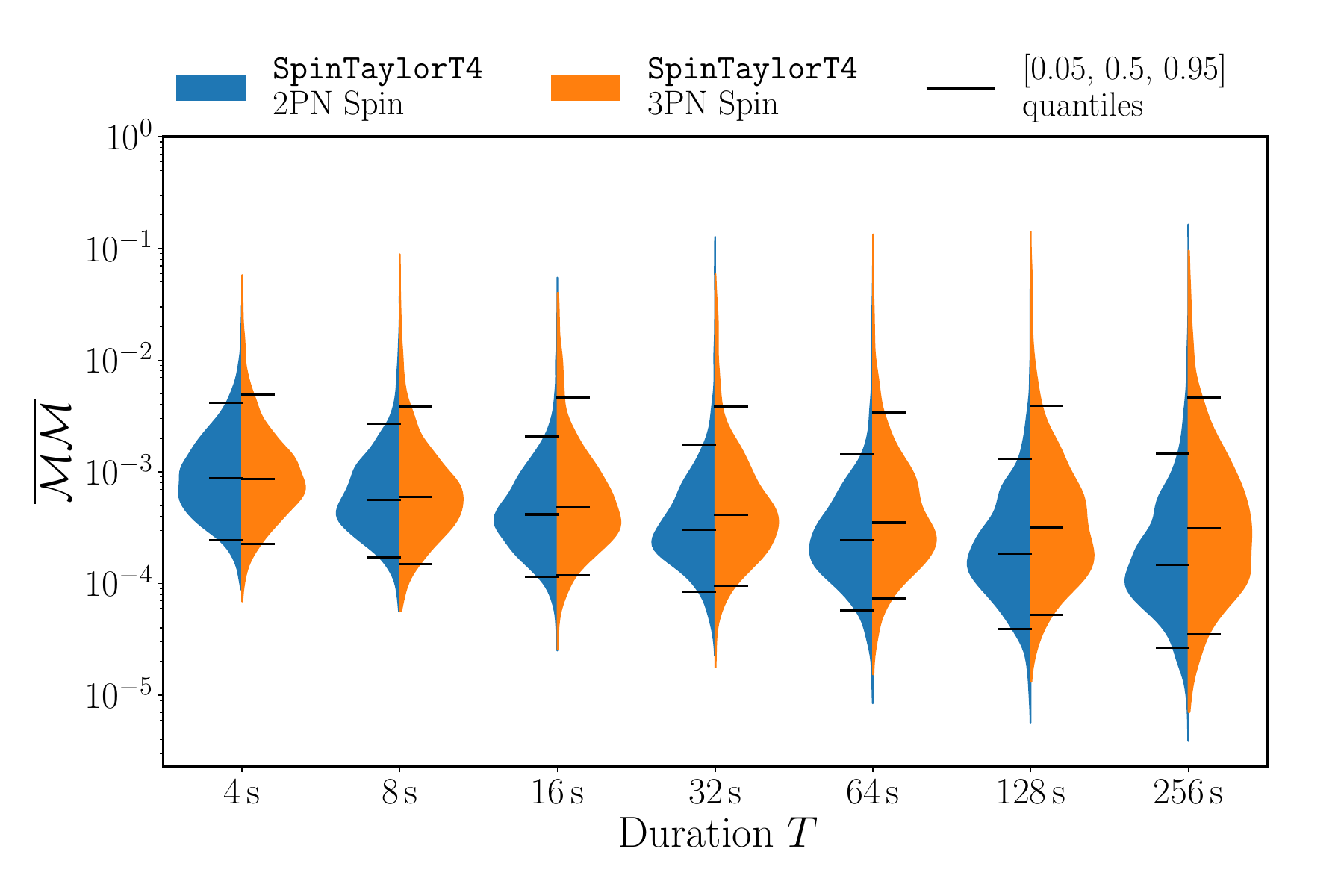}
\caption{\justifying Violin plots showing the distribution of mismatches when comparing \pyEFPEHM and \STfour, including spin-dependent terms at 2PN (left) and 3PN (right), as a function of the segment duration $T$ or, equivalently, the chirp mass range $\mathcal{M}_c$ specified in Table~\ref{table:ChirpMassRanges}. Each distribution consists of 2000 random samples drawn from the parameter distributions described in Sec.~\ref{sec:validate:wf:params}, considering quasi-circular and precessing binaries, and using a maximum frequency of $f_\mathrm{max} = 0.8 f_\mathrm{ISCO}$.
}
\label{fig:MM_violins_prec_SpinTaylorT4_PN_spin_4_6}
\end{figure}

Fig~\ref{fig:MM_violins_prec_SpinTaylorT4_PN_spin_4_6} shows the mismatches between \pyEFPEHM and \STfour. To ensure a consistent comparison, we restrict both models to the same PN orders, i.e., 3.5PN in the non-spinning phasing, corresponding to the highest order available in the \texttt{lal} implementation of \STfour~\cite{lalsuite_code,Sturani:2015STA,Isoyama:2020lls} used here, and 1PN in the GW amplitudes, which is the highest order implemented in \pyEFPEHM and includes the ten 1PN GW modes. For the spin-dependent contributions, we consider both 2PN and 3PN truncations. The 2PN case corresponds to the PN order at which the MSA was rigorously derived in previous literature~\cite{Klein:2021jtd,Morras:2025nlp}, while 3PN is the highest order implemented in the \texttt{lal} \STfour model.

We find that the ``2PN Spin'' and ``3PN Spin'' cases yield comparable mismatches, validating the MSA approximation derived in Sec.~\ref{sec:HighPN_MSA}. In both cases, mismatches decrease with increasing signal duration. This behavior contrasts with the analogous comparison between \pyEFPE and \STfour presented in Ref.~\cite{Morras:2025nlp}, where in the ``3PN Spin'' case mismatches increased for longer signals due to the restriction of the MSA to 2PN order in \pyEFPE and the consequent neglect of 2.5PN spin effects that accumulate over long inspirals. This agreement is particularly noteworthy given that the comparison shown in Fig.~\ref{fig:MM_violins_prec_SpinTaylorT4_PN_spin_4_6} includes higher-order GW modes, which are more sensitive to precession modelling due to their stronger angular dependence compared to the dominant $(l,m)=(2,2)$ mode.

The overall small mismatches indicate that the combined effects of differing spin supplementary conditions and the SUA approximation of the frequency-domain waveform are subdominant. The observed decrease in mismatch with increasing signal duration further suggests that, as expected, these differences become less relevant as the signal becomes increasingly dominated by the early inspiral.

Nevertheless, as in Fig.~\ref{fig:equalized_pyEFPEHM_vs_pyEFPE_MM_Speedup_violins}, Fig.~\ref{fig:MM_violins_prec_SpinTaylorT4_PN_spin_4_6} also exhibits small tails extending to higher mismatches. These are likewise caused by failures of the MSA approximation in systems undergoing transitional precession or passing through configurations where the total and orbital angular momenta become exactly aligned. Extending the current MSA framework to handle such cases will therefore be important for improving the robustness of the waveform model.

\subsubsection{Comparing with \vfive models}
\label{sec:validate:wf:SEOB}

In this section, we compare \pyEFPEHM against state-of-the-art \texttt{SEOBNR} waveform models, namely \vfiveEHM~\cite{Gamboa:2024hli,Gamboa:2024imd} for eccentric spin-aligned binaries, and \vfivePHM~\cite{Pompili:2023tna,Khalil:2023kep,Ramos-Buades:2023ehm,Estelles:2025zah} for quasi-circular precessing binaries. These models are based on the effective one-body (EOB) formalism~\cite{Buonanno:1998gg} and are calibrated against numerical relativity (NR) simulations, enabling them to accurately describe the full inspiral-merger-ringdown signal. Although they are comparatively more computationally expensive than many analytical approximants, \vfive models are among the most accurate waveforms currently available, typically yielding the smallest mismatches against NR simulations~\cite{Gamboa:2024hli,Estelles:2025zah}. They therefore provide a good benchmark to test the accuracy of \pyEFPEHM.

\begin{figure}[t!]
\centering  
\includegraphics[width=0.5\textwidth]{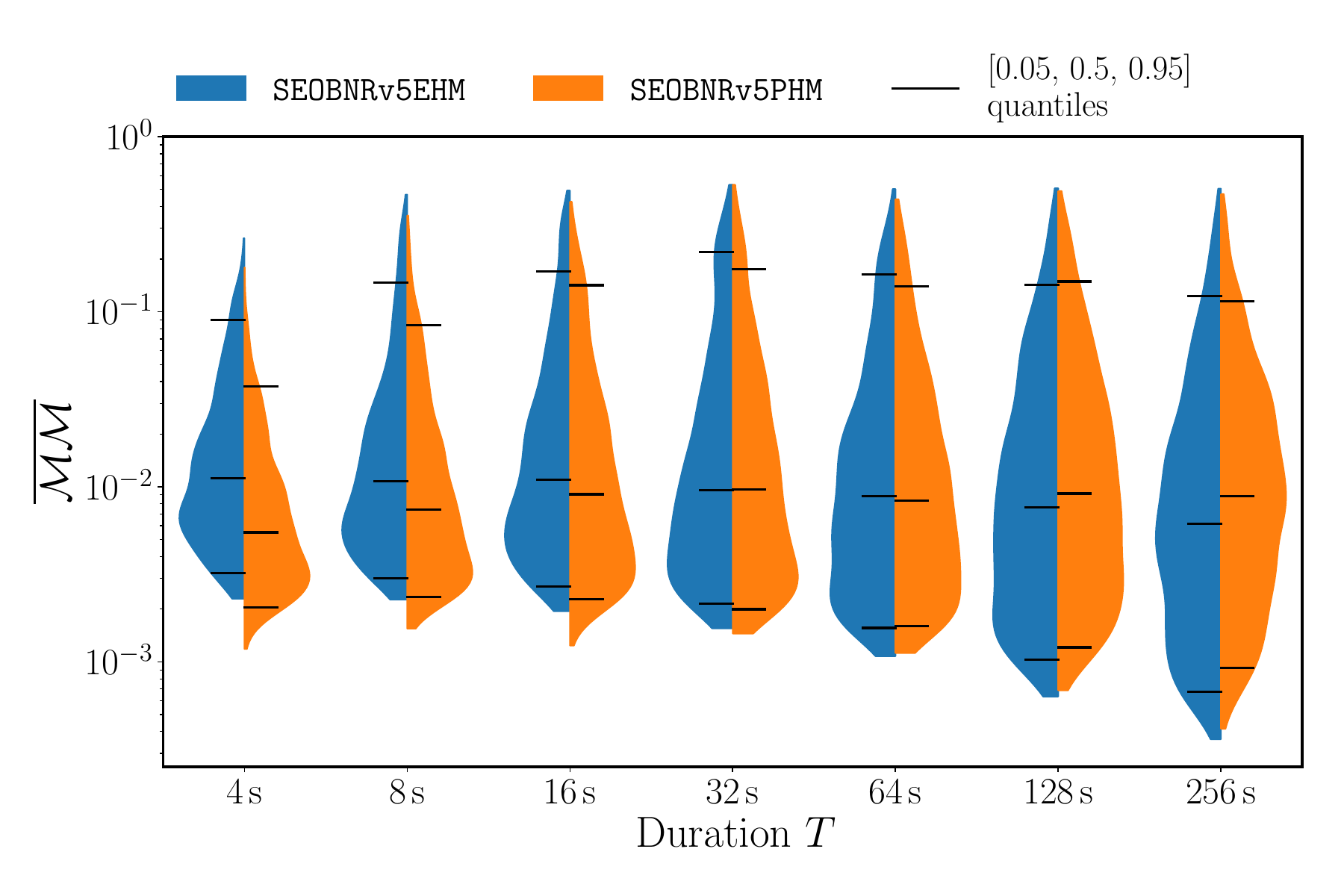}
\caption{\justifying Violin plots showing the distribution of mismatches when comparing \pyEFPEHM with \vfiveEHM in the eccentric spin-aligned limit (left) and with \vfivePHM in the quasi-circular precessing limit (right), as a function of the segment duration $T$ or, equivalently, the chirp mass range $\mathcal{M}_c$ specified in Table~\ref{table:ChirpMassRanges}. Each distribution consists of 2000 random samples drawn from the parameter distributions described in Sec.~\ref{sec:validate:wf:params}, using a maximum frequency of $f_\mathrm{max} = 0.8 f_\mathrm{MECO}$.
}
\label{fig:MM_violins_SEOBNRv5_EHM_PHM}
\end{figure}

Fig.~\ref{fig:MM_violins_SEOBNRv5_EHM_PHM} shows the mismatches between \pyEFPEHM and the \vfiveEHM and \vfivePHM models. To ensure a consistent comparison, we compute mismatches using only the GW modes that are implemented in both \pyEFPEHM and the corresponding \vfive model. In both the \vfiveEHM and \vfivePHM cases, this includes the $(l,m) = (2,1)$, $(2,2)$, $(3,2)$, $(3,3)$, and $(4,4)$ modes. Because \pyEFPEHM is an inspiral-only model, we compute mismatches up to a maximum frequency of $f_\mathrm{max} = 0.8\,f_\mathrm{MECO}$, above which the PN approximation is expected to become inaccurate. We note, however, that a small fraction of the merger–ringdown signal in the \texttt{SEOBNR} $(2,1)$ mode may still leak into the comparison, since this mode has approximately half the frequency of the dominant $(2,2)$ mode.

We find that the mismatches between \pyEFPEHM and both \vfiveEHM and \vfivePHM are comparable, indicating that the eccentric spin-aligned and quasi-circular precessing limits achieve a similar level of modeling accuracy, with the medians of the distributions lying mostly around $10^{-2}$. Moreover, the mismatch values do not depend strongly on the signal duration, suggesting that the differences between the waveforms do not accumulate significantly over long inspirals.

These mismatches are notably larger than those found in the comparison with \STfour (Fig.~\ref{fig:MM_violins_prec_SpinTaylorT4_PN_spin_4_6}). This is expected since \pyEFPEHM and \STfour share the same T4 PN expansion in the phasing~\cite{Buonanno:2009zt,Morras:2025nlp}, and in that comparison are additionally configured to include the same PN orders, naturally leading to close agreement. The \vfive models, by contrast, include PN contributions to different orders resummed within the EOB framework~\cite{Buonanno:1998gg}, and are further calibrated against NR simulations. In the late inspiral, where the convergence of the PN expansion degrades, this leads to a significant dephasing between the models~\cite{Buonanno:2009zt}.

\begin{figure}[t!]
\centering  
\includegraphics[width=0.5\textwidth]{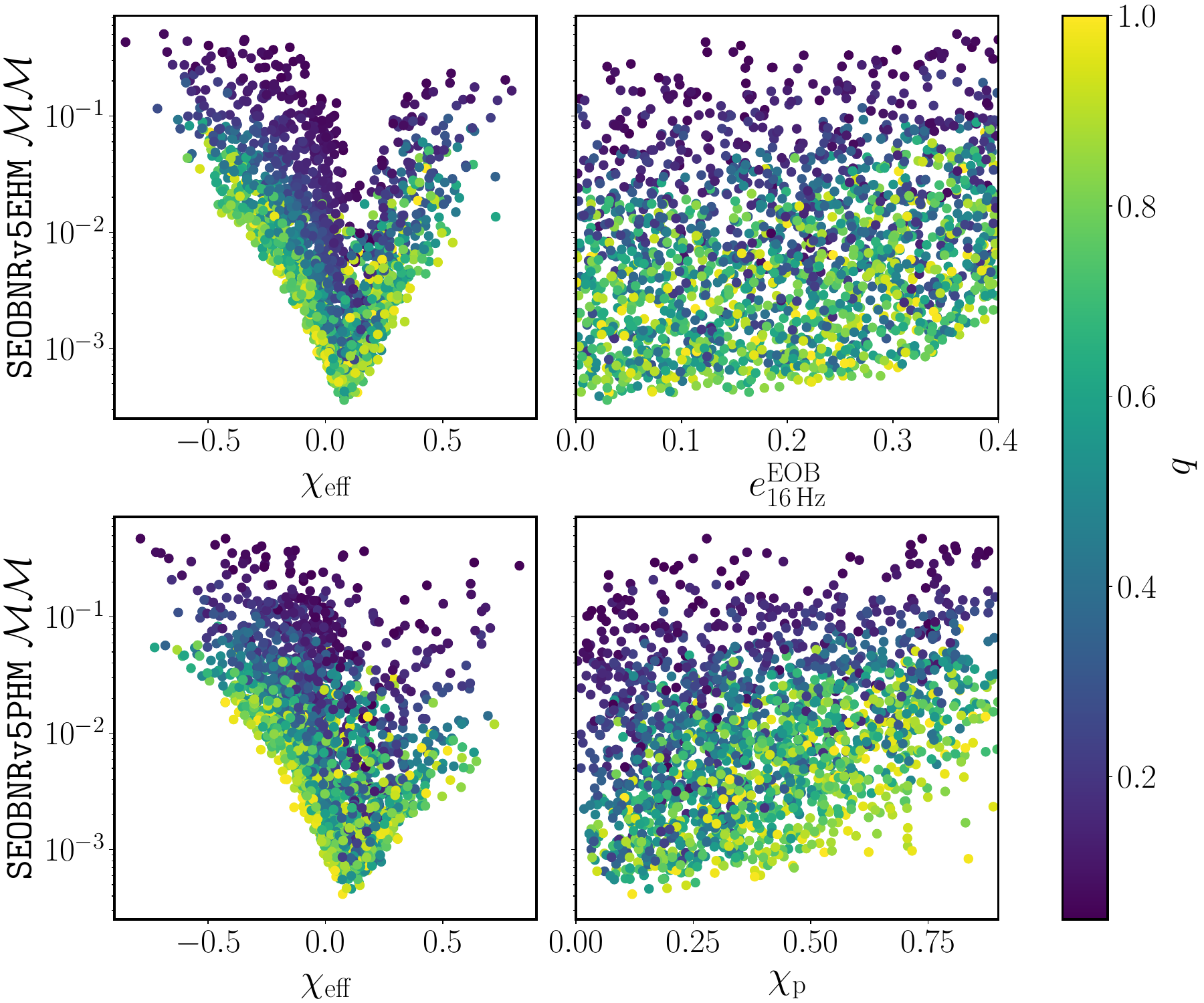}
\caption{\justifying Scatter plots of the mismatches between \pyEFPEHM and \vfive models as a function of different source parameters. The top row shows the mismatch with \vfiveEHM as a function of the effective inspiral spin parameter $\chi_\mathrm{eff}$ (top left) and the initial \vfiveEHM eccentricity $e_{16\,\mathrm{Hz}}^\mathrm{EOB}$ (top right). The bottom row shows the mismatch with \vfivePHM as a function of $\chi_\mathrm{eff}$ (bottom left) and the effective precession spin parameter $\chi_\mathrm{p}$~\cite{Schmidt:2014iyl} (bottom right panel). In all panels, the color of the points indicates the mass ratio $q=m_2/m_1$. The points correspond to the 2000 samples used for the $T=256\,\mathrm{s}$ ($\mathcal{M}_c \in [0.95, 1.4]\,M_\odot$) distributions shown in Fig.~\ref{fig:MM_violins_SEOBNRv5_EHM_PHM}.
}
\label{fig:SEOBNRv5EHM_SEOBNRv5PHM_T_256_N_2000_mismatch_distributions}
\end{figure}

However, the distributions exhibit a long tail toward larger mismatches, extending above $10^{-1}$. To identify the origin of this tail, we show in Fig.~\ref{fig:SEOBNRv5EHM_SEOBNRv5PHM_T_256_N_2000_mismatch_distributions} the mismatches corresponding to the $T=256\,\mathrm{s}$ signals ($\mathcal{M}_c \in [0.95, 1.4]\,M_\odot$) used to construct the distributions in Fig.~\ref{fig:MM_violins_SEOBNRv5_EHM_PHM}, as a function of different source parameters. We observe the same qualitative trends for all signal durations considered in this study, and show the $T=256\,\mathrm{s}$ case as a representative example. In particular, we find that the largest mismatches correspond to systems with very unequal masses ($m_2/m_1 \lesssim 0.1$) and large spins aligned with the orbital angular momentum ($|\chi_\mathrm{eff}| \gtrsim 0.5$). By contrast, the mismatch shows only a mild positive correlation with the eccentricity or the magnitude of the in-plane spin components, indicating that these effects do not currently dominate the waveform differences.

\subsubsection{Comparing with \TEOBDali}
\label{sec:validate:wf:TEOB}

In this section, we compare \pyEFPEHM against the \TEOBDali~\cite{Nagar:2024oyk,Nagar:2024dzj,Albanesi:2025txj} waveform model. Like the \texttt{SEOBNR} family, \TEOBDali is based on the EOB formalism and calibrated against NR simulations. While it has been shown to be less accurate than \vfive models~\cite{Gamboa:2024hli,Estelles:2025zah}, \TEOBDali aims to incorporate a broad range of two-body physics, enabling waveform generation for generic compact binaries evolving along arbitrary orbits, including eccentric and precessing configurations. As such, it can be a useful model for testing \pyEFPEHM over a wider region of parameter space, since \pyEFPEHM is likewise designed to model inspiral waveforms for generic binaries on arbitrary orbits.

\begin{figure}[t!]
\centering  
\includegraphics[width=0.5\textwidth]{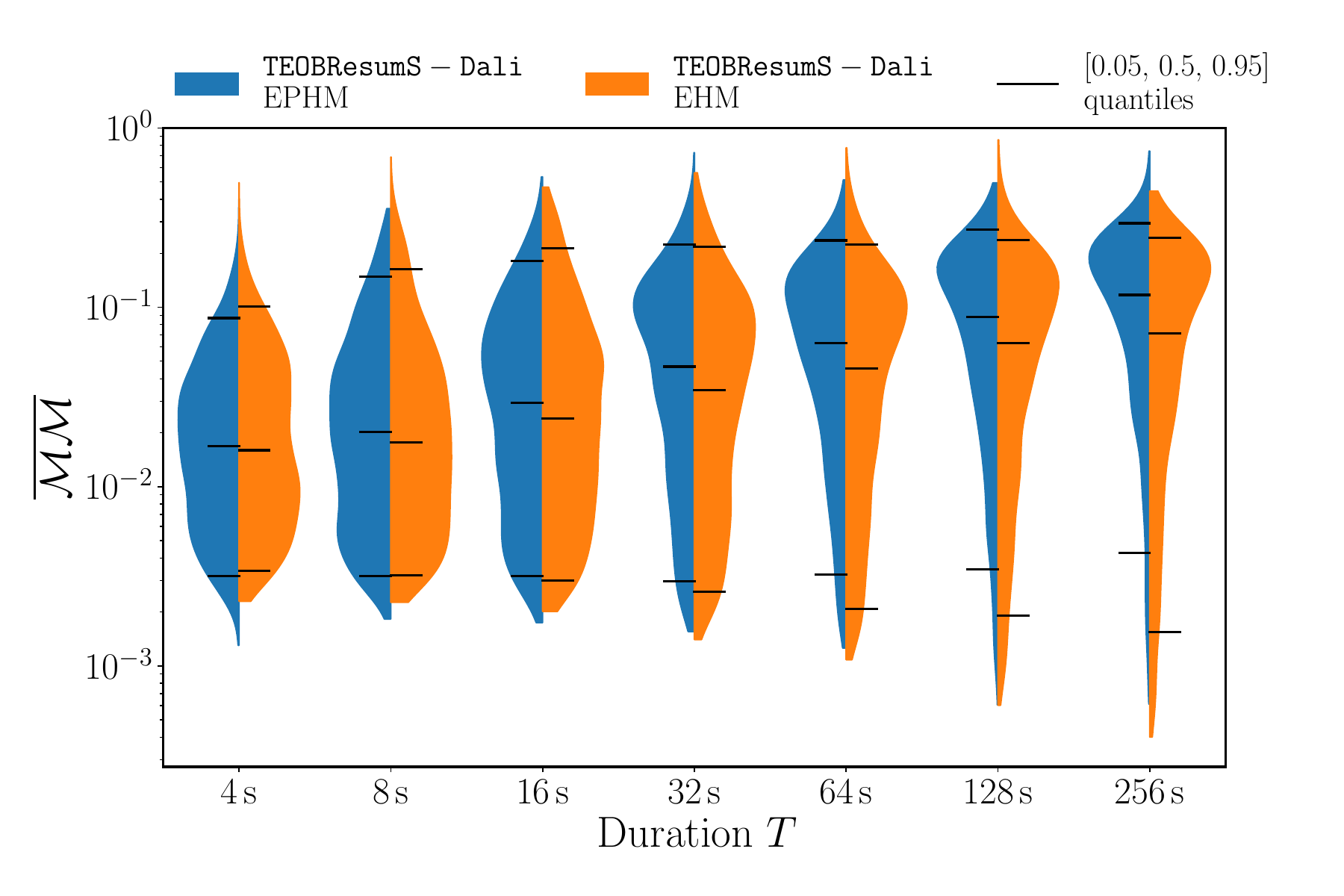}
\caption{\justifying  Violin plots showing the distribution of mismatches when comparing \pyEFPEHM and \TEOBDali, in the eccentric precessing case (left) and in the eccentric spin-aligned case (right), as a function of the segment duration $T$ or, equivalently, the chirp mass range $\mathcal{M}_c$ specified in Table~\ref{table:ChirpMassRanges}. Each distribution consists of 2000 random samples drawn from the parameter distributions described in Sec.~\ref{sec:validate:wf:params}, using a maximum frequency of $f_\mathrm{max} = 0.8 f_\mathrm{ISCO}$.
}
\label{fig:MM_violins_TEOBResumS_EPHM_EHM}
\end{figure}

Fig.~\ref{fig:MM_violins_TEOBResumS_EPHM_EHM} shows the mismatches between \pyEFPEHM and \TEOBDali~\cite{Albanesi:2025txj,Nagar:2024dzj,Nagar:2024oyk} in the eccentric precessing and eccentric spin-aligned cases. As in the \vfive comparisons, we ensure consistency by computing mismatches using only the GW modes implemented in both models. In this case, this includes all modes available in \pyEFPEHM except the $m=0$ modes, which are not modeled by \TEOBDali, namely $(l,|m|) = (2,2)$, $(2,1)$, $(3,3)$, $(3,2)$, $(3,1)$, $(4,4)$, and $(4,2)$.

We find that the mismatches with \TEOBDali are comparable in both cases, indicating that the inclusion of eccentric precession does not introduce significantly larger errors relative to the eccentric spin-aligned baseline. However, in contrast to the comparisons with the \vfive models, while the mismatches are still relatively small for short signal durations, they grow rapidly as the signal duration increases. This behavior can be traced to the treatment of radiation reaction in the eccentric sector of \TEOBDali, where the GW flux is fully modeled only at Newtonian order~\cite{Chiaramello:2020ehz}. As a result, \TEOBDali introduces errors in the binary phasing already at 1PN order. As discussed in Sec.~\ref{sec:HighPN_MSA}, such differences lead to a dephasing that accumulates strongly during the early inspiral and can be arbitrarily large for arbitrarily long signals.

\subsubsection{Exploring the \pyEFPEHM runtime}
\label{sec:validate:wf:pyEFPEHM}

\begin{figure}[t!]
\centering  
\includegraphics[width=0.5\textwidth]{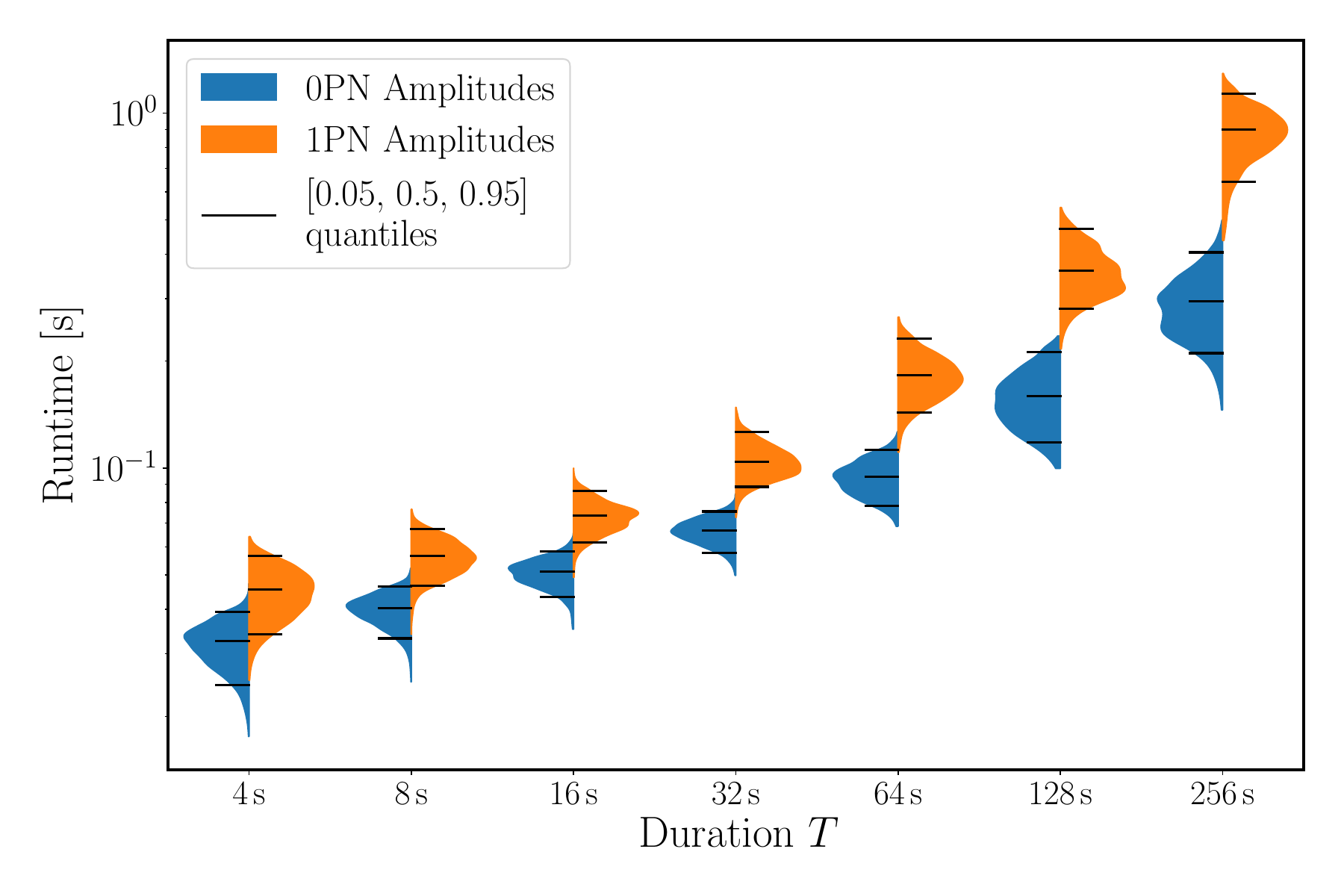}
\caption{\justifying Violin plots showing the distribution of the \pyEFPEHM runtime for waveforms computed with 0PN amplitudes (left) and 1PN amplitudes (right) as a function of the segment duration $T$, or equivalently the chirp mass range $\mathcal{M}_c$ specified in Table~\ref{table:ChirpMassRanges}. Each distribution consists of 2000 random samples drawn from the parameter distributions described in Sec.~\ref{sec:validate:wf:params}, considering eccentricity and precession, and using a maximum frequency of $f_\mathrm{max} = 4096\,\mathrm{Hz}$. The timings were measured on a single core of an \textit{Intel Core Ultra 7 265U} laptop processor.
}
\label{fig:Amps_0PN_vs_Amps_1PN_runtime_violins}
\end{figure}

\begin{figure}[t!]
\centering  
\includegraphics[width=0.49\textwidth]{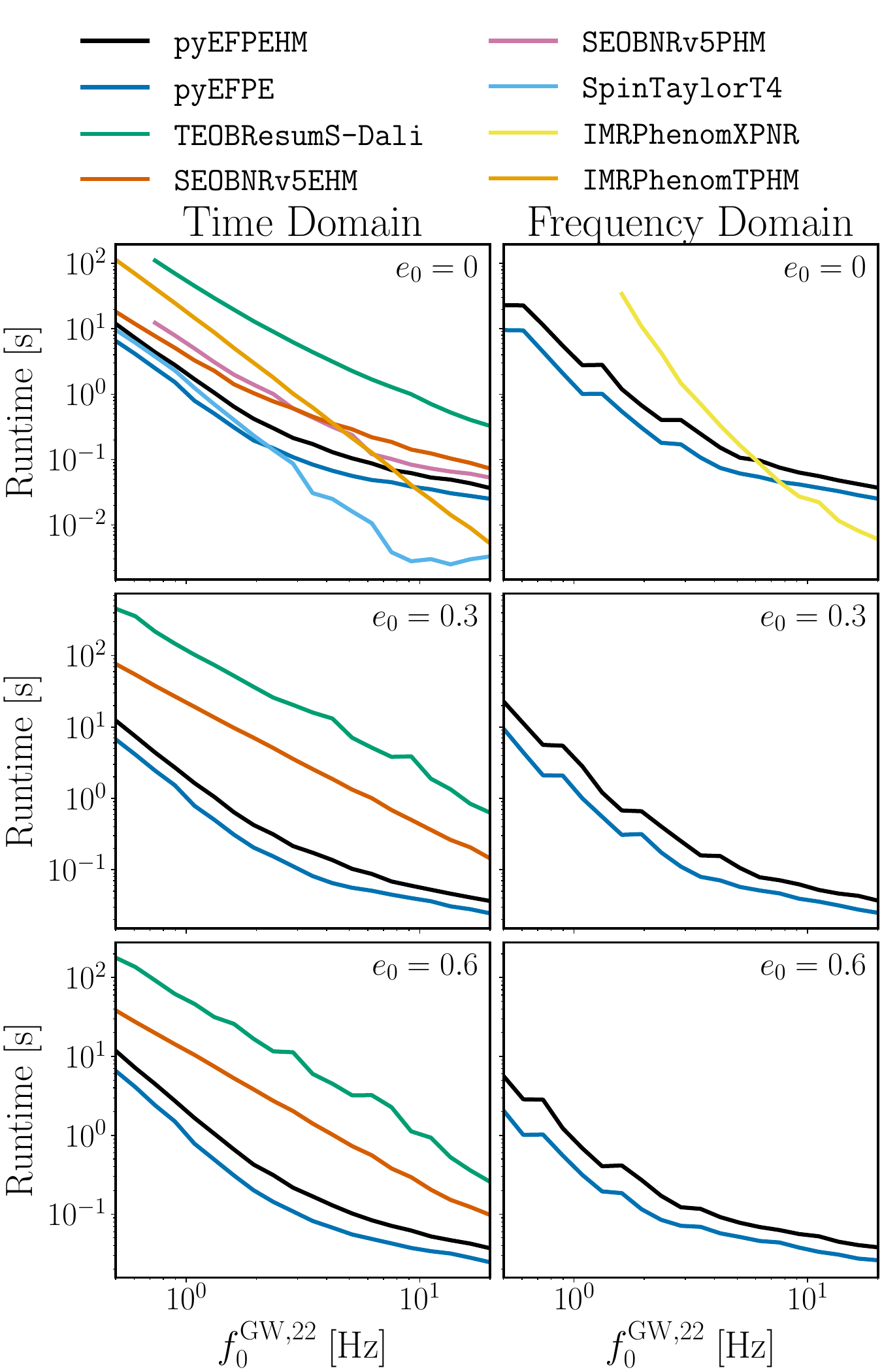}
\caption{\justifying Generation runtime for a selection of waveform models as a function of the minimum waveform generation frequency. The left and right columns of panels show timings for time- and frequency-domain waveform generation, respectively, while the top, middle and bottom rows correspond to eccentricities $e_0 = 0$, $e_0 = 0.3$ and $e_0 = 0.6$. In all cases we fix the system to have primary mass $m_1 = 20\,M_\odot$, secondary mass $m_2 = 10\,M_\odot$, primary spin $\bm{\chi}_{1,0} = (-0.44, -0.26, 0.48)$ and secondary spin $\bm{\chi}_{2,0} = (-0.31, 0.01, -0.84)$, chosen as a representative precessing configuration. For time-domain waveform generation we use a fixed sampling rate of $512\,\mathrm{Hz}$, while for frequency-domain waveforms we use a maximum frequency of $f_\mathrm{max} = 256\,\mathrm{Hz}$ and a frequency resolution of $\Delta f = 1/T$, where $T$ is the waveform duration in seconds rounded up to the nearest power of two. All timings were measured on a single core of an \textit{Intel Core Ultra 7 265U} laptop processor with 32 GB of RAM. For some models, waveform generation failed at low frequencies due to either internal errors or memory limitations, and those models are not shown below the corresponding frequency.
}
\label{fig:plot_wf_timings_N_5}
\end{figure}

While in Sec.~\ref{sec:validate:wf:pyEFPE} we studied the relative \pyEFPEHM runtime with respect to \pyEFPE for settings chosen to reproduce the predecessor, in this section we explore the absolute \pyEFPEHM runtime using more default configurations.

In Fig.~\ref{fig:Amps_0PN_vs_Amps_1PN_runtime_violins} we show the \pyEFPEHM runtime when computing the amplitudes at 0PN and 1PN. As expected, the 1PN amplitude case is significantly more computationally expensive than the 0PN case, with the difference increasing for longer signals, ranging from a factor of $\sim 1.3$ to $\sim 3$. This scaling arises mostly because the 0PN waveform contains only two GW modes ($(l,m) = (2,0), (2,1)$), while the 1PN waveform includes ten modes, substantially increasing the computational cost. Nevertheless, even when including the ten 1PN GW modes, \pyEFPEHM remains highly efficient, with waveform generation taking between $\sim 0.03\,\mathrm{s}$ and $\sim 1 \,\mathrm{s}$ across the parameter space accessible to the LVK.

In Fig.~\ref{fig:plot_wf_timings_N_5} we show waveform generation runtimes for a selection of state-of-the-art waveform models. Even though the models shown do not all contain the same physics (e.g. eccentricity, precession, or merger-ringdown), this comparison helps to better contextualize the computational cost of \pyEFPEHM.

We observe that, for the longest waveform durations, \pyEFPE and \pyEFPEHM are among the most computationally efficient models. This is partly because both models analytically describe the physics on the orbital and precession timescales, needing only to integrate ordinary differential equations (ODEs) on the radiation reaction timescale, making the initialisation cost weakly dependent of signal length. The other models shown integrate ODEs on either the precession or orbital timescale (in the case of EOB), meaning that ODE integration can remain a leading cost even for long signals.

We note that Fig.~\ref{fig:plot_wf_timings_N_5} shows timings for waveforms evaluated on dense grids, where the cost is dominated by the large number of evaluation points. Several data analysis methods, such as multibanding~\cite{Morisaki:2021ngj}, relative binning~\cite{Zackay:2018qdy,Cornish:2021lje}, reduced order quadratures~\cite{Canizares:2013ywa,Smith:2016qas}, or time-frequency representations~\cite{Cornish:2020odn,Bandopadhyay:2025fyx}, instead require waveform evaluations on sparse grids, significantly reducing the evaluation cost. Such methods will be essential for efficiently analysing data from next-generation detectors. In such cases, the initialisation cost becomes even more relevant, further favouring models such as \pyEFPE and \pyEFPEHM whose initialisation cost is weakly dependent of signal length.

Finally, we note that waveform runtimes depend heavily on code implementation. Both \pyEFPE and \pyEFPEHM are written in Python, where interpreter overhead introduces significant slowdowns, particularly in the ODE integration, where we estimate a factor of $\sim \ord{10}$ slowdown relative to compiled code. This partly explains why \pyEFPE and \pyEFPEHM are slower than some other models for short signals, where these interpreter costs dominate while other models benefit from compiled implementations. While we chose Python for ease of development and to facilitate modification of the waveform model, improving the implementation is an important avenue for future work as \pyEFPEHM matures.

\subsection{Comparison with numerical relativity}
\label{sec:validate:NR}

While the comparisons presented in Sec.~\ref{sec:validate:wf} between \pyEFPEHM and other analytical waveform models are valuable for assessing performance across a wide and densely sampled region of parameter space, comparisons with numerical relativity (NR) simulations are widely regarded as the gold standard for validating GW waveform models, since NR simulations provide solutions to the full Einstein equations without relying on perturbative approximations, yielding highly accurate waveforms.

The primary limitation of NR lies in its extreme computational cost, with a single simulation typically requiring hundreds of thousands of CPU hours, and producing waveforms spanning only $\ord{10-100}$ orbital cycles~\cite{Scheel:2025jct}. As a result, the number of available NR simulations remains limited. While several thousands of public binary black hole simulations have been produced, they still cover only a small fraction of the full nine-dimensional parameter space of generic binaries, with particularly sparse sampling of long, eccentric, and precessing systems~\cite{Scheel:2025jct}.

\subsubsection{Mismatch comparisons}
\label{sec:validate:NR:MM}

Here, we compare \pyEFPEHM with a selection of waveforms from the SXS collaboration’s third catalog of binary black hole simulations~\cite{Scheel:2025jct}. 
As in Sec.~\ref{sec:validate:wf}, waveform comparisons are performed using the mismatch. However, because SXS simulations do not provide a gauge-invariant initial average orbital frequency~\cite{Shaikh:2023ypz,Bonino:2024xrv}, we additionally minimize over the initial frequency $f_0^{\mathrm{GW},22}$ used to generate the \pyEFPEHM waveform. Consequently, throughout this section the mismatch is defined as

\begin{equation}
    \overline{\mathcal{MM}}(h_1, h_2) = \min_{\substack{\phi_0, t_0, \kappa, \\ \phi_S, e_0, \ell_0, \\ f_0^{\mathrm{GW},22}}} \mathcal{MM}(h_1, h_2)  \, ,
    \label{eq:MinMisMatchDef_ecc_NR}
\end{equation}

\noindent where, as in Sec.~\ref{sec:validate:wf}, the minimization over the polarization angle $\kappa$ is performed analytically~\cite{Harry:2016ijz}, the minimization over the reference time $t_0$ is efficiently carried out using the FFT, and the minimization over the reference phase $\phi_0$, the in-plane spin rotation angle $\phi_S$, the initial eccentricity $e_0$, the mean anomaly $\ell_0$, and the initial frequency $f_0^{\mathrm{GW},22}$ is performed numerically using a differential evolution algorithm~\cite{Storn:1997uea}.

As an initial guess for the value of $f_0^{\mathrm{GW},22}$, we use the Newtonian-order prediction based on the instantaneous orbital frequency $f^\mathrm{inst}_\mathrm{orb}$, eccentricity $e$, and mean anomaly $\ell$ reported at the reference simulation time, namely

\begin{subequations}
 \label{eq:f0GW_guess}
\begin{align}
    f_0^{\mathrm{GW},22} & = 2 f^\mathrm{inst}_\mathrm{orb} \frac{(1 - e \cos{u})^2}{\sqrt{1-e^2}} \, , \\
    \ell & = u - e \sin{u} \, .
\end{align}    
\end{subequations}

The GW strain obtain from a given NR simulation depends on the direction from which it is observed, specified by the inclination angle $\iota$, the azimuthal angle $\phi$, and the polarization angle $\kappa$. For different choices of $(\iota, \phi, \kappa)$, the mismatch can vary due to effects such as mismodeling of higher-order modes or precession. To account for this dependence, we compute an averaged mismatch over these extrinsic parameters,

\begin{equation}
\overline{\mathcal{MM}}_\mathrm{avg} = \frac{1}{4\pi^2} \int_{-1}^1 \d(\cos\iota) \int_0^\pi \d \phi \int_0^\pi \d \kappa \,  \overline{\mathcal{MM}} \, ,
\label{eq:MinMisMatch_avg_def}
\end{equation}

\noindent which corresponds to assuming isotropically distributed source orientations and polarization angles.

\begin{table*}[t!]
    \centering
    \begin{tabular}{|c|c c c c c|c|c|c|c|c|}
        \hline
         & & & & & & \multicolumn{5}{c|}{$\overline{\mathcal{MM}}_\mathrm{avg}$} \\ \cline{7-11}
         SXS ID & $M [M_\odot]$ & $q$ & $\chi_\mathrm{eff}$ & $\chi_\mathrm{p}$ & $e_\mathrm{SXS}$ & \pyEFPEHM & \pyEFPE & \texttt{Dali} & \texttt{v5EHM} & \texttt{v5PHM} \\ \hline
         SXS:BBH:2619 & $17.0$ & $0.546$ & $-0.227$ & $0.328$ & $6.30 \times 10^{-5}$ & 0.023 & 0.14 &  $9.7 \times 10^{-4}$ & ----- & $3.4 \times 10^{-4}$ \\
         SXS:BBH:2621 & $17.0$ & $0.560$ & $0.0394$ & $0.494$ & $2.22 \times 10^{-4}$ & 0.0026 & 0.031 & 0.0087 & ----- & $9.5 \times 10^{-4}$ \\ \hline
         SXS:BBH:2538 & $32.2$ & $0.333$ & $4.72 \times 10^{-7}$ & $1.35 \times 10^{-9}$ & $0.306$ & 0.020 & 0.057 & 0.0024 & 0.0011 & ----- \\
         SXS:BBH:2549 & $34.4$ & $0.250$ & $-1.73 \times 10^{-8}$ & $4.20 \times 10^{-10}$ & $0.505$ & 0.080 & 0.15 & 0.038 & 0.019 & ----- \\
         SXS:BBH:2561 & $40.2$ & $0.125$ & $-5.29 \times 10^{-7}$ & $1.81 \times 10^{-9}$ & $0.305$ & 0.13 & 0.21 & 0.010 & 0.0042 & ----- \\
         SXS:BBH:3951 & $15.0$ & $0.500$ & $9.66 \times 10^{-8}$ & $7.73 \times 10^{-11}$ & $0.612$ & 0.080 & 0.17 & 0.14 & 0.024 & ----- \\ \hline
         SXS:BBH:0088 & $39.6$ & $1.000$ & $-1.01 \times 10^{-4}$ & $0.500$ & $0.0742$ & 0.0027 & 0.0032 & $4.1 \times 10^{-4}$ & ----- & ----- \\
         SXS:BBH:4286 & $51.0$ & $0.250$ & $-5.72 \times 10^{-4}$ & $0.950$ & $0.103$ & 0.019 & 0.035 & 0.036 & ----- & ----- \\
         SXS:BBH:4290 & $67.5$ & $0.0833$ & $0.0251$ & $0.799$ & $0.103$ &  0.034 & 0.080 & 0.062& ----- & ----- \\ \hline
    \end{tabular}
    \caption{\justifying  NR simulations from the SXS collaboration studied in this work. We list the SXS ID, the total mass $M$ used in the comparisons (in solar masses), the mass ratio $q=m_2/m_1$, the effective inspiral spin parameter $\chi_\mathrm{eff}$ (see Eq.~\eqref{eq:chi_eff_def}), the effective precession spin parameter $\chi_\mathrm{p}$~\cite{Schmidt:2014iyl}, the eccentricity $e_\mathrm{SXS}$ as reported by the SXS collaboration, and the average mismatch (see Eq.~\eqref{eq:MinMisMatch_avg_def}) against \pyEFPEHM, \pyEFPE, \TEOBDali (\texttt{Dali}), \vfiveEHM (\texttt{v5EHM}) and \vfivePHM (\texttt{v5PHM}). The quantities $q$, $\chi_\mathrm{eff}$, $\chi_\mathrm{p}$ and $e_\mathrm{SXS}$ are reported at the reference time of the simulation. Horizontal lines separate quasi-circular precessing (SXS:BBH:2619 and SXS:BBH:2621), eccentric non-spinning (SXS:BBH:2538, SXS:BBH:2549, SXS:BBH:2561, and SXS:BBH:3951), and eccentric precessing simulations (SXS:BBH:0088, SXS:BBH:4286, and SXS:BBH:4290).
    }
    \label{table:sxs_sim_info}
\end{table*}

Since \pyEFPEHM is an inspiral-only waveform model, meaningful comparisons with NR require simulations that are as long as possible. We therefore restrict our analysis to the longest available SXS simulations in each relevant region of parameter space, listed in Table~\ref{table:sxs_sim_info}, and chosen to probe \pyEFPEHM in different physical limits. These include two quasi-circular precessing simulations (SXS:BBH:2619 and SXS:BBH:2621~\cite{Sun:2024kmv}), four highly eccentric non-spinning simulations (SXS:BBH:2538, SXS:BBH:2549, SXS:BBH:2561, and SXS:BBH:3951~\cite{Ramos-Buades:2022lgf})\footnote{No long, highly eccentric spin-aligned simulations are available in the SXS collaboration’s third catalog of binary black hole simulations~\cite{Scheel:2025jct}.}, and three eccentric precessing simulations (SXS:BBH:0088, SXS:BBH:4286, and SXS:BBH:4290).

As in Sec.~\ref{sec:validate:wf:params}, mismatches are computed starting from a minimum frequency of $f_\mathrm{min} = 20\,\mathrm{Hz}$ and using the Advanced LIGO A+ noise PSD~\cite{KAGRA:2013rdx,ObservingScenariosPSDs}. To include as much of the inspiral as possible in the comparison, each NR simulation is rescaled to a total physical mass $M$ such that the initial average GW frequency estimate $f_0^{\mathrm{GW},22}$, computed from Eq.~\eqref{eq:f0GW_guess} at the reference time of the simulation, is $19\,\mathrm{Hz}$. This value is chosen to lie slightly below $f_\mathrm{min}$ in order to avoid edge effects in the mismatch computation. The resulting total masses for each simulation are reported in Table~\ref{table:sxs_sim_info}. Finally, mismatches are computed up to a maximum frequency of $f_\mathrm{max} = 0.8 f_\mathrm{ISCO}$, to minimize contamination from merger-ringdown effects in the comparison. 

Since NR waveforms approximate the exact solution of the two-body problem within numerical accuracy, we include the full available GW signal in the comparison, incorporating all modes up to $l=8$ and memory contributions~\cite{Scheel:2025jct}. To assess the level of improvement provided by \pyEFPEHM and to place its performance in context, we compute mismatches against NR under identical conditions for \pyEFPE, \TEOBDali, \vfiveEHM, and \vfivePHM, restricting comparisons to the parameter regions where each model is applicable (spin-aligned systems for \vfiveEHM and quasi-circular systems for \vfivePHM).

Despite \pyEFPEHM containing only analytical PN information, the orientation-averaged mismatches against NR reported in Table~\ref{table:sxs_sim_info} are relatively small. The observed mismatch trends in the NR comparisons closely follow those found in Sec.~\ref{sec:validate:wf:SEOB} when comparing against \vfive, which closely track NR. In particular, the mismatches obtained with \pyEFPEHM increase for more unequal-mass systems and for larger values of the aligned spin and eccentricity. This behavior is expected, since decreasing $q$ and increasing $e$ cause the system to probe the strong-field, high-velocity regime for a longer fraction of the inspiral, where the PN expansion is less accurate.

Nonetheless, comparing \pyEFPEHM with its predecessor \pyEFPE, we observe that \pyEFPEHM yields smaller mismatches against NR in all cases, being about one order of magnitude lower for the very long quasi-circular simulations tested, a factor of two for the eccentric non-spinning and for the unequal mass eccentric precessing. The smallest improvement is observed for the mildly eccentric-precessing case of SXS:BBH:0088, for which both models yield mismatches $\overline{\mathcal{MM}}_\mathrm{avg} \sim 10^{-3}$. This may be attributed to the equal-mass nature of the system, which suppresses higher-order mode contributions and reduces the impact of higher-order PN corrections.

Comparisons with \TEOBDali show that the relative performance of the two models depends on the configuration. For quasi-circular simulations, \TEOBDali yields smaller mismatches for SXS:BBH:2619, while \pyEFPEHM performs better for SXS:BBH:2621. For eccentric systems, \TEOBDali achieves smaller mismatches for the less eccentric and more unequal-mass cases. In contrast, \pyEFPEHM performs better for the long and highly eccentric system SXS:BBH:3951, consistent with the discussion in Sec.~\ref{sec:validate:wf:TEOB} regarding the absence of higher-order PN terms in the eccentric radiation reaction of \TEOBDali. \pyEFPEHM also yields smaller mismatches for the eccentric precessing simulations SXS:BBH:4286 and SXS:BBH:4290, which may indicate the importance of incorporating eccentric corrections in the precession dynamics, not included self-consistently in \TEOBDali~\cite{Gamba:2024cvy,Albanesi:2025txj}.

Finally, \vfiveEHM and \vfivePHM achieve smaller mismatches than both \pyEFPEHM and \TEOBDali in all cases considered, although their applicability is restricted to spin-aligned eccentric and quasi-circular precessing systems, respectively. Given the much higher accuracy of the \vfive models~\cite{Gamboa:2024hli,Estelles:2025zah}, this suggests that the mismatch trends reported in Sec.~\ref{sec:validate:wf:SEOB} are representative of those expected in comparisons with NR across a broader region of parameter space.

\subsubsection{Waveform comparisons}
\label{sec:validate:NR:WF}

\begin{figure*}
    \centering
    \includegraphics[width=\textwidth]{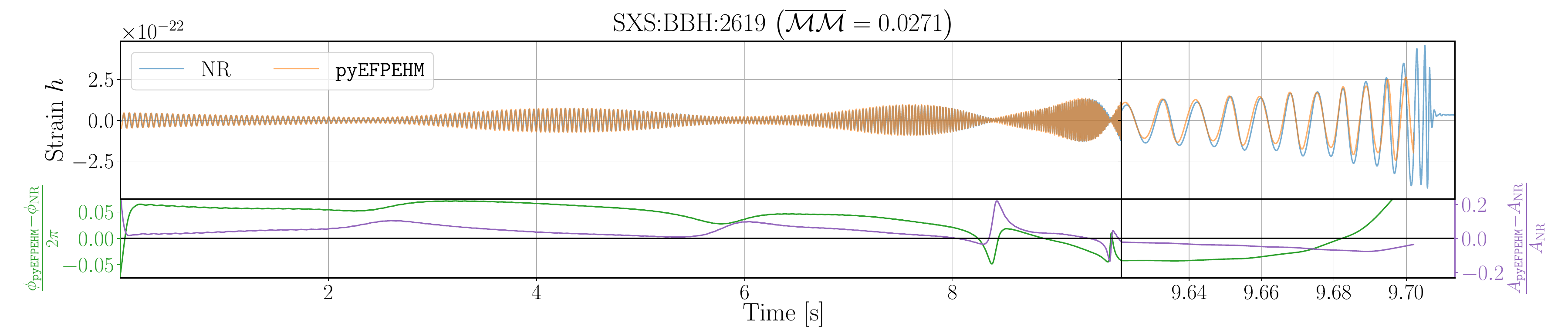}
    \includegraphics[width=\textwidth]{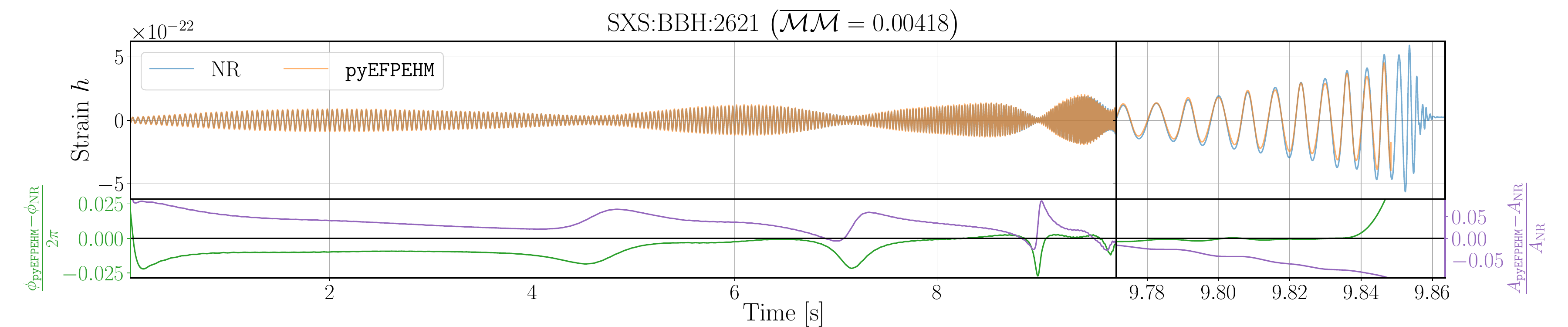}
    \caption{\justifying Comparisons between NR and \pyEFPEHM waveforms for the quasi-circular precessing simulations SXS:BBH:2619 and SXS:BBH:2621, shown for the inclination, phase, and polarization angles that yield the largest mismatch against \pyEFPEHM. Each case is labeled with the corresponding SXS ID and the mismatch for the configuration displayed. The top panels show the time-domain GW strains, while the bottom panels show the phase and relative amplitude differences, computed using Eqs.~\eqref{eq:dphi_cwt} and \eqref{eq:A_cwt}. The \pyEFPEHM waveforms shown are generated using the time-domain implementation up to the ISCO. A $2\,\mathrm{Hz}$ high-pass filter is applied to both waveforms to reduce the visual impact of GW memory.
    }
    \label{fig:sxs_bbh_max_MM_td_wfs_QC}
\end{figure*}

\begin{figure*}
    \centering
    \includegraphics[width=\textwidth]{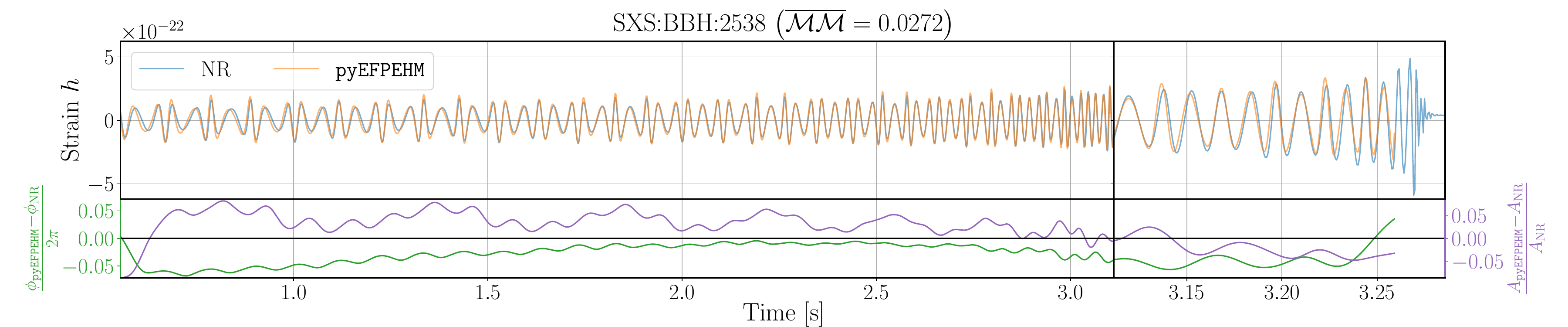}
    \includegraphics[width=\textwidth]{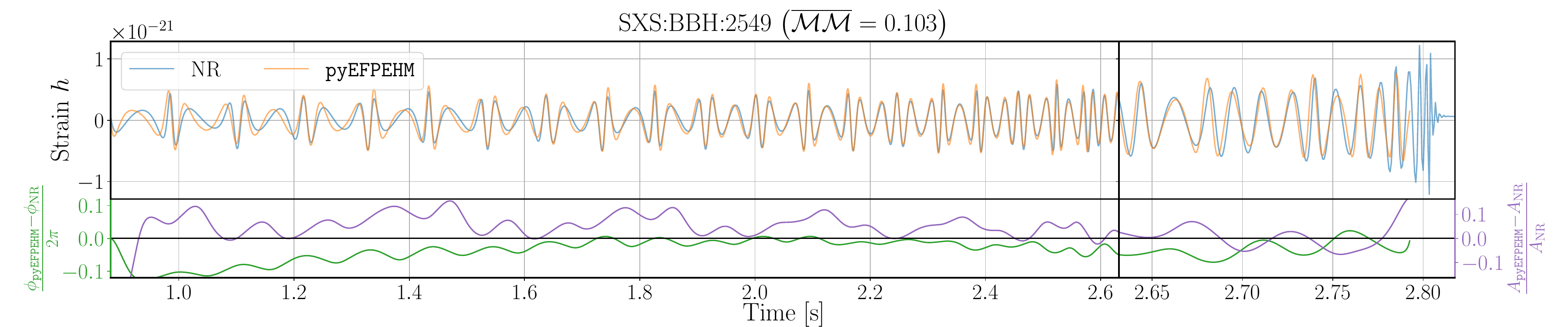}
    \includegraphics[width=\textwidth]{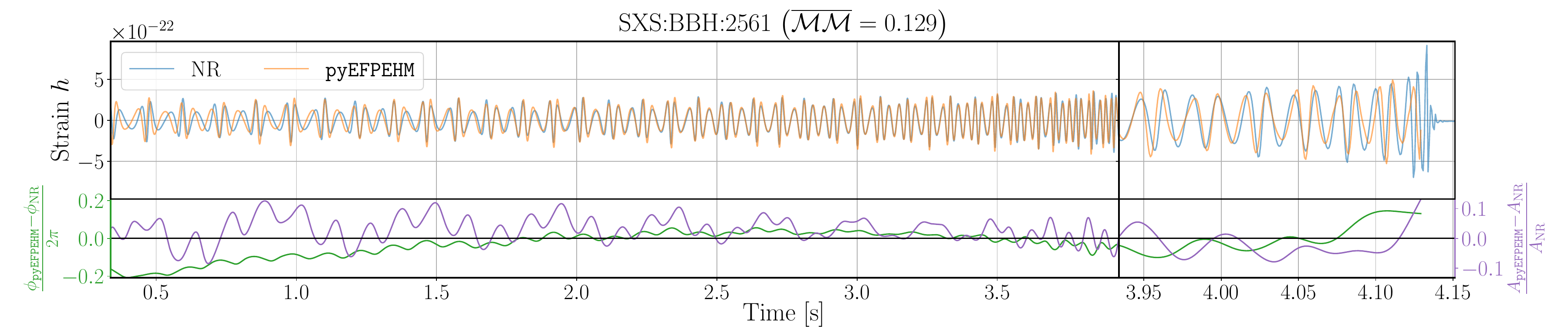}
    \includegraphics[width=\textwidth]{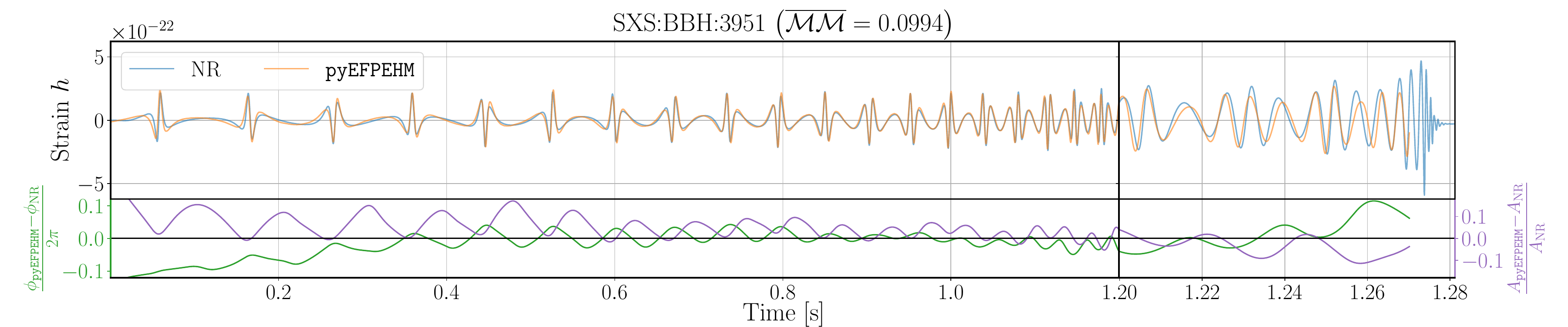}
    \caption{\justifying Comparisons between NR and \pyEFPEHM waveforms for the eccentric non-spinning simulations SXS:BBH:2538, SXS:BBH:2549, SXS:BBH:2561, and SXS:BBH:3951, shown for the inclination, phase, and polarization angles that yield the largest mismatch against \pyEFPEHM. Each case is labeled with the corresponding SXS ID and the mismatch for the configuration displayed. The top panels show the time-domain GW strains, while the bottom panels show the phase and relative amplitude differences, computed using Eqs.~\eqref{eq:dphi_cwt} and \eqref{eq:A_cwt}. The \pyEFPEHM waveforms shown are generated using the time-domain implementation up to the ISCO. A $2\,\mathrm{Hz}$ high-pass filter is applied to both waveforms to reduce the visual impact of GW memory.}
    \label{fig:sxs_bbh_max_MM_td_wfs_E}
\end{figure*}

\begin{figure*}
    \centering
    \includegraphics[width=\textwidth]{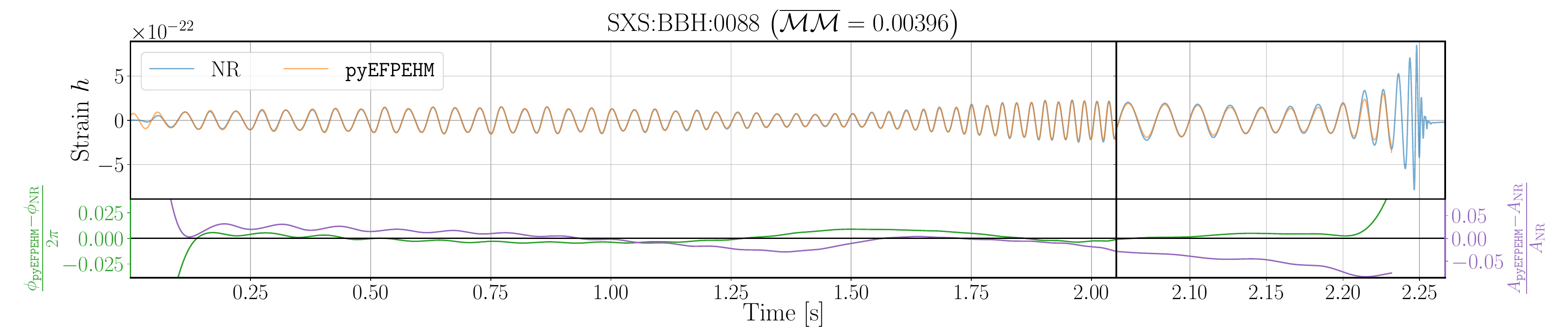}
    \includegraphics[width=\textwidth]{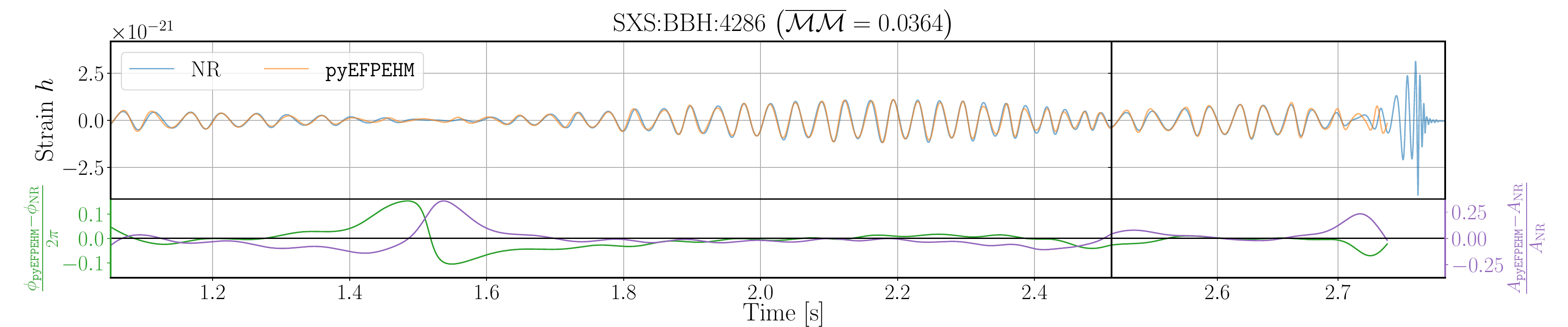}
    \includegraphics[width=\textwidth]{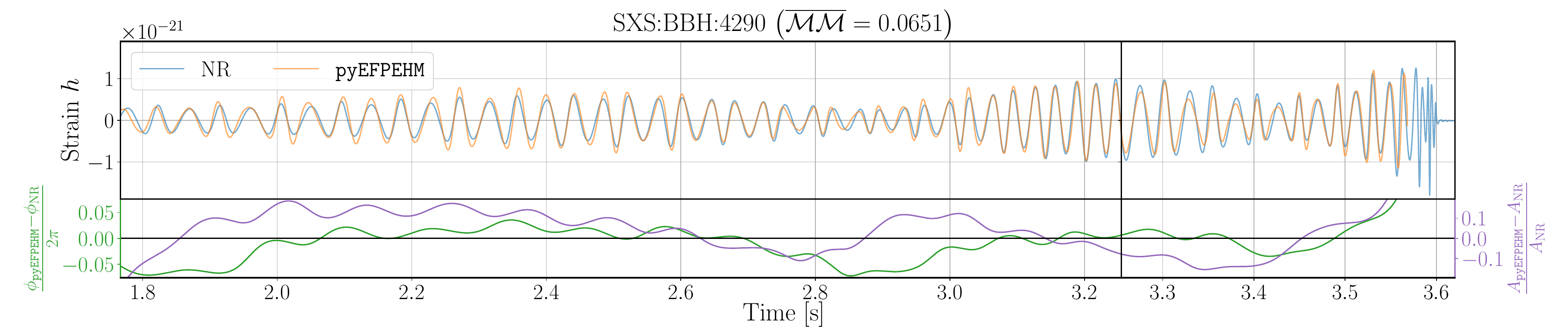}
    \caption{\justifying Comparisons between NR and \pyEFPEHM waveforms for the eccentric precessing simulations SXS:BBH:0088, SXS:BBH:4286, and SXS:BBH:4290, shown for the inclination, phase, and polarization angles that yield the largest mismatch against \pyEFPEHM. Each case is labeled with the corresponding SXS ID and the mismatch for the configuration displayed. The top panels show the time-domain GW strains, while the bottom panels show the phase and relative amplitude differences, computed using Eqs.~\eqref{eq:dphi_cwt} and \eqref{eq:A_cwt}. The \pyEFPEHM waveforms shown are generated using the time-domain implementation up to the ISCO. A $2\,\mathrm{Hz}$ high-pass filter is applied to both waveforms to reduce the visual impact of GW memory.}
    \label{fig:sxs_bbh_max_MM_td_wfs_EP}
\end{figure*}

To better understand the origin of the mismatches, in Figs.~\ref{fig:sxs_bbh_max_MM_td_wfs_QC}, \ref{fig:sxs_bbh_max_MM_td_wfs_E}, and \ref{fig:sxs_bbh_max_MM_td_wfs_EP} we show the GW strains for the inclination, phase, and polarization angles that yield the largest mismatch against \pyEFPEHM for each simulation. While mismatch calculations are performed using the frequency-domain implementation of \pyEFPEHM based on the SUA up to $f_\mathrm{max}=0.8 f_\mathrm{ISCO}$, the waveforms shown in these figures are generated using the time-domain implementation up to the ISCO. To reduce the visual impact of GW memory, which can obscure the comparison, we apply a $2\,\mathrm{Hz}$ high-pass filter to both waveforms.

Since for most configurations the model and NR waveforms are very similar and difficult to distinguish by eye, in the bottom panels we also show their phase and relative amplitude differences. Due to eccentricity, the signals are not quasi-monochromatic, and defining an instantaneous amplitude and phase is not straightforward. To tackle this, we construct a time-frequency representation of the signals using a continuous wavelet transform,

\begin{equation}
    H(t, f) = \int_{-\infty}^\infty  \d t' h(t') w_Q^{*}(t'-t;f) \, ,
    \label{eq:cwt}
\end{equation}

\noindent where we employ Morlet wavelets with constant quality factor $Q$,

\begin{equation}
    w_Q(t, f) = \frac{\sqrt{2 \pi} f}{Q} \exp\left\{\rmi 2 \pi f t - \frac{(2 \pi f t)^2}{2 Q^2}\right\} \, .
    \label{eq:fd_morlet_wavelets}
\end{equation}

The quality factor $Q$ controls the effective number of cycles each wavelet has, and therefore the trade-off between time and frequency resolution. We adopt $Q=8$ as an intermediate choice. From this representation we define a time-dependent amplitude as

\begin{equation}
    A(t)= \int_{f_\mathrm{min}}^{f_\mathrm{max}}  \d f \frac{1}{f} |H(t, f)|^2 \, ,
    \label{eq:A_cwt}
\end{equation}

\noindent where the $1/f$ factor accounts for the constant-$Q$ transform. Similarly, we define the phase difference between two signals as

\begin{equation}
    \phi_1(t) - \phi_2(t)= \arg\left\{\int_{f_\mathrm{min}}^{f_\mathrm{max}}  \d f \frac{1}{f} H_1(t, f) H_2^{*}(t, f) \right\} \, .
    \label{eq:dphi_cwt}
\end{equation}

For the frequency integrals we use the same range $[20\,\mathrm{Hz},0.8 f_\mathrm{ISCO}]$ adopted in the mismatch calculations.

For the quasi-circular precessing simulations shown in Fig.~\ref{fig:sxs_bbh_max_MM_td_wfs_QC}, \pyEFPEHM exhibits an amplitude evolution consistent with NR up to the late inspiral, and for SXS:BBH:2621, the phase also remains coherent. In contrast, for SXS:BBH:2619, which has larger anti-aligned spin, the phase evolution is consistent at early times but begins to deviate as the system approaches merger.

For the eccentric non-spinning simulations shown in Fig.~\ref{fig:sxs_bbh_max_MM_td_wfs_E}, several common features are observed. In particular, \pyEFPEHM has a smaller phase than NR at early times, the two agree well at intermediate times, and discrepancies reappear as the system approaches merger. This behavior is likely related to the fact that the eccentricity in \pyEFPEHM is determined by minimizing the mismatch, which preferentially enforces agreement in the frequency range contributing most strongly to the signal-to-noise ratio. We also find that both amplitude and phase differences become more pronounced for more unequal-mass configurations, as illustrated by SXS:BBH:2561, which has the smallest mass ratio ($q=0.125$) among the eccentric simulations. This trend highlights the importance of higher-order eccentric corrections, which are currently included only up to 3PN order in the phasing and 1PN order in the amplitudes.

Finally, for the eccentric precessing simulations shown in Fig.~\ref{fig:sxs_bbh_max_MM_td_wfs_EP}, the agreement is generally very good for the more comparable-mass systems (SXS:BBH:0088 and SXS:BBH:4286), while it deteriorates for the case of SXS:BBH:4290, which has a mass ratio of $q=1/12$. Nonetheless, meaningful comparisons in the eccentric precessing regime are currently limited by the sparse coverage of the SXS catalog, which at present contains only relatively short ($N_\mathrm{orbits} \lesssim 40$) and mildly eccentric ($e \lesssim 0.1$) simulations.

\subsection{Parameter estimation}
\label{sec:validate:PE}

Bayesian parameter estimation (PE) for compact binary coalescences~\cite{Veitch:2014wba, Thrane:2018qnx} is one of the primary applications of waveform models such as \pyEFPEHM. The computational efficiency demonstrated in Sec.~\ref{sec:validate:wf:pyEFPEHM} suggests that Bayesian PE with \pyEFPEHM should be feasible for LVK-like analyses. Similarly, the mismatches obtained against other analytical models and NR in Secs.~\ref{sec:validate:wf} and~\ref{sec:validate:NR} can be used to estimate the signal-to-noise ratio up to which PE is not expected to be significantly biased~\cite{Lindblom:2008cm}, using the indistinguishability criterion

\begin{equation}
    \mathcal{MM} \lesssim \frac{N_p}{2 \rho^2} \, ,
    \label{eq:indistinguishable_MM}
\end{equation}

\noindent where $\rho$ denotes the SNR of the signal and $N_p$ is the number of parameters being inferred. In the eccentric precessing case considered here, PE requires exploring a 17-dimensional parameter space, including the two component masses, the six spin components, the orbital eccentricity and mean anomaly, the inclination angle, the coalescence phase and time, the luminosity distance, the sky location (right ascension and declination), and the polarization angle. Nonetheless, the accuracy of parameter recovery cannot be inferred from mismatches alone~\cite{Thompson:2025hhc}. While mismatches quantify overall waveform agreement, parameter biases depend on the detailed structure of waveform errors and on whether these errors can be absorbed by shifts in the inferred parameters. Furthermore, for such a high-dimensional parameter space, a typical PE analysis requires $\ord{10^8}$ likelihood evaluations, making PE a demanding test of the numerical stability and robustness of the waveform model. Therefore, performing PE provides a stringent and informative validation of the waveform model.

Here, we perform PE using \texttt{bilby}~\cite{Ashton:2018jfp,Smith:2019ucc,Romero-Shaw:2020owr}, employing its implementation of the \texttt{dynesty}~\cite{Speagle:2020dqf} nested sampling algorithm. To assess the PE performance of \pyEFPEHM in a controlled setting, we perform zero-noise injections~\cite{Rodriguez:2013oaa}. This approach avoids both the statistical fluctuations associated with specific noise realizations and the additional uncertainty inherent in real-event analyses, for which the true source parameters are unknown. Synthetic signals are injected into the LIGO Hanford (H1), LIGO Livingston (L1), and Virgo detectors, assuming the LIGO A+ and Virgo AdV+ sensitivity curves projected for O5~\cite{KAGRA:2013rdx,ObservingScenariosPSDs}. The data are analyzed from a minimum frequency of $f_\mathrm{min}=20\,\mathrm{Hz}$, which is also taken as the initial frequency of the \pyEFPEHM waveform. Consequently, the initial eccentricity $e_0$ is defined at $20\,\mathrm{Hz}$.

We adopt broad, uninformative priors that are uniform in the component masses, spin magnitudes, and coalescence time, and isotropic in sky location as well as in the binary and spin orientations. The luminosity distance prior assumes sources are uniformly distributed in comoving volume, using the \textit{Planck15} cosmology~\cite{Planck:2015fie}. When considering eccentricity, we further impose flat priors on the initial eccentricity $e_0 \in [0, 0.5]$ and on the mean anomaly $\ell_0 \in [0, 2\pi]$. The full details of the priors used are listed in Table~\ref{tab:priors} of Appendix~\ref{sec:appendix:PE_extra}.

\subsubsection{\pyEFPEHM injection -- recovery}
\label{sec:validate:PE:pyEFPE}

As in Ref.~\cite{Morras:2025nlp}, we begin with a basic consistency test in which a signal generated with \pyEFPEHM is recovered using the same waveform model. Although such self-injections do not probe waveform modeling systematics, they provide a validation of the numerical stability and internal consistency of both the waveform implementation and the PE pipeline. In particular, they allow us to verify that the true parameters are recovered without bias when the assumed signal model is the one injected.

We consider an injection with detector-frame component masses $m_1 = 10\,M_\odot$ and $m_2 = 2.0\,M_\odot$, and a moderate initial eccentricity $e_0 = e_{20\,\mathrm{Hz}} = 0.2$. We analyze $32\,\mathrm{s}$ of data up to a maximum frequency of $300\,\mathrm{Hz}$, chosen such that $f_\mathrm{max} \approx 0.8\,f_\mathrm{ISCO}$. Additional details on the injected signal are provided in App.~\ref{sec:appendix:PE_extra}.

\begin{figure}[t!]
\centering  
\includegraphics[width=0.5\textwidth]{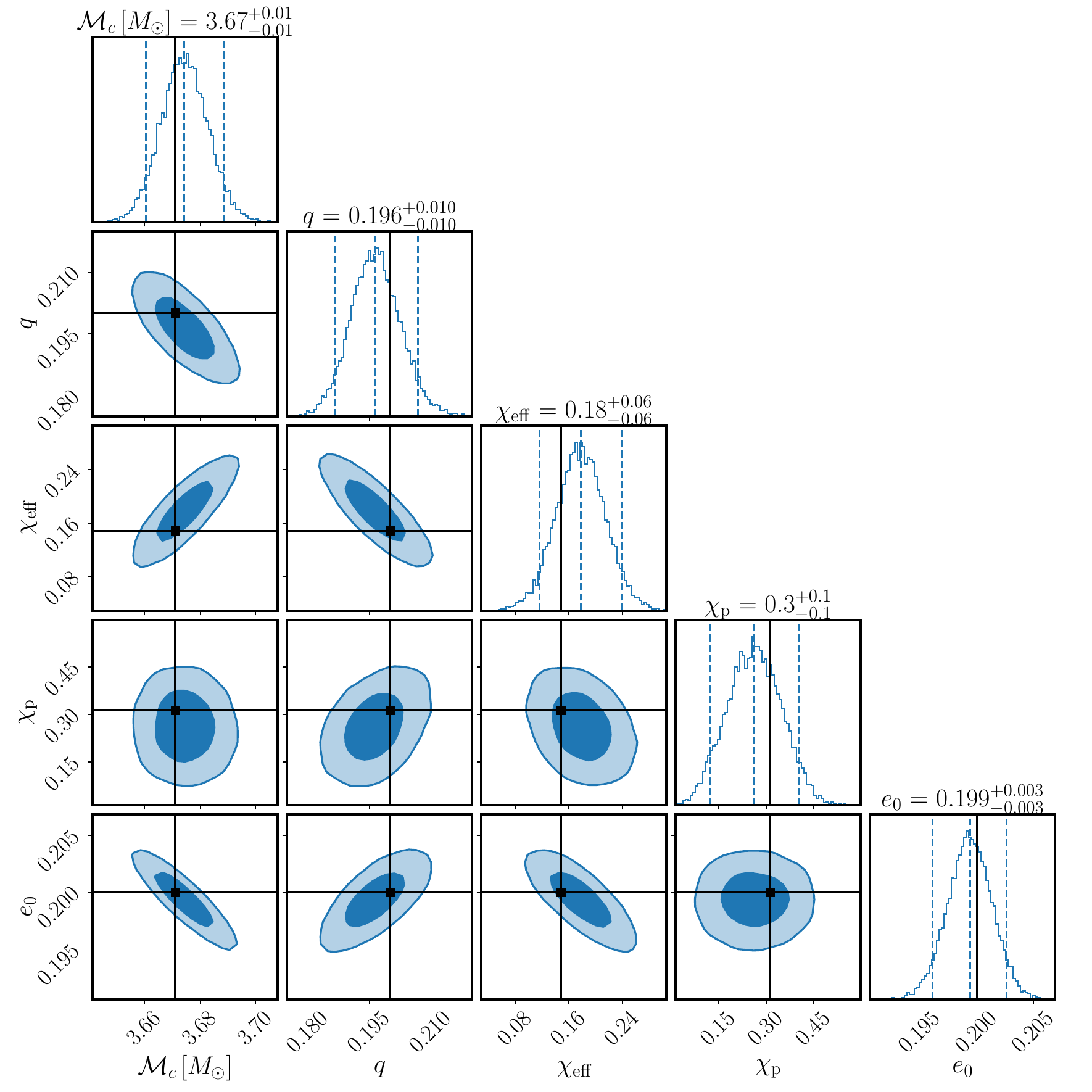}
\caption{\justifying Corner plot showing the joint posterior distributions of selected intrinsic parameters from the \pyEFPEHM injection--recovery study, including the chirp mass $\mathcal{M}_c$, mass ratio $q$, effective inspiral spin parameter $\chi_\mathrm{eff}$, effective precession spin parameter $\chi_\mathrm{p}$, and initial eccentricity $e_0$. The diagonal panels show the marginal distributions for each parameter, together with the median and 90\% credible intervals. The off-diagonal panels display the bivariate correlations between parameter pairs, with contours indicating the $50\%$ and $90\%$ credible regions. The black lines denote the injected parameter values.}
\label{fig:inj_EFPE_rec_EFPE_intrinsic_important}
\end{figure}

In Fig.~\ref{fig:inj_EFPE_rec_EFPE_intrinsic_important} we show a corner plot of the posterior distributions for the most relevant intrinsic parameters. As a proxy for the measurement of precession, we use the effective precession spin parameter $\chi_\mathrm{p}$ defined in Ref.~\cite{Schmidt:2014iyl} as

\begin{equation}
    \chi_\mathrm{p} = \max\left\{ \chi_{1\perp,0}, \frac{q (4 q + 3)}{4 + 3 q} \chi_{2\perp,0} \right\} \, ,
    \label{eq:chi_p_def}
\end{equation}

\noindent where $\chi_{i \perp,0} = \sqrt{\chi_{i x,0}^2 + \chi_{i y,0}^2}$ is the initial magnitude of the in-plane spin. Posterior distributions for a more complete set of parameters are shown in Fig.~\ref{fig:inj_EFPE_rec_EFPE_all_important} in App.~\ref{sec:appendix:PE_extra}.

In both Figs.~\ref{fig:inj_EFPE_rec_EFPE_intrinsic_important} and \ref{fig:inj_EFPE_rec_EFPE_all_important}, we find that the recovered posteriors are in good agreement with the injected values. The distributions are approximately Gaussian and centered near the true parameters, providing a validation of the numerical stability and internal consistency of the waveform model and the PE pipeline.

Despite the moderate network signal-to-noise ratio of $18.4$, the intrinsic parameters are tightly constrained, with sub-percent uncertainties in $\mathcal{M}_c$, $q$, and $\chi_\mathrm{eff}$, and strong constraints on both $\chi_\mathrm{eff}$ and $\chi_\mathrm{p}$. As discussed in Ref.~\cite{Morras:2025nlp} and observed in Ref.~\cite{Morras:2025xfu}, eccentric binaries exhibit multiple eccentric harmonics. Requiring these harmonics to remain phase-coherent enables independent measurements of both the accumulated periastron advance phase $\delta\lambda$ and the orbital phase $\lambda$ with $\ord{1\,\mathrm{rad}}$ precision. Since both phases accumulate to large values over the inspiral, they can be determined with high relative precision. Moreover, because $\delta\lambda$ and $\lambda$ depend differently on the intrinsic parameters, this provides additional information to break parameter degeneracies and significantly improve intrinsic parameter constraints.

In contrast to the analogous analysis performed in Ref.~\cite{Morras:2025nlp} using \pyEFPE, Fig.~\ref{fig:inj_EFPE_rec_EFPE_all_important} shows that with \pyEFPEHM the extrinsic parameters, in particular the inclination $\iota_0$ and luminosity distance $d_L$, are also well constrained. This improvement is due to the inclusion of higher-order modes, which in quasi-circular binaries are known to not only improve the mass ratio and spin constraints, but also break the distance--inclination degeneracy~\cite{Ohme:2013nsa,Hannam:2013uu,Usman:2018imj,Mills:2020thr}, and similarly enhance parameter estimation for eccentric binaries.

\subsubsection{\vfivePHM injection -- \pyEFPEHM recovery}
\label{sec:validate:PE:v5PHM}

In this section, we assess how well \pyEFPEHM recovers the parameters of a precessing, quasi-circular binary ($e_0 = 0$). This constitutes an important consistency check for an eccentric waveform model, as it tests whether the model spuriously infers nonzero eccentricity when analyzing quasi-circular signals.

For this test, we inject an \vfivePHM signal with the same parameters as the \pyEFPEHM injection of Sec.~\ref{sec:validate:PE:pyEFPE}, but setting the eccentricity to zero, resulting in a network signal-to-noise ratio of $22.5$. We choose to inject the \vfivePHM waveform because this model, which we already compared to \pyEFPEHM using mismatch studies in Sec.~\ref{sec:validate:wf:SEOB}, provides an accurate representation of numerical-relativity waveforms~\cite{Estelles:2025zah}.

We analyze this zero-noise injection using three different signal models. The first uses \vfivePHM itself, providing a reference for the expected recovery when the assumed signal model matches the injected one. The second uses \pyEFPEHM with the eccentricity fixed to zero $(e_0 = 0)$, allowing a direct comparison under identical priors and analysis settings. The third uses \pyEFPEHM with a uniform prior on the initial eccentricity, $e_0 \in [0, 0.5]$, in order to assess the behavior of the eccentric model when recovering a quasi-circular signal.

\begin{figure}[t!]
\centering  
\includegraphics[width=0.5\textwidth]{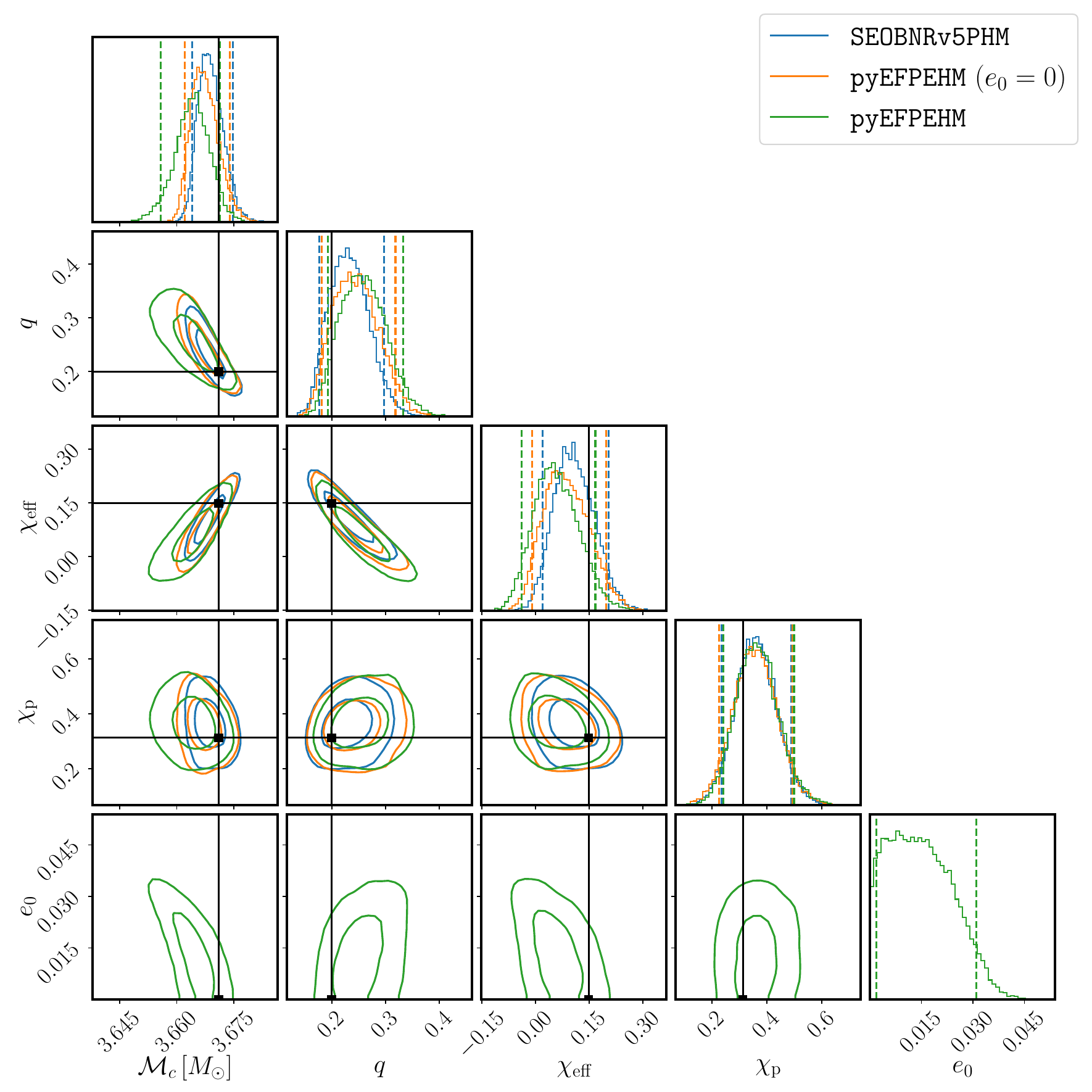}
\caption{\justifying Corner plot showing the joint posterior distributions of selected intrinsic parameters for the \vfivePHM injection, recovered using \vfivePHM, \pyEFPEHM with the eccentricity fixed to zero (labeled \pyEFPEHM~$(e_0 = 0)$), and \pyEFPEHM with a uniform prior on the initial eccentricity $e_0 \in [0, 0.5]$ (labeled \pyEFPEHM). The parameters shown are the chirp mass $\mathcal{M}_c$, mass ratio $q$, effective inspiral spin parameter $\chi_\mathrm{eff}$, effective precession spin parameter $\chi_\mathrm{p}$, and initial eccentricity $e_0$. The diagonal panels display the marginal distributions for each parameter, while the off-diagonal panels show the bivariate correlations, with contours indicating the $50\%$ and $90\%$ credible regions. The black lines denote the injected parameter values.}
\label{fig:inj_v5PHM_32s_intrinsic_important}
\end{figure}

In Fig.~\ref{fig:inj_v5PHM_32s_intrinsic_important}, we show the posterior distributions for the most relevant intrinsic parameters from these three PE analyses. Posterior distributions for a more complete set of parameters are shown in Fig.~\ref{fig:inj_v5PHM_32s_all_important} in App.~\ref{sec:appendix:PE_extra}.

We find that, despite the relatively high signal-to-noise ratio of $22.5$, the posterior distributions obtained with the two quasi-circular recoveries (\vfivePHM and \pyEFPEHM~$(e_0 = 0)$) are statistically consistent with each other and with the injected parameter values. Small deviations are visible in the posteriors of the mass ratio $q$ and effective inspiral spin $\chi_\mathrm{eff}$, which are likely attributable to the purely PN description in \pyEFPEHM becoming less accurate in the regime of extreme mass ratios and high spins, as already observed in Sec.~\ref{sec:validate:wf}.

Importantly, when recovering the signal with the eccentric \pyEFPEHM model, the inferred eccentricity is consistent with zero and constrained to be $e_0 < 0.03$ at $90\%$ credible level. Therefore, for this precessing quasi-circular injection, \pyEFPEHM does not spuriously identify eccentricity.

Figure~\ref{fig:inj_v5PHM_32s_intrinsic_important} also shows the well known correlation between the eccentricity and the chirp mass~\cite{Favata:2021vhw,Morras:2025nlp}. Although the eccentricity posterior is compatible with zero, the prior enforces $e_0 \geq 0$, which results in posterior support at small positive values of $e_0$. This is compensated by a shift of the chirp mass toward lower values in order to preserve the overall duration and phase evolution of the waveform.

\section{Conclusion}
\label{sec:conclusion}

Accurate and computationally efficient waveform models for eccentric and precessing binaries are essential for probing formation channels of compact objects, such as dynamical captures and binary-binary interactions~\cite{Wen:2002km,Samsing:2013kua,Rodriguez:2018pss,Zevin:2018kzq,Tagawa:2019osr,Sedda:2020wzl}, hierarchical mergers~\cite{Gerosa:2021mno}, and binaries in which the Lidov--Kozai mechanism~\cite{Zeipel:1910,Lidov:1962,Kozai:1962} plays a central role, such as in stellar triples~\cite{Antonini:2017tgo,Trani:2021tan,Stegmann:2025clo,Antonini:2017ash} or driven by perturbations from a massive black hole at the center of the host galaxy~\cite{Antonini:2012ad,Stephan:2016kwj}. As detector sensitivities improve, future observations will probe further into the inspiral and detect higher signal-to-noise ratio events, making physical effects such as eccentricity, spin precession, and higher-order modes increasingly relevant for GW astronomy. Nonetheless, waveforms including all of these effects at the same time remain limited in accuracy and computational speed.

In this work we introduce \pyEFPEHM, an accurate and computationally efficient PN waveform model for the inspiral of eccentric precessing compact binaries, capable of generating GW strains in both the time and frequency domains. \pyEFPEHM constitutes a significant improvement over its predecessor \pyEFPE, extending both the physical content and the accuracy of the model.

To improve the phasing accuracy of \pyEFPEHM, we extended the eccentric PN evolution equations, currently known only up to 3PN order, by incorporating all known quasi-circular PN corrections. As shown in Sec.~\ref{sec:HighPNQC}, above 2.5PN order the quasi-circular contribution dominates at each PN order. These additions include adiabatic (non-spinning) tidal contributions up to 7.5PN order~\cite{Dones:2024odv} and adiabatic spin-tidal effects at 6.5PN order~\cite{Abdelsalhin:2018reg}, enabling applications to binaries containing neutron stars.

In a similar spirit, in Sec.~\ref{sec:HighPN_MSA} we extended the multiple-scale analysis (MSA) solution of the spin-precession equations to higher PN orders by adding quasi-circular corrections up to 4PN order to the 2PN eccentric equations used in \pyEFPE. Finally, in Sec.~\ref{sec:HMs} we incorporated eccentric waveform amplitude corrections up to 1PN order, including the GW multipoles $(l,|m|) = (2,2)$, $(2,1)$, $(2,0)$, $(3,3)$, $(3,2)$, $(3,1)$, $(3,0)$, $(4,4)$, $(4,2)$, and $(4,0)$.

In Sec.~\ref{sec:validate} we thoroughly tested and validated \pyEFPEHM. In particular, we showed that despite the inclusion of significantly richer physics, \pyEFPEHM remains computationally efficient. Improvements in the code implementation make it up to twice as fast as \pyEFPE for long signals when only the leading-order amplitudes are included, and only about $30\%$ slower than \pyEFPE when all multipoles are used.

In Sec.~\ref{sec:validate:wf} we assessed the accuracy of \pyEFPEHM through mismatch studies against other waveform models, including \pyEFPE, \STfour, \vfiveEHM, \vfivePHM, and \TEOBDali, as well as against numerical relativity. These studies suggest that \pyEFPEHM provides an accurate description of the inspiral of eccentric precessing compact binaries up to very close to merger. However, the accuracy in the late inspiral degrades for systems with very unequal masses ($m_2/m_1 \lesssim 0.1$), significant aligned spins ($|\chi_\mathrm{eff}| \gtrsim 0.5$), and very high eccentricities ($e \gtrsim 0.6$), where the PN expansion is expected to become less reliable.

Finally, in Sec.~\ref{sec:validate:PE} we show the capability of \pyEFPEHM to perform full Bayesian parameter estimation on simulated data from current ground-based GW detector networks, and to accurately recover the parameters of signals described by both \pyEFPEHM and \vfivePHM.

While \pyEFPEHM represents a substantial improvement in physical content and accuracy, several extensions are needed for future applications. To improve inspiral accuracy, higher-order PN amplitude corrections~\cite{Mishra:2015bqa,Boetzel:2019nfw,Khalil:2021txt,Henry:2023tka}, including non-linear memory~\cite{Christodoulou:1991cr,Blanchet:1992br,Favata:2011qi} and mode asymmetries~\cite{Boyle:2014ioa,Ghosh:2023mhc}, should be incorporated. To better model the late inspiral, especially for extreme mass ratios and large aligned spins, self-force information~\cite{Honet:2025lmk,Mathews:2025txc} and calibration to EOB and NR waveforms can be incorporated~\cite{Pratten:2020fqn,Hamilton:2025xru}. Finally, the most significant current limitation is the lack of a merger–ringdown model, restricting \pyEFPEHM to low-mass systems. Nevertheless, previous studies suggest that the merger–ringdown phase could likely be captured by fitting physically motivated ans\"atze to NR and extreme-mass-ratio simulations~\cite{Varma:2019csw,Pratten:2020fqn,Pompili:2023tna}.

Therefore, despite its current limitations, \pyEFPEHM represents a significant step toward accurate, efficient, and physically complete waveform models for gravitational waves from compact binary coalescences, enabling future studies of their formation channels and dynamics.

\section*{Code Availability}

The repository containing the waveform model, along with scripts to reproduce the figures in this paper, will be made available at Ref.~\cite{pyEFPEHM_repo}. 

\section*{Acknowledgments}

We thank Maria Haney for her helpful comments and suggestions as internal LIGO reviewer.
GM's and AB's research is supported in part by the European Research Council (ERC) Horizon Synergy Grant “Making Sense of the Unexpected in the Gravitational-Wave Sky” grant agreement no. GWSky–101167314.
G.P. is very grateful for support from a Royal Society University Research Fellowship URF{\textbackslash}R1{\textbackslash}221500 and RF{\textbackslash}ERE{\textbackslash}221015.
G.P. and P.S. acknowledge support from STFC grants ST/V005677/1 and ST/Y00423X/1, and UK Space Agency grant UKRI/ST/B000971/1. 
P.S. also acknowledges support from a Royal Society Research Grant RG{\textbackslash}R1{\textbackslash}241327.
We also acknowledge the computational resources provided by the Max Planck Institute for Gravitational Physics (Albert Einstein Institute), Potsdam, in particular, the Hypatia cluster. The material presented in this paper is based upon work supported by National Science Foundation’s (NSF) LIGO Laboratory, which is a major facility fully funded by the NSF.
This manuscript has LIGO document number P2600118.


\onecolumngrid
\appendix

\section{Explicit form of high-order post-Newtonian quasi-circular corrections to the phasing}
\label{sec:appendix:PN_QC}

In this Appendix, we list the quasi-circular (QC) corrections beyond 3PN order that have been included in \pyEFPEHM, following the notation of Ref.~\cite{Morras:2025nlp}. As in Ref.~\cite{Morras:2025nlp}, we work in harmonic coordinates~\cite{Choquet-Bruhat:2009xil}, but adopt the Newton-Wigner~\cite{Khalil:2023kep,Pryce:1948pf,Newton:1949cq} spin supplementary condition (SSC) instead of the covariant Tulczyjew-Dixon~\cite{Tulczyjew:1959zza,Dixon:1970zza} SSC. Both SSCs yield identical coefficients for the lower PN orders detailed in Ref.~\cite{Morras:2025nlp}. Guided by the structure of these lower-order terms and the expected form of higher-order corrections, we apply the following substitutions to the QC expressions available in the literature:

\begin{subequations}
\label{eq:QC_to_Ecc}
\begin{align}
    x & \longrightarrow y^2 = \frac{x}{1 - e^2} \, , \label{eq:QC_to_Ecc:x} \\
    \log{x} & \longrightarrow 2 \log{\frac{2 y (1-e^2)^{3/2}}{1 + \sqrt{1 - e^2}}} \, , \label{eq:QC_to_Ecc:logx} \\
    M \frac{\d}{\d t} & \longrightarrow \D = \frac{M}{(1 - e^2)^{3/2}} \frac{\d}{\d t} \, , \label{eq:QC_to_Ecc:d_dt}
\end{align}
\end{subequations}

\noindent where $x = (M \omega)^{2/3}$ is the PN expansion parameter commonly used in the QC case, and the expressions on the left- and right-hand sides agree in the QC limit. Using these rules, high-order QC corrections can be incorporated into the evolution equation for $y$, which we write as

\begin{align}
  \label{eq:RR_eqs_appendix}
 \D y = \nu y^9 \left(a_0 + \sum_{n=2}^{9} a_n  y^n \right)\, .
\end{align}

We begin with corrections for systems without spin components perpendicular to the orbital angular momentum. Beyond 3PN, the known QC corrections include the 4.5PN non-spinning terms~\cite{Blanchet:2023bwj}

\begin{subequations}
\label{eq:aNSQC}
\begin{align}
    a_7^\mathrm{NS,QC} =& \pi  \left(-\frac{883}{126} + \frac{71735}{189} \nu  + \frac{73196}{189} \nu^2 \right) \, , \label{eq:aNSQC:7} \\
    a_8^\mathrm{NS,QC} =& \frac{3959271176713}{3972969000} - \frac{5776}{315}\pi^2 + \frac{1995856}{11025} \gamma_E + \frac{2140336}{11025} \log{2} - \frac{9477}{49}\log{3} \nonumber\\
    & + \left(-\frac{317589100793}{61122600}-\frac{1472377}{2520}  \pi^2 -\frac{372800}{441} \gamma_E - \frac{8561344}{11025} \log{2} + \frac{37908}{49} \log{3}\right) \nu \nonumber\\
    & + \left(\frac{504221849}{51030}-\frac{22099}{60} \pi^2 \right) \nu^2-\frac{1909807}{9720} \nu^3+\frac{917}{180} \nu^4 + \left(-\frac{1995856}{11025}+\frac{372800}{441} \nu \right)\log\left[ \frac{1 + \sqrt{1 - e^2}}{8 y \left(1 - e^2 \right)^{3/2}} \right] \, , \label{eq:aNSQC:8} \\
    a_9^\mathrm{NS,QC} =& \pi \Bigg\{ \frac{343801320119}{116424000} - \frac{219136}{525} \gamma_E + \left(-\frac{516333533}{83160}+\frac{3608}{15} \pi^2 \right) \nu - \frac{6821669}{6480} \nu^2 - \frac{24107249}{41580} \nu^3 \nonumber\\
    & + \frac{219136}{525} \log\left[ \frac{1 + \sqrt{1 - e^2}}{8 y \left(1 - e^2 \right)^{3/2}} \right] \Bigg\} \, , \label{eq:aNSQC:9}
\end{align}
\end{subequations}

\noindent 4PN spin-orbit corrections~\cite{Cho:2022syn}

\begin{subequations}
\label{eq:aSOQC}
\begin{align}
    a_7^\mathrm{SO,QC} =& \left(-\frac{4323559}{5670} + \frac{87341}{42} \nu - \frac{17840}{27} \nu^2 \right) \chi_\mathrm{eff} + \left(-\frac{1932041}{5670}+\frac{40289}{90} \nu - \frac{10819}{135} \nu^2 \right) \delta \mu  \delta \chi \, , \label{eq:aSOQC:7} \\
    a_8^\mathrm{SO,QC} =& \pi \left[ \left(-\frac{30542}{45} + \frac{26536}{15} \nu \right) \chi_\mathrm{eff} + \left(-\frac{93914}{315} + \frac{34303}{105} \nu \right) \delta \mu \delta \chi \right] \, , \label{eq:aSOQC:8}
\end{align}
\end{subequations}

\noindent 4PN aligned spin-spin corrections~\cite{Cho:2022syn}

\begin{subequations}
\label{eq:aSSQC}
\begin{align}
    a_7^\mathrm{SS,QC} =& \pi  \left[(128+32 \delta q_S) \chi_\mathrm{eff}^2+64 \delta q_A \chi_\mathrm{eff} \delta \chi +\left(\frac{4}{5} + 32 \delta q_S\right) \delta \chi^2 \right] \, , \label{eq:aSSQC:7} \\
    a_8^\mathrm{SS,QC} =& \Bigg\{\frac{14931877}{5670} + \frac{1931813}{11340} \delta q_S +\left(-\frac{45809}{15} - \frac{17305}{84} \delta q_S\right) \nu + \left(\frac{12536}{45} + \frac{3134}{45} \delta q_S\right) \nu^2 \nonumber\\
    & + \left(\frac{81919}{1260} - \frac{1327}{20} \nu \right) \delta \mu  \delta q_A\Bigg\} \chi_\mathrm{eff}^2 +\Bigg\{\left(\frac{1931813}{5670} - \frac{17305}{42} \nu +\frac{6268}{45} \nu^2\right) \delta q_A \nonumber\\
    & + \left[\frac{1597856}{945} + \frac{81919}{630} \delta q_S + \left(-\frac{19949}{15}-\frac{1327}{10} \delta q_S \right) \nu \right] \delta \mu \Bigg\} \chi_\mathrm{eff}  \delta \chi + \Bigg\{\frac{144727}{540} + \frac{1931813}{11340} \delta q_S \nonumber\\
    & + \left(-\frac{765353}{756} - \frac{17305}{84} \delta q_S \right) \nu +\left(\frac{71707}{180} + \frac{3134}{45} \delta q_S\right) \nu^2+\left(\frac{81919}{1260} - \frac{1327}{20} \nu \right) \delta \mu  \delta q_A\Bigg\} \delta \chi^2 \, , \label{eq:aSSQC:8}
\end{align}
\end{subequations}

\noindent 3.5PN aligned cubic-in-spin corrections~\cite{Marsat:2014xea}

\begin{align}
    a_7^\mathrm{SSS,QC} =& \left[-\frac{4688}{15}-\frac{1436}{15} \delta q_S - \frac{88}{5} \delta o_S \right] \chi_\mathrm{eff}^3 + \left[\left(-\frac{1496}{15} - \frac{374}{15} \delta q_S \right) \delta \mu - \frac{2078}{15}  \delta q_A -\frac{264}{5} \delta o_A \right] \chi_\mathrm{eff}^2 \delta \chi \nonumber\\
    & + \left[-\frac{43}{15} + \frac{152}{15} \delta q_S - \frac{748}{15} \delta q_A \delta \mu  - \frac{264}{5} \delta o_S \right] \chi_\mathrm{eff}  \delta \chi^2+\left[\left(-\frac{3}{5}-\frac{374}{15} \delta q_S \right) \delta \mu +\frac{794}{15} \delta q_A - \frac{88}{5} \delta o_A \right] \delta \chi^3 \, , \label{eq:aSSSQC:7} 
\end{align}

\noindent 7.5PN non-spinning quasi-circular adiabatic tidal corrections~\cite{Dones:2024odv}

\begin{subequations}
\label{eq:aTQC}
\begin{align}
    a_{10}^\mathrm{T,QC} =& \frac{144}{5} \Lambda^{(2)}_{A,\mathrm{av}} + \frac{12}{5} \Lambda^{(2)}_{B,\mathrm{av}} \, , \label{eq:aTQC:10} \\
    a_{12}^\mathrm{T,QC} =& \left(\frac{4421}{140} - \frac{571}{10} \nu \right) \Lambda^{(2)}_{A,\mathrm{av}}+\left(\frac{38}{35} + \frac{751}{10} \nu \right) \Lambda^{(2)}_{B,\mathrm{av}} + \frac{4148}{15}  \Sigma^{(2)}_{A,\mathrm{av}} -\frac{4}{15} \Sigma^{(2)}_{B,\mathrm{av}} \, , \label{eq:aTQC:12} \\
    a_{13}^\mathrm{T,QC} =& \pi  \left(\frac{576}{5} \Lambda^{(2)}_{A,\mathrm{av}} + \frac{48}{5} \Lambda^{(2)}_{B,\mathrm{av}} \right) \, , \label{eq:aTQC:13} \\
    a_{14}^\mathrm{T,QC} =& \left(\frac{6993499}{15120} + \frac{18677}{105} \nu - \frac{1239}{10} \nu^2\right) \Lambda^{(2)}_{A,\mathrm{av}}+\left(\frac{21011}{630} + \frac{378769}{560} \nu - \frac{5231}{12} \nu^2 \right) \Lambda^{(2)}_{B,\mathrm{av}} + 60 \Lambda^{(3)}_{A,\mathrm{av}} \nonumber \\
    & + \left(\frac{74966}{105} - \frac{11854}{15} \nu \right) \Sigma^{(2)}_{A,\mathrm{av}}+\left(-\frac{299}{315} + \frac{15506}{15} \nu \right) \Sigma^{(2)}_{B,\mathrm{av}} \, , \label{eq:aTQC:14} \\
    a_{15}^\mathrm{T,QC} =& \pi  \left\{\left(\frac{62199}{280} - \frac{12469}{20} \nu \right) \Lambda^{(2)}_{A,\mathrm{av}} + \left(\frac{152}{35} + \frac{5139}{20} \nu \right) \Lambda^{(2)}_{B,\mathrm{av}} + \frac{5528}{5} \Sigma^{(2)}_{A,\mathrm{av}} - \frac{8}{15} \Sigma^{(2)}_{B,\mathrm{av}}\right\} \, , \label{eq:aTQC:15}
\end{align}
\end{subequations}

\noindent and 6.5PN quasi-circular adiabatic spin-tidal corrections~\cite{Abdelsalhin:2018reg,Castro:2022mpw}

\begin{align}
    a_{13}^\mathrm{ST,QC} =& \left(-\frac{1292}{5} \chi - \frac{456}{5} \delta \mu  \delta \chi \right) \Lambda^{(2)}_{A,\mathrm{av}}+\frac{2751}{20} \delta \chi  \Lambda^{(2)}_{A,\mathrm{diff}} + \left( - \frac{64}{5} \chi + \frac{43}{5} \delta\mu \delta\chi \right) \Lambda^{(2)}_{B,\mathrm{av}} - \frac{51}{5} \delta \chi  \Lambda^{(2)}_{B,\mathrm{diff}} - \frac{986}{3} \chi  \Sigma^{(2)}_{A,\mathrm{av}} \nonumber \\
    & + \frac{4936}{15} \delta \chi  \Sigma^{(2)}_{A,\mathrm{diff}} + \frac{2}{15} \chi  \Sigma^{(2)}_{B,\mathrm{av}} + \frac{4}{15} \delta \chi  \Sigma^{(2)}_{B,\mathrm{diff}} - \frac{856}{5} \big(\chi \Lambda^{(23)}_{A,\mathrm{av}}+\delta \chi  \Lambda^{(23)}_{A,\mathrm{diff}}\big)+\frac{272}{5} \big(\chi \Lambda^{(32)}_{A,\mathrm{av}}+\delta \chi  \Lambda^{(32)}_{A,\mathrm{diff}}\big)  \nonumber \\
    & + \frac{833}{15} \big(\chi  \Sigma^{(23)}_{A,\mathrm{av}} + \delta \chi  \Sigma^{(23)}_{A,\mathrm{diff}}\big)-\frac{204}{5} \big(\chi  \Sigma^{(32)}_{A,\mathrm{av}}+\delta \chi  \Sigma^{(32)}_{A,\mathrm{diff}}\big)-8 \big(\chi  \Lambda^{(23)}_{B,\mathrm{av}}+\delta \chi  \Lambda^{(23)}_{B,\mathrm{diff}}\big)-\frac{1}{15} \big(\chi  \Sigma^{(23)}_{B,\mathrm{av}}+\delta \chi  \Sigma^{(23)}_{B,\mathrm{diff}}\big) \, , \label{eq:aSTQC:13} 
\end{align}

\noindent where $\gamma_E = 0.577\ldots$ is the Euler-Mascheroni constant, and, extending the notation of Ref.~\cite{Morras:2025nlp}, we have defined

\begin{subequations}
\label{eq:SymQOParams}
\begin{align}
    \delta q_S =& q_1 + q_2 - 2\, , \label{eq:SymQOParams:dqS} \\
    \delta q_A =& q_1 - q_2\, , \label{eq:SymQOParams:dqA} \\
    \delta o_S =& o_1 + o_2 - 2\, , \label{eq:SymQOParams:doS} \\
    \delta o_A =& o_1 - o_2\, , \label{eq:SymQOParams:doA}
\end{align}
\end{subequations}

\noindent with $q_i$ and $o_i$ the spin quadrupole and octupole parameters of binary component $i = \{1, 2\}$, respectively (sometimes denoted $\kappa_i$ and $\lambda_i$ in the literature). For black holes, $q_i = o_i = 1$. To simplify tidal terms we have defined

\begin{subequations}
\begin{align}
    K^{(l)}_{A,\mathrm{av}}   =& 2^{2 l} \frac{m_1^{2 l} m_2 K^{(l)}_1 + m_2^{2 l} m_1 K^{(l)}_2}{(m_1 + m_2)^{2 l + 1}}  \, , &
    K^{(l)}_{A,\mathrm{diff}} =& 2^{2 l} \frac{m_1^{2 l} m_2 K^{(l)}_1 - m_2^{2 l} m_1 K^{(l)}_2}{(m_1 + m_2)^{2 l + 1}} \, , \\
    K^{(l)}_{B, \mathrm{av}}   =& 2^{2 l} \frac{m_1^{2 l + 1} K^{(l)}_1 + m_2^{2 l + 1} K^{(l)}_2}{(m_1 + m_2)^{2 l + 1}}  \, , &
    K^{(l)}_{B, \mathrm{diff}} =& 2^{2 l} \frac{m_1^{2 l + 1} K^{(l)}_1 - m_2^{2 l + 1} K^{(l)}_2}{(m_1 + m_2)^{2 l + 1}}  \, , \\
    K^{(l l')}_{A,\mathrm{av}}   =& 2^{l + l' -1} \frac{m_1^{l + l' -1} m_2 K^{(l l')}_1 + m_2^{l + l' -1} m_1 K^{(l l')}_2}{(m_1 + m_2)^{l + l'}}  \, , &
    K^{(l l')}_{A,\mathrm{diff}} =& 2^{l + l' -1} \frac{m_1^{l + l' -1} m_2 K^{(l l')}_1 - m_2^{l + l' -1} m_1 K^{(l l')}_2}{(m_1 + m_2)^{l + l'}} \, , \\
    K^{(l l')}_{B, \mathrm{av}}   =& 2^{l + l' -1} \frac{m_1^{l + l'} K^{(l l')}_1 + m_2^{l + l'} K^{(l l')}_2}{(m_1 + m_2)^{l + l'}}  \, , &
    K^{(l l')}_{B, \mathrm{diff}} =& 2^{l + l' -1} \frac{m_1^{l + l'} K^{(l l')}_1 - m_2^{l + l'} K^{(l l')}_2}{(m_1 + m_2)^{l + l'}}  \, ,    
\end{align}    
\end{subequations}

\noindent where $K^{(\bm{l})}_i$ represent the different dimensionless tidal deformabilities of body $i$, which in our case include the mass quadrupole ($\Lambda^{(2)}_i$), mass octupole ($\Lambda^{(3)}_i$), and current quadrupole ($\Sigma^{(2)}_i$). In addition, we account for the spin-induced couplings where a magnetic quadrupolar (octupolar) tidal moment induces a mass octupole (quadrupole) through the deformabilities $\Lambda^{(32)}_i$ ($\Lambda^{(23)}_i$), and an electric quadrupolar (octupolar) tidal moment induces a current octupole (quadrupole) through $\Sigma^{(32)}_i$ ($\Sigma^{(23)}_i$). We define tidal deformabilities in terms of the dimensionless tidal Love numbers~\cite{Hinderer:2007mb}

\begin{subequations}
\begin{align}
    \Lambda^{(l)}_i    &= \frac{2}{(2 l - 1)!!} k_i^{(l)} \left( \frac{c^2 R_i}{G m_i} \right)^{2 l + 1} \, , \\
    \Sigma^{(l)}_i     &= \frac{l - 1}{4 (l+1) (2 l - 1)!!} j_i^{(l)} \left( \frac{c^2 R_i}{G m_i} \right)^{2 l + 1} \, , \\
    \Lambda^{(l l')}_i &=  k_i^{(l l')} \left( \frac{c^2 R_i}{G m_i} \right)^{l + l' - 1} \, , \\
    \Sigma^{(l l')}_i  &=  j_i^{(l l')} \left( \frac{c^2 R_i}{G m_i} \right)^{l + l' - 1} \, ,
\end{align}
\end{subequations}

\noindent where $R_i$ is the radius of body $i$, and all deformabilities vanish for black holes.

While the non-spinning and aligned-spin corrections to the evolution of $y$ listed above are expected to be exact at $O(e^0)$ in eccentricity, the same is not true when the objects in the binary have spin components perpendicular to the orbital angular momentum, $\bm{s}_{i \perp}$. 
The reason is that such components are expected to introduce quasi-circular $\ord{e^0}$ corrections to the eccentricity evolution $\D e^2$, modifying the residual eccentricity of Eq.~\eqref{eq:emin_spin} and leading to additional $O(e^0)$ corrections to the evolution of $y$. 
Based on 2PN results~\cite{Klein:2018ybm}, the corrections to $\D y$ arising from $\D e^2$ are expected to be much smaller than the pure QC contributions. Nonetheless, we use a tilde ( $\tilde{}$ ) on the coefficients below to indicate that they are likely incomplete. From 3PN to 4PN, we have~\cite{Khalil:2023kep}

\begin{subequations}
\label{eq:aSSpQC}
\begin{align}
    \tilde{a}^\mathrm{SS\perp,QC}_6 =& \left[\frac{3251}{42}+\frac{6373}{90} \nu \right] s_\perp^2 + \left[\frac{62}{15} - \frac{2539}{105} \delta q_S + \left(-\frac{181}{45} + \frac{172}{5} \delta q_S\right) \nu -\frac{443}{30} \delta q_A \delta \mu \right] (s_{1\perp}^2 + s_{2\perp}^2) \nonumber\\
    & + \left[\left(-\frac{599}{15} - \frac{443}{30} \delta q_S \right) \delta \mu + \left(-\frac{2539}{105} + \frac{172}{5} \nu \right) \delta q_A\right] (s_{1\perp}^2 - s_{2\perp}^2) \, , \label{eq:aSSpQC:6} \\
    \tilde{a}^\mathrm{SS\perp,QC}_7 =& -\frac{96}{5} \pi  \Big[ 2 s_\perp^2+\delta q_S (s_{1\perp}^2 + s_{2\perp}^2)+\delta q_A (s_{1\perp}^2 - s_{2\perp}^2) \Big] \, , \label{eq:aSSpQC:7} \\
    \tilde{a}^\mathrm{SS\perp,QC}_8 =& \bigg\{-\frac{9355721}{22680} - \frac{195697}{280} \nu - \frac{162541}{1080} \nu^2 \bigg\} s_\perp^2 + \bigg\{-\frac{563}{60} - \frac{1931813}{11340} \delta q_S +\left(\frac{13427}{420} + \frac{17305}{84} \delta q_S \right) \nu \nonumber \\ 
    & + \left(\frac{12109}{540} - \frac{3134}{45} \delta q_S\right) \nu^2+\left(-\frac{81919}{1260} + \frac{1327}{20} \nu \right) \delta q_A \delta \mu \bigg\} (s_{1\perp}^2 + s_{2\perp}^2) \nonumber\\
    & + \bigg\{\left[-\frac{6302}{105}-\frac{81919}{1260} \delta q_S + \left(\frac{1801}{10}+\frac{1327}{20} \delta q_S \right) \nu \right] \delta \mu +\left[-\frac{1931813}{11340}+\frac{17305}{84} \nu -\frac{3134}{45} \nu^2\right] \delta q_A \bigg\} (s_{1\perp}^2 - s_{2\perp}^2) \, , \label{eq:aSSpQC:8}
\end{align}
\end{subequations}

\noindent where $s_\perp^2 = \left\Vert \bm{s}_{1\perp} + \bm{s}_{2\perp}  \right\Vert^2$ and we note that the 3.5PN term ($\tilde{a}^\mathrm{SS\perp,QC}_7$) includes only partial information about the fully spinning case~\cite{Khalil:2023kep}. 

When computing the precession average of these coefficients using the MSA~\cite{Morras:2025nlp}, it is convenient to express the quantities appearing in Eq.~\eqref{eq:aSSpQC} as

\begin{subequations}
\label{eq:sp2Ssp2A_MSA}
\begin{align}
    s_{1\perp}^2 + s_{2\perp}^2 =& s_1^2 + s_2^2 - \frac{\chi_\mathrm{eff}^2  + \delta\chi^2}{2} \, , \label{eq:sp2Ssp2A_MSA:S} \\
    s_{1\perp}^2 - s_{2\perp}^2 =& s_1^2 - s_2^2 - \chi_\mathrm{eff} \delta\chi \, . \label{eq:sp2Ssp2A_MSA:A}
\end{align}
\end{subequations}

\section{Explicit post-Newtonian expressions for the precession equations}
\label{sec:appendix:PN_prec_expressions}

In Sec.~\ref{sec:HighPN_MSA} we kept the discussion independent of the explicit post-Newtonian form of the precession equation, to make it valid to when higher order eccentric corrections are derived, or when using different spin supplementary conditions. Nonetheless, in this appendix we list the explicit expressions that have been used in \pyEFPEHM. In particular, we use the orbit average spin-precession equations, with eccentric effects up to 2PN and containing only the quasi-circular part up to 4PN. As in Appendix~\ref{sec:appendix:PN_QC}, we work in harmonic coordinates with the Newton-Wigner spin supplementary condition, and we apply Eq.~\eqref{eq:QC_to_Ecc} to the quasi-circular expressions to better represent the expected shape of the eccentric ones. 

We obtain the 4PN precession frequencies of Eq.~\eqref{eq:raw_prec_eqs} from Ref.~\cite{Khalil:2023kep}, removing the terms that would change the norm of $\uvec{l}_N$ as described in Eq.~\eqref{eq:Dln_of_Dlnraw}. That is

\begin{subequations}
\label{eq:prec_freqs}
\begin{align}
    \bm{\Omega}_{l_N} =& \Bigg\{\frac{7}{4}-\frac{3 y \chi_\mathrm{eff}}{2}+y^2 \left(-\frac{9}{8}-\frac{19 \nu }{8}\right)+y^3 \left[\left(\frac{203}{32}+\frac{13 \nu }{8}\right) \chi_\mathrm{eff}-\frac{31 \delta \mu  \delta \chi }{32}\right] + y^4 \left(-\frac{27}{32}-\frac{207 \nu }{32}+\frac{127 \nu ^2}{96}\right) \nonumber\\
    & + y^5 \left[\left(-\frac{255}{128}-\frac{2281 \nu }{384}-\frac{61 \nu ^2}{96}\right) \chi_\mathrm{eff}+\left(\frac{195}{128}+\frac{911 \nu }{384}\right) \delta \mu  \delta \chi \right]\Bigg\} (\bm{s}_1 + \bm{s}_2)  \nonumber \\
    & + \Bigg\{ \frac{\delta \mu }{4}+y^2 \delta \mu  \left(\frac{9}{8}-\frac{\nu }{8}\right)+y^3 \left[-\frac{49}{32}  \delta \mu  \chi_\mathrm{eff}+\left(\frac{5}{32}+\frac{3 \nu }{8}\right) \delta \chi \right]+y^4 \delta \mu  \left(\frac{27}{32}-\frac{81 \nu }{32}+\frac{\nu ^2}{96}\right) \nonumber\\
    & + y^5 \left[\left(\frac{177}{128}+\frac{1121 \nu }{384}\right) \delta\mu  \chi_\mathrm{eff} + \left(-\frac{117}{128}+\frac{545 \nu }{384}-\frac{77 \nu ^2}{96}\right) \delta \chi \right] \Bigg\} (\bm{s}_1 - \bm{s}_2) \nonumber \\
    & + \nu y^5 \left\{ \frac{28}{5} (\bm{s}_1 + \bm{s}_2)\times\uvec{l}_N  + \frac{4}{5} \delta \mu (\bm{s}_1 - \bm{s}_2)\times\uvec{l}_N \right\}\, , \label{eq:prec_freqs:lN} \\
    \bm{\Omega}_1 = &  \Bigg\{ \frac{7}{4} + \frac{\delta \mu }{4}-\frac{3 y \chi_\mathrm{eff}}{2} + y^2 \left[\frac{33}{16}-\frac{61 \nu }{48}+\delta \mu  \left(\frac{15}{16}-\frac{\nu }{48}\right)\right]+ y^3 \bigg[\left(-\frac{5}{12}+\frac{3 \nu }{4}\right) \chi_\mathrm{eff}+\left(\frac{1}{12}+\frac{2 \nu }{3}\right) \delta \chi \nonumber\\
    & -\frac{31}{24} \delta \mu  (\chi_\mathrm{eff}+\delta \chi )\bigg] + y^4 \left[\frac{135}{32}-\frac{367 \nu }{32}+\frac{29 \nu ^2}{96}+\delta \mu  \left(\frac{81}{32}-\frac{55 \nu }{32}-\frac{\nu ^2}{96}\right)\right] \nonumber\\
    & + y^5 \left[\left(-\frac{103}{32}+\frac{937 \nu }{144}-\frac{\nu ^2}{16}\right) \chi_\mathrm{eff}+\left(-\frac{31}{32}+\frac{65 \nu }{18}-\frac{7 \nu ^2}{9}\right) \delta \chi +\left(-\frac{47}{16}+\frac{319 \nu }{144}\right) \delta \mu  (\chi_\mathrm{eff}+\delta \chi )\right] \Bigg\} \uvec{l}_N \nonumber \\
    & + y \Bigg\{\frac{1}{2}-\frac{\nu  y^2}{4}+y^4 \left(\frac{3}{8}+\frac{49 \nu }{16}+\frac{\nu ^2}{48}\right)\Bigg\}\bm{s}_2
    \, , \label{eq:prec_freqs:s1} 
\end{align}    
\end{subequations}

\noindent where $\bm{\Omega}_2$ can be obtained from $\bm{\Omega}_1$ by exchanging $1 \leftrightarrow 2$, that is $\bm{s}_1 \leftrightarrow \bm{s}_2$, $\delta\mu \leftrightarrow -\delta\mu$ and $\delta\chi \leftrightarrow -\delta\chi$.

The coefficients $a_J$, $b_J$ and $c_J$ appearing in Eq.~\eqref{eq:J_def} for the total PN angular momentum, are given by~\cite{Khalil:2023kep}

\begin{subequations}
\label{eq:J_vec_coefs}
\begin{align}
    a_J = & \frac{\nu}{y}  \Bigg\{ 1 + y^2 \Bigg[\frac{3}{2}+\frac{\nu }{6}+y \left(-\frac{49}{24} \chi_\mathrm{eff}-\frac{7 \delta \mu  \delta \chi }{24}-y \bm{s}_1 \cdot \bm{s}_2\right)+y^2 \left(\frac{27}{8}-\frac{19 \nu }{8}+\frac{\nu ^2}{24}+\chi_\mathrm{eff}^2-\frac{s_1^2+s_2^2}{2}\right) \nonumber\\
    & + y^3 \left(\left(-\frac{121}{32}+\frac{671 \nu }{288}\right) \chi_\mathrm{eff}+\left(-\frac{55}{32}+\frac{11 \nu }{288}\right) \delta \mu  \delta \chi +\frac{2 \nu}{3} y \bm{s}_1 \cdot \bm{s}_2\right)\Bigg] \Bigg\} \, , \label{eq:J_vec_coefs:aJ} \\
    b_J = & \frac{1}{2} \Bigg\{ 1-\nu  y^2 \Bigg[\frac{7}{4}-y \chi_\mathrm{eff}+y^2 \left(\frac{33}{16}-\frac{61 \nu }{48}\right)+y^3 \left(\left(-\frac{119}{48}+\frac{\nu }{2}\right) \chi_\mathrm{eff}+\frac{7 \delta \mu  \delta \chi }{48}\right)\Bigg] \Bigg\} \, , \label{eq:J_vec_coefs:bJ} \\
    c_J = & \frac{1}{2} \delta \mu  \Bigg\{1 - \nu  y^2 \Bigg[\frac{1}{4}+y^2 \left(\frac{15}{16}-\frac{\nu }{48}\right)+y^3 \left(\frac{7 \chi_\mathrm{eff}}{48}+\frac{\delta \mu  \delta \chi }{48}\right)\Bigg]\Bigg\} \, . \label{eq:J_vec_coefs:cJ}
\end{align}
\end{subequations}

The coefficients $A_\chi$, $B_\chi$, $A_{\delta\chi}$ and $B_{\delta\chi}$ that appear in Eq.~\eqref{eq:eff_spins_prec_eqs} for $\D\chi$ and $\D\delta\chi$ are given by

\begin{subequations}
\label{eq:eff_spins_prec_eqs_coefs}
\begin{align}
    A_\chi = & -\frac{3}{4} \delta \mu + y \left[\frac{49 \delta \mu  \chi_\mathrm{eff}}{48}+\left(-\frac{5}{48}-\frac{\nu }{4}\right) \delta \chi \right] + y^2 \left(-\frac{7}{16}+\frac{65 \nu }{24}\right) \delta \mu \nonumber\\
    & +y^3 \left[\left(-\frac{59}{64}-\frac{1121 \nu }{576}\right) \delta \mu  \chi_\mathrm{eff}+\left(\frac{39}{64}-\frac{545 \nu }{576}+\frac{77 \nu ^2}{144}\right) \delta \chi \right] \, , \label{eq:eff_spins_prec_eqs_coefs:Achi} \\
    B_\chi = & -\frac{28}{5} \left(s_1^2+s_2^2+2 \bm{s}_1 \cdot \bm{s}_2-\chi_\mathrm{eff}^2\right)-\frac{4}{5} \delta \mu  (s_1^2-s_2^2-\chi_\mathrm{eff} \delta \chi ) \, , \label{eq:eff_spins_prec_eqs_coefs:Bchi} \\
    A_{\delta\chi} = & 1-y \chi_\mathrm{eff}+y^2 \left(-\frac{3}{4}-\frac{3 \nu }{2}\right)+y^3 \left[\left(\frac{203}{48}+\frac{13 \nu }{12}\right) \chi_\mathrm{eff}-\frac{31 \delta \mu  \delta \chi }{48}\right]+y^4 \left(-\frac{11}{16}-\frac{16 \nu }{3}+\frac{7 \nu ^2}{8}\right) \nonumber \\
    & + y^5 \left[\left(-\frac{85}{64}-\frac{2281 \nu }{576}-\frac{61 \nu ^2}{144}\right) \chi_\mathrm{eff}+\left(\frac{65}{64}+\frac{911 \nu }{576}\right) \delta \mu  \delta \chi \right] \, , \label{eq:eff_spins_prec_eqs_coefs:Adchi} \\
    B_{\delta\chi} = & - \frac{4}{5} \delta \mu  \left(s_1^2+s_2^2-2 \bm{s}_1 \cdot \bm{s}_2-\delta \chi ^2\right)-\frac{28}{5} (s_1^2 - s_2^2 - \chi_\mathrm{eff} \delta \chi) \, . \label{eq:eff_spins_prec_eqs_coefs:Bdchi}
\end{align}
\end{subequations}

Finally, the coefficients $\kappa_{\D \Delta_{J^2}}$ and $\delta_{\D \Delta_{J^2}}$ appearing in Eq.~\eqref{eq:DDJ2_SP} for $\D \Delta_{J^2}$ are given by

\begin{subequations}
\label{eq:coefs_J2_PN}
\begin{align}
    \kappa_{\D \Delta_{J^2}} = & 1 + y^2 \left(- \frac{9}{4} - \frac{\nu}{6} \right) +  y^3  \left(\frac{61 \chi _\mathrm{eff}}{12}+\frac{7 \delta \mu  \delta \chi }{12}\right) +y^4 \left\{-\frac{135}{8}+\frac{27 \nu }{8}-\frac{7 \nu ^2}{16}+\frac{3}{2} \left[(\bm{s}_1 + \bm{s}_2)^2 - 2 \chi_\mathrm{eff}^2\right]\right\}\, , \label{eq:coefs_J2_PN:kappa} \\
    \delta_{\D \Delta_{J^2}} = & \frac{1}{6} \nu \delta\mu \, . \label{eq:coefs_J2_PN:delta}
\end{align}    
\end{subequations}

\section{Extra parameter estimation information and results}
\label{sec:appendix:PE_extra}

All PE analyses in this paper are performed using \texttt{bilby}~\cite{Ashton:2018jfp,Smith:2019ucc,Romero-Shaw:2020owr}. Sampling is carried out with \texttt{bilby}'s implementation of the \texttt{dynesty} nested sampler~\cite{Speagle:2020dqf}, using the \texttt{acceptance-walk} sampling scheme with three parallel runs, $n_\mathrm{live}=1000$, and $n_\mathrm{accept}=60$.

Quasi-circular injections are specified by 15 parameters: the detector-frame component masses $m_1$ and $m_2$; the dimensionless spin magnitudes $a_1$ and $a_2$; the tilt angles between each spin and the orbital angular momentum $\theta_1$ and $\theta_2$; the azimuthal angle between the spin vectors $\phi_{12}$; the azimuthal angle between the total and orbital angular momenta $\phi_{JL}$; the angle between the total angular momentum and the line of sight $\theta_{JN}$; the reference orbital phase $\phi_\mathrm{ref}=\lambda_0$; the luminosity distance $d_L$; the right ascension $\mathrm{ra}$; the declination $\mathrm{dec}$; the polarization angle $\psi$; and the geocentric coalescence time $t_c$. For eccentric injections, two additional parameters are introduced: the initial eccentricity $e_0$ and the initial mean anomaly $\ell_0$.

Table~\ref{tab:injected_params} lists the parameters of the two injections analyzed in Sec.~\ref{sec:validate:PE}, together with the signal-to-noise ratio (SNR) in each detector and the corresponding network SNR. Table~\ref{tab:priors} lists the prior distributions used in \texttt{bilby} for all PE analyses in this paper. Table~\ref{tab:runtimes} summarizes the computational cost of each PE analysis, including the total run duration, number of likelihood evaluations, and wall-clock time per likelihood evaluation.

\begin{figure}[t!]
\centering  
\includegraphics[width=\textwidth]{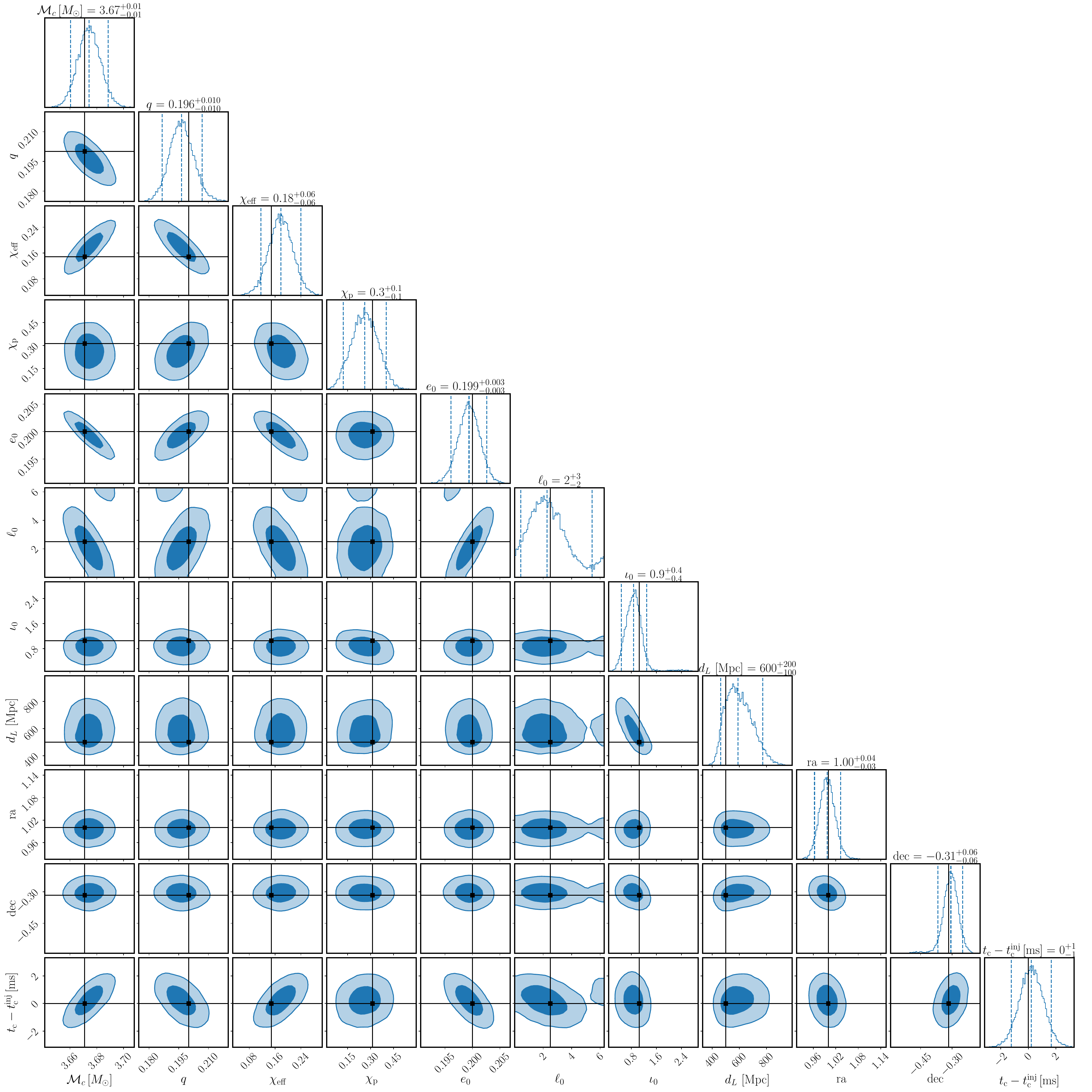}
\caption{\justifying Corner plot showing the joint posterior distributions of selected parameters from the \pyEFPEHM injection--recovery study, including the chirp mass $\mathcal{M}_c$, mass ratio $q$, effective inspiral spin parameter $\chi_\mathrm{eff}$, effective precession spin parameter $\chi_\mathrm{p}$, initial eccentricity $e_0$, mean anomaly $\ell_0$, inclination $\iota_0$, luminosity distance $d_L$, right ascension $\mathrm{ra}$, declination $\mathrm{dec}$ and difference between measured and injected coalescence times $t_c - t_c^\mathrm{inj}$. The diagonal panels show the marginal distributions for each parameter, together with the median and 90\% credible intervals. The off-diagonal panels display the bivariate correlations between parameter pairs, with contours indicating the $50\%$ and $90\%$ credible regions. The black lines denote the injected parameter values.}
\label{fig:inj_EFPE_rec_EFPE_all_important}
\end{figure}

\begin{figure}[t!]
\centering  
\includegraphics[width=\textwidth]{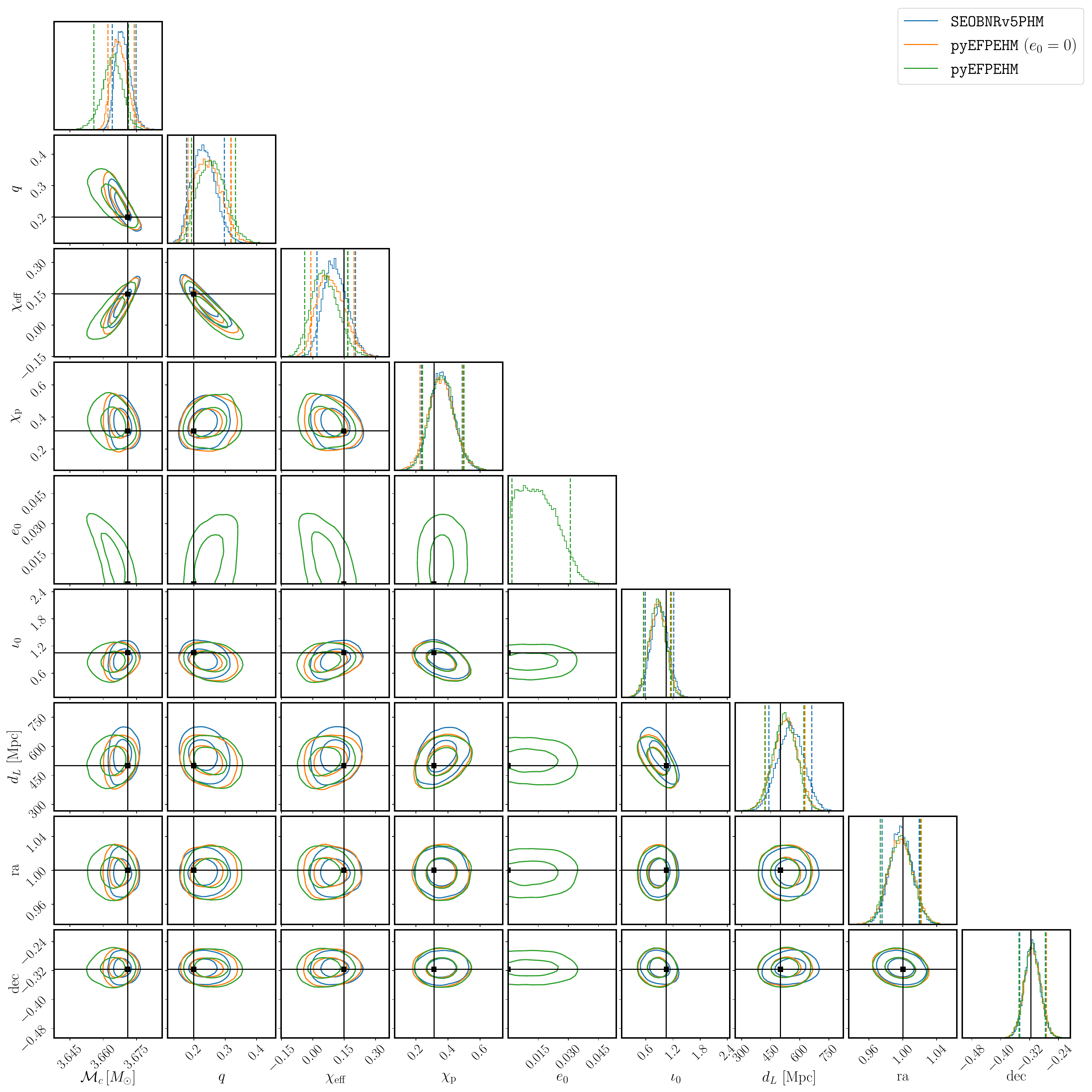}
\caption{\justifying Corner plot showing the joint posterior distributions of selected parameters for the \vfivePHM injection, recovered using \vfivePHM, \pyEFPEHM with the eccentricity fixed to zero (labeled \pyEFPEHM~$(e_0 = 0)$), and \pyEFPEHM with a uniform prior on the initial eccentricity $e_0 \in [0, 0.5]$ (labeled \pyEFPEHM). The parameters shown are the chirp mass $\mathcal{M}_c$, mass ratio $q$, effective inspiral spin parameter $\chi_\mathrm{eff}$, effective precession spin parameter $\chi_\mathrm{p}$, initial eccentricity $e_0$, inclination $\iota_0$, luminosity distance $d_L$, right ascension $\mathrm{ra}$ and declination $\mathrm{dec}$. The diagonal panels display the marginal distributions for each parameter, while the off-diagonal panels show the bivariate correlations, with contours indicating the $50\%$ and $90\%$ credible regions. The black lines denote the injected parameter values.}
\label{fig:inj_v5PHM_32s_all_important}
\end{figure}

Finally, in Figs.~\ref{fig:inj_EFPE_rec_EFPE_all_important} and \ref{fig:inj_v5PHM_32s_all_important} we present the corner plots of the posterior distributions for the most relevant parameters obtained from the \pyEFPEHM and \vfivePHM parameter estimation analyses, respectively.

\begin{table}[h]
    \centering
    \begin{tabular}{c|c|c}
         & \pyEFPEHM injection & \vfivePHM injection \\
         \hline
         $m_1 \, [M_\odot]$ & 10 & 10 \\
         $m_2 \, [M_\odot]$ & 2 & 2 \\
         $a_1$ & 0.4 & 0.4 \\
         $a_2$ & 0.6 & 0.6 \\
         $\theta_1 \, [\mathrm{rad}]$ & 0.9 & 0.9 \\
         $\theta_2 \, [\mathrm{rad}]$ & 2.2 & 2.2 \\
         $\phi_{12} \, [\mathrm{rad}]$ & 3.0 & 3.0 \\
         $\phi_{JL} \, [\mathrm{rad}]$ & 3.3 & 3.3 \\
         $\theta_{JN} \, [\mathrm{rad}]$ & 1.0 & 1.0 \\
         $d_L  \, [\mathrm{Mpc}]$ & 500 & 500 \\
         $\phi_\mathrm{ref} \, [\mathrm{rad}]$ & 0.9 & 0.9 \\
         $\mathrm{ra} \, [\mathrm{rad}]$ & 1.0 & 1.0 \\
         $\mathrm{dec} \, [\mathrm{rad}]$ & -0.316 & -0.316 \\
         $\psi \, [\mathrm{rad}]$ & 0.6 & 0.6 \\
         $t_c^\mathrm{GPS}\, [\mathrm{s}]$ & 1262276684 & 1262276684 \\
         $e_0$ & 0.2 & -- \\
         $\ell_0 \, [\mathrm{rad}]$ & 2.5 & -- \\
         \hline
         H1 SNR & 12.38 & 15.17 \\
         L1 SNR & 12.78 & 15.51 \\
         V1 SNR & 4.73 & 6.10 \\
         Network SNR & 18.4 & 22.5 \\
         \hline
    \end{tabular}
    \caption{Values of the parameters for the two injections analyzed in Sec.~\ref{sec:validate:PE}. We also list the SNR in each detector (H1, L1 and V1), computed with their projected O5 sensitivities~\cite{KAGRA:2013rdx,ObservingScenariosPSDs}, also used in the PEs of Sec.~\ref{sec:validate:PE}. Finally, we show the total network SNR, obtained by combining in quadrature the individual detector SNRs.}
    \label{tab:injected_params}
\end{table}

\begin{table}[h]
    \centering
    \begin{tabular}{c|c|c|c}
         Parameter & Distribution & Minimum & Maximum \\
         \hline
         $\mathcal{M}_c \, [M_\odot]$  & Uniform in components & 3.4      & 4.0      \\
         $q$                            & Uniform in components & 0.05     & 1        \\
         $a_1$                          & Uniform               & 0        & 0.9      \\
         $a_2$                          & Uniform               & 0        & 0.9      \\
         $\theta_1 \, [\mathrm{rad}]$   & Sine                  & 0        & $\pi$    \\
         $\theta_2 \, [\mathrm{rad}]$   & Sine                  & 0        & $\pi$    \\
         $\phi_{12} \, [\mathrm{rad}]$  & Uniform (periodic)    & 0        & $2\pi$   \\
         $\phi_{JL} \, [\mathrm{rad}]$  & Uniform (periodic)    & 0        & $2\pi$   \\
         $d_L \, [\mathrm{Mpc}]$        & Uniform source frame  & 10       & 5000     \\
         $\mathrm{ra} \, [\mathrm{rad}]$  & Uniform (periodic)  & 0        & $2\pi$   \\
         $\mathrm{dec} \, [\mathrm{rad}]$ & Cosine              & $-\pi/2$ & $\pi/2$  \\
         $\theta_{JN} \, [\mathrm{rad}]$  & Sine                & 0        & $\pi$    \\
         $\psi \, [\mathrm{rad}]$        & Uniform (periodic)   & 0        & $\pi$    \\
         $\phi_\mathrm{ref} \, [\mathrm{rad}]$ & Uniform (periodic) & 0   & $2\pi$   \\
         $e_0$                           & Uniform               & 0        & 0.5      \\
         $\ell_0 \, [\mathrm{rad}]$      & Uniform (periodic)    & 0        & $2\pi$   \\
         \hline
    \end{tabular}
    \caption{Prior distributions used in \texttt{bilby} for all parameter estimation analyses in this paper. The eccentricity $e_0$ and mean anomaly $\ell_0$ priors apply only to the eccentric analyses.}
    \label{tab:priors}
\end{table}

\begin{table}[h]
    \centering
    \begin{tabular}{c|c|c|c|c}
         injection & \multicolumn{1}{c|}{\pyEFPEHM} & \multicolumn{3}{c}{\vfivePHM} \\
         \hline
         recovery & \pyEFPEHM & \pyEFPEHM & \pyEFPEHM $(e_0=0)$ & \vfivePHM \\
         \hline
         Run duration                             & $3\,\mathrm{d} \; 20\,\mathrm{h}$ & $3\,\mathrm{d} \; 1\,\mathrm{h}$ & $2\,\mathrm{d} \; 21\,\mathrm{h}$ & $4\,\mathrm{d} \; 3\,\mathrm{h}$ \\
         Likelihood evaluations                   & $7.7 \times 10^7$ & $6.6 \times 10^7$ & $6.8 \times 10^7$ & $7.6 \times 10^7$ \\
         Thread-seconds per likelihood evaluation   & $0.28\,\mathrm{s}$ & $0.25\,\mathrm{s}$ & $0.23\,\mathrm{s}$ & $0.30\,\mathrm{s}$ \\
         \hline
    \end{tabular}
    \caption{Computational cost of the parameter estimation analyses described in Sec.~\ref{sec:validate:PE}. All analyses were performed using 64 threads on a single \textit{AMD EPYC 7513 CPU}.}
    \label{tab:runtimes}
\end{table}

\twocolumngrid

\FloatBarrier

\bibliography{Refs}

\end{document}